\documentclass[aps,prb,twocolumn,longbibliography,superscriptaddress,amsmath,amssymb,amsfonts,citeautoscript,floatfix]{revtex4-2}
\usepackage[utf8]{inputenc}
\usepackage[english]{babel}

\usepackage{graphicx}
\usepackage{dcolumn}
\usepackage{float}
\usepackage{amsmath}
\usepackage{amssymb}
\usepackage{bm}
\usepackage[urlcolor=blue,colorlinks=true,citecolor=red,linkcolor=red,pdfstartview={FitH},bookmarks=false]{hyperref}
\usepackage{xcolor}
\usepackage{listings}
\usepackage{enumitem}
\usepackage{tikz}
\usetikzlibrary{decorations.pathreplacing,calligraphy}
\usepackage[normalem]{ulem} 

\def\tA{{t_{\mathrm{A}}}}
\def\tB{{t_{\mathrm{B}}}}
\def\tS{{t_{\mathrm{S}}}}
\def\rhoS{{\rho_{\mathrm{S}}}}

\def\tint{{t_{\mathrm{int}}}}
\def\tintStar{{t_{\mathrm{int}}^*}}

\def\NS{{N_{\mathrm{S}}}}

\def\HQ{{H_{\mathrm{Q}}}}
\def\HS{{H_{\mathrm{S}}}}
\def\HT{{H_{\mathrm{T}}}}

\def\muS{{\mu_{\mathrm{S}}}}

\def\Ic{{I_{\mathrm{c}}}}
\def\vF{{v_{\mathrm{F}}}}

\def\thetac{{\theta_{\mathrm{c}}}}
\def\Emin{{E_{\mathrm{min}}}}

\def\DeltaQ{{\Delta_{\mathrm{Q}}}}

\DeclareMathOperator{\sgn}{sgn}

\newcommand{\newtext}[1]{{{#1}}}

\allowdisplaybreaks
\begin{document}
\title{Josephson effect in a Fibonacci quasicrystal}

\author{Anna Sandberg}
\affiliation{Department of Physics and Astronomy, Uppsala University, Box 516, S-751 20 Uppsala, Sweden}
\affiliation{Department of Physics, Stockholm University, AlbaNova University Center, SE-106 91 Stockholm, Sweden}

\author{Oladunjoye A. Awoga}
\affiliation{Solid State Physics and NanoLund,Lund University, Box 118, 22100 Lund, Sweden}

\author{Annica M. Black-Schaffer}
\affiliation{Department of Physics and Astronomy, Uppsala University, Box 516, S-751 20 Uppsala, Sweden}

\author{Patric Holmvall}
\email[e-mail:]{patric.holmvall@physics.uu.se}
\affiliation{Department of Physics and Astronomy, Uppsala University, Box 516, S-751 20 Uppsala, Sweden}

\date{\today}

\begin{abstract}
Quasiperiodicity has recently been proposed to enhance superconductivity and its proximity effect.
At the same time, there has been significant experimental progress in the fabrication of quasiperiodic structures, also in reduced dimensions.
Motivated by these developments, we use microscopic tight-binding theory to investigate the DC Josephson effect through a ballistic Fibonacci chain attached to two superconducting leads.
The Fibonacci chain is one of the most studied examples of quasicrystals, hosting a rich multifractal spectrum, containing topological gaps with different winding numbers.
We study how the Andreev bound states (ABS), current-phase relation, and the critical current depend on the quasiperiodic degrees of freedom, from short to long junctions.
While the current-phase relation shows a traditional $2\pi$ sinusoidal or sawtooth profile, we find that the ABS obtain quasiperiodic oscillations and that the Andreev reflection is qualitatively altered, leading to quasiperiodic oscillations in the critical current as a function of junction length.
Surprisingly, despite earlier proposals of quasiperiodicity enhancing superconductivity \newtext{compared to crystalline junctions}, we do not in general find that it enhances the critical current.
However, we find significant \newtext{current} enhancement for reduced interface transparency due to the modified Andreev reflection.
Furthermore, by varying the chemical potential, e.g.~by an applied gate voltage, we find a fractal oscillation between superconductor-normal metal-superconductor (SNS) and superconductor-insulator-superconductor (SIS) behavior.
Finally, we show that the winding of the subgap states leads to an equivalent winding in the critical current, such that the winding numbers, and thus the topological invariant, can be determined.
\end{abstract}

\maketitle

\section{Introduction}
\label{sec:intro}
Quasicrystals~\cite{shechtman.84,levine.steinhardt.84,levine.steinhardt.86,socolar.steinhardt.86,stadnik.98}, and more generally aperiodic systems~\cite{macia.06,kempkes.19,canyellas.23,iliasov.24}, provide fascinating platforms to study exotic and topologically non-trivial behavior in physics~\cite{deguchi.15,fulga.16,sakai.17,kamiya.18,sakai.19,araujo.19,shiino.21,jagannathan.21,nagai.24}.
Quasicrystals are neither periodic nor random disordered, but instead quasiperiodic as they exhibit a deterministic long-range order through discrete scale invariance and a non-crystallographic rotation symmetry~\cite{penrose.74,hiller.85,janot.97}.
In contrast to randomly disordered systems, quasicrystals therefore have clearly distinguished Bragg peaks in diffraction experiments~\cite{goldman.93,baake.grimm.2012}.
This peculiar combination leads to fascinating spectral and electronic properties~\cite{kohmoto.83,ostlund.83,ostlund.pandit.84,kohmoto.oono.84,tang.kohmoto.86,kalugin.kitaev.levitov.86,evangelou.87,kohmoto.sutherland.tang.87,sutherland.kohmoto.87,luck.89,hiramoto.92}, such as multifractality with critical states and wave functions that are neither localized nor extended~\cite{kalugin.katz.14,mace.jagannathan.piechon.16,mace.17}, hyperuniformity~\cite{torquato.18,baake.grimm.19}, pseudogaps~\cite{hafner.92,zijlstra.00,stadnik.01,rotenberg.04}, topological gaps~\cite{bellissard.89,bellissard.92,mace.jagannathan.piechon.17,yamamoto.22}, and topological invariants that are otherwise only possible in higher-dimensional systems~\cite{kraus.12,kraus.zilberberg.12,verbin.13,kraus.13,verbin.15,flicker.wezel.15,kraus.16,dareau.17,huang.liu.18,petrides.18,huang.liu.19,chen.20,rai.21,fan.huang.21,sykes.barnett.22}.

The influence of quasiperiodicity on different ordered states and transport phenomena has also attracted significant interest~\cite{tamura.10,kraus.13,deguchi.15,fulga.16,sakai.17,kamiya.18,sakai.19,araujo.19,khosravian.21,jagannathan.21,shiino.21,ghanta.23,moustaj.23,fukushima.23,jiang.23}.
For instance, it has recently been proposed that quasiperiodicity can enhance the superconducting order parameter and transition temperature~\cite{fan.21,zhang.22,oliveira.23,sun.24,wang.24}, its proximity effect~\cite{rai.19,rai.20}, topological superconductivity~\cite{kobialka.24}, persistent currents in normal metal rings~\cite{qiu.14,roy.23}, and cause enhanced or anomalous transport phenomena~\cite{mayou.93,pierce.guo.poon.94,roche.trambly.mayou.97,jeon.21,jeon.kim.lee.22,wang.zhao.23}.
The enhancement has been directly linked to the underlying topological~\cite{rai.19} and critical states~\cite{fan.21,zhang.22,oliveira.23,sun.24} of the quasicrystal, similar to how criticality may enhance superconductivity also in disordered systems~\cite{arrigoni.fradkin.kivelson.04,martin.podolsky.kivelson.05,feigelman.07,feigelman.10,burmistrov.12,mayoh.15,zhao.19,burmistrov.21}.
However, while disordered systems typically exhibit such critical behavior only at the Anderson localization transition~\cite{schreiber.91,evers.mirlin.08}, critical behavior is ubiquitous in quasicrystals~\cite{jagannathan.tarzia.20}.

From an experimental standpoint, quasiperiodic systems are also becoming increasingly accessible. Significant progress in quasicrystal growth~\cite{nagao.15,wolf.21} and synthetic engineering with atomic precision~\cite{eigler.90,stroscio.91,gomes.12,polini.13,drost.17,khajetoorians.19,schneider.20,huda.20,kuster.22,freeney.22} have recently enabled creation of quasiperiodic structures in reduced dimensions~\cite{guidoni.97,ledieu.04,sharma.05,ledieu.08,smerdon.08,talapin.09,forster.13,jia.16,bandres.16,collins.17,yan.19,xin.22,dang.23} and in moir{\'e} structures~\cite{lubin.12,joon-ahn.18,mahmood.21,uri.23,ghadimi.24}.
A prime example is the Fibonacci quasicrystal~\cite{jagannathan.21}, which is closely related to the dodecagonal and icosahedral quasicrystals~\cite{ledieu.04}, and hosts a multifractal spectrum of topological gaps with subgap winding states~\cite{rai.21}.
Its implementation as a one-dimensional (1D) atomic chain, the so-called Fibonacci chain, is also relevant to 3D systems~\cite{wang.24} where it naturally appears or can be experimentally engineered in e.g.~stacked materials~\cite{merlin.85,todd.86,bajema.merlin.87,cohn.88,zhu.97}.
Synthetic Fibonacci chains have additionally been realized in the context of photonics~\cite{negro.03,kraus.12,verbin.13,verbin.15,yuan.18}, phononics~\cite{steurer.sutter.07}, polaritonics~\cite{tanese.14,baboux.17,goblot.20}, cold atoms~\cite{singh.15}, dielectric chains or circuits~\cite{reisner.23,franca.24}, and magnonics~\cite{lisiecki.19}.
In these systems, the Fibonacci chain spectrum and topology have often directly been measured~\cite{kraus.12,kraus.zilberberg.12,verbin.13,kraus.13,verbin.15,flicker.wezel.15,kraus.16,dareau.17,huang.liu.18,petrides.18,huang.liu.19,chen.20,rai.21,fan.huang.21,sykes.barnett.22}.

\begin{figure}[t!]
    \begin{tikzpicture}[scale=0.35]
        \filldraw[white] (-2,-2.3) rectangle (22,3.5);
        \node[left] at (-1,3.7) {(a)};
        \filldraw[gray!40,very thick] (-1,-2.3) rectangle (5,2.3);
        \filldraw[gray!20,very thick] (5,-2.3) rectangle (15,2.3);
        \filldraw[gray!40,very thick] (15,-2.3) rectangle (21,2.3);
        \draw[very thick,dotted] (0,0)--(-1.5,0);
        \draw[very thick,dotted] (10,0)--(12,0);
        \draw[very thick,dotted] (20,0)--(21.5,0);
        \filldraw[red] (0,0) circle (0.1);
        \filldraw[red] (2,0) circle (0.1);
        \filldraw[red] (4,0) circle (0.1);
        \filldraw[red] (6,0) circle (0.1);
        \filldraw[red] (8,0) circle (0.1);
        \filldraw[red] (10,0) circle (0.1);
        \filldraw[red] (12,0) circle (0.1);
        \filldraw[red] (14,0) circle (0.1);
        \filldraw[red] (16,0) circle (0.1);
        \filldraw[red] (18,0) circle (0.1);
        \filldraw[red] (20,0) circle (0.1);
        \draw[very thick, ->] (0,-0.15) .. controls (0.5,-0.8) and (1.5,-0.8) .. (2,-0.15);
        \draw[very thick, ->] (2,-0.15) .. controls (1.5,-0.8) and (0.5,-0.8) .. (0,-0.15);
        \draw[very thick, ->] (2,-0.15) .. controls (2.5,-0.8) and (3.5,-0.8) .. (4,-0.15);
        \draw[very thick, ->] (4,-0.15) .. controls (3.5,-0.8) and (2.5,-0.8) .. (2,-0.15);
        \draw[very thick,dashed, ->] (4,-0.15) .. controls (4.5,-0.8) and (5.5,-0.8) .. (6,-0.15);
        \draw[very thick,dashed, ->] (6,-0.15) .. controls (5.5,-0.8) and (4.5,-0.8) .. (4,-0.15);
        
        \draw[blue!90,very thick, ->] (6,-0.15) .. controls (6.5,-0.8) and (7.5,-0.8) .. (8,-0.15);
        \draw[blue!90,very thick, ->] (8,-0.15) .. controls (7.5,-0.8) and (6.5,-0.8) .. (6,-0.15);
        \draw[orange!90,very thick, ->] (8,-0.15) .. controls (8.5,-0.8) and (9.5,-0.8) .. (10,-0.15);
        \draw[orange!90,very thick, ->] (10,-0.15) .. controls (9.5,-0.8) and (8.5,-0.8) .. (8,-0.15);
        \draw[blue!90,very thick, ->] (12,-0.15) .. controls (12.5,-0.8) and (13.5,-0.8) .. (14,-0.15);
        \draw[blue!90,very thick, ->] (14,-0.15) .. controls (13.5,-0.8) and (12.5,-0.8) .. (12,-0.15);
        \draw[very thick,dashed, ->] (14,-0.15) .. controls (14.5,-0.8) and (15.5,-0.8) .. (16,-0.15);
        \draw[very thick,dashed, ->] (16,-0.15) .. controls (15.5,-0.8) and (14.5,-0.8) .. (14,-0.15);
        \draw[very thick, ->] (16,-0.15) .. controls (16.5,-0.8) and (17.5,-0.8) .. (18,-0.15);
        \draw[very thick, ->] (18,-0.15) .. controls (17.5,-0.8) and (16.5,-0.8) .. (16,-0.15);
        \draw[very thick, ->] (18,-0.15) .. controls (18.5,-0.8) and (19.5,-0.8) .. (20,-0.15);
        \draw[very thick, ->] (20,-0.15) .. controls (19.5,-0.8) and (18.5,-0.8) .. (18,-0.15);
        \node[below] at (1,-0.8) {$\tS$};
        \node[below] at (3,-0.8) {$\tS$};
        \node[below] at (5,-0.8) {$\tint$};
        \node[below,blue!90] at (7,-0.8) {$\tA$};
        \node[below,orange!90] at (9,-0.8) {$\tB$};
        \node[below,blue!90] at (13,-0.8) {$\tA$};
        \node[below] at (15,-0.8) {$\tint$};
        \node[below] at (17,-0.8) {$\tS$};
        \node[below] at (19,-0.8) {$\tS$};
        \node[above] at (0,0.2) {$\Delta$};
        \node[above] at (2,0.2) {$\Delta$};
        \node[above] at (4,0.2) {$\Delta$};
        \node[above] at (16,0.2) {$\Delta$};
        \node[above] at (18,0.2) {$\Delta$};
        \node[above] at (20,0.2) {$\Delta$};
        \node[above] at (5.8,1.1) {$\mu$};
        \node[above] at (-0.2,1.1) {$\mu_S$};
        \node[above] at (15.8,1.1) {$\mu_S$};
        \node[above] at (2,2.2) {\textbf{SC lead}};
        \node[above] at (10,2.2) {\textbf{Fibonacci chain}};
        \node[above] at (18,2.2) {\textbf{SC lead}};
        \draw [decorate, decoration = {brace}] (14,-2.5) --  (6,-2.5);
        \node[below] at (10,-2.7) {L};
    \end{tikzpicture}
    \\[0.2cm]
    \begin{tikzpicture}[scale=0.65]
        \filldraw[white] (-1.2,-1) rectangle (11.5,0.3);
        \node[left] at (-1,0.3) {(b)};
        \filldraw[red] (0,0) circle (0.1);
        \filldraw[red] (2,0) circle (0.1);
        \filldraw[red] (4,0) circle (0.1);
        \filldraw[red] (6,0) circle (0.1);
        \filldraw[red] (8,0) circle (0.1);
        \filldraw[red] (10,0) circle (0.1);
        \draw[blue!90,very thick, ->] (0,-0.15) .. controls (0.5,-0.8) and (1.5,-0.8) .. (2,-0.15);
        \draw[blue!90,very thick, ->] (2,-0.15) .. controls (1.5,-0.8) and (0.5,-0.8) .. (0,-0.15);
        \draw[orange!90,very thick, ->] (2,-0.15) .. controls (2.5,-0.8) and (3.5,-0.8) .. (4,-0.15);
        \draw[orange!90,very thick, ->] (4,-0.15) .. controls (3.5,-0.8) and (2.5,-0.8) .. (2,-0.15);
        \draw[blue!90,very thick, ->] (4,-0.15) .. controls (4.5,-0.8) and (5.5,-0.8) .. (6,-0.15);
        \draw[blue!90,very thick, ->] (6,-0.15) .. controls (5.5,-0.8) and (4.5,-0.8) .. (4,-0.15);
        \draw[blue!90,very thick, ->] (6,-0.15) .. controls (6.5,-0.8) and (7.5,-0.8) .. (8,-0.15);
        \draw[blue!90,very thick, ->] (8,-0.15) .. controls (7.5,-0.8) and (6.5,-0.8) .. (6,-0.15);
        \draw[orange!90,very thick, ->] (8,-0.15) .. controls (8.5,-0.8) and (9.5,-0.8) .. (10,-0.15);
        \draw[orange!90,very thick, ->] (10,-0.15) .. controls (9.5,-0.8) and (8.5,-0.8) .. (8,-0.15);
    
        \node[blue!90,below] at (1,-0.8) {$\tA$};
        \node[orange!90,below] at (3,-0.8) {$\tB$};
        \node[blue!90,below] at (5,-0.8) {$\tA$};
        \node[blue!90,below] at (7,-0.8) {$\tA$};
        \node[orange!90,below] at (9,-0.8) {$\tB$};
    \end{tikzpicture}
    \\[0.2cm]
    \begin{tikzpicture}[scale=0.25]
        \filldraw[white] (-0.3,-2.1) rectangle (31.5,2);
        \node[left] at (-0.3,2) {(c)};
        \filldraw[blue!10] (-0.3,-2.1) rectangle (10,0.8);
        \filldraw[orange!10] (10,-2.1) rectangle (20,0.8);
        \filldraw[blue!10] (20,-2.1) rectangle (30.3,0.8);
        \filldraw[red] (0,0) circle (0.1);
        \filldraw[red] (2,0) circle (0.1);
        \filldraw[red] (4,0) circle (0.1);
        \filldraw[red] (6,0) circle (0.1);
        \filldraw[red] (8,0) circle (0.1);
        \filldraw[red] (10,0) circle (0.1);
        \draw[blue!90,very thick, ->] (0,-0.15) .. controls (0.5,-0.8) and (1.5,-0.8) .. (2,-0.15);
        \draw[blue!90,very thick, ->] (2,-0.15) .. controls (1.5,-0.8) and (0.5,-0.8) .. (0,-0.15);
        \draw[orange!90,very thick, ->] (2,-0.15) .. controls (2.5,-0.8) and (3.5,-0.8) .. (4,-0.15);
        \draw[orange!90,very thick, ->] (4,-0.15) .. controls (3.5,-0.8) and (2.5,-0.8) .. (2,-0.15);
        \draw[blue!90,very thick, ->] (4,-0.15) .. controls (4.5,-0.8) and (5.5,-0.8) .. (6,-0.15);
        \draw[blue!90,very thick, ->] (6,-0.15) .. controls (5.5,-0.8) and (4.5,-0.8) .. (4,-0.15);
        \draw[blue!90,very thick, ->] (6,-0.15) .. controls (6.5,-0.8) and (7.5,-0.8) .. (8,-0.15);
        \draw[blue!90,very thick, ->] (8,-0.15) .. controls (7.5,-0.8) and (6.5,-0.8) .. (6,-0.15);
        \draw[orange!90,very thick, ->] (8,-0.15) .. controls (8.5,-0.8) and (9.5,-0.8) .. (10,-0.15);
        \draw[orange!90,very thick, ->] (10,-0.15) .. controls (9.5,-0.8) and (8.5,-0.8) .. (8,-0.15);
        
        \node[blue!90,below] at (1,-0.8) {$\tA$};
        \node[orange!90,below] at (3,-0.8) {$\tB$};
        \node[blue!90,below] at (5,-0.8) {$\tA$};
        \node[blue!90,below] at (7,-0.8) {$\tA$};
        \node[orange!90,below] at (9,-0.8) {$\tB$};
        
        \filldraw[red] (12,0) circle (0.1);
        \filldraw[red] (14,0) circle (0.1);
        \filldraw[red] (16,0) circle (0.1);
        \filldraw[red] (18,0) circle (0.1);
        \filldraw[red] (20,0) circle (0.1);
        \draw[blue!90,very thick, ->] (10,-0.15) .. controls (10.5,-0.8) and (11.5,-0.8) .. (12,-0.15);
        \draw[blue!90,very thick, ->] (12,-0.15) .. controls (11.5,-0.8) and (10.5,-0.8) .. (10,-0.15);
        \draw[orange!90,very thick, ->] (12,-0.15) .. controls (12.5,-0.8) and (13.5,-0.8) .. (14,-0.15);
        \draw[orange!90,very thick, ->] (14,-0.15) .. controls (13.5,-0.8) and (12.5,-0.8) .. (12,-0.15);
        \draw[blue!90,very thick, ->] (14,-0.15) .. controls (14.5,-0.8) and (15.5,-0.8) .. (16,-0.15);
        \draw[blue!90,very thick, ->] (16,-0.15) .. controls (15.5,-0.8) and (14.5,-0.8) .. (14,-0.15);
        \draw[blue!90,very thick, ->] (16,-0.15) .. controls (16.5,-0.8) and (17.5,-0.8) .. (18,-0.15);
        \draw[blue!90,very thick, ->] (18,-0.15) .. controls (17.5,-0.8) and (16.5,-0.8) .. (16,-0.15);
        \draw[orange!90,very thick, ->] (18,-0.15) .. controls (18.5,-0.8) and (19.5,-0.8) .. (20,-0.15);
        \draw[orange!90,very thick, ->] (20,-0.15) .. controls (19.5,-0.8) and (18.5,-0.8) .. (18,-0.15);
        
        \node[blue!90,below] at (11,-0.8) {$\tA$};
        \node[orange!90,below] at (13,-0.8) {$\tB$};
        \node[blue!90,below] at (15,-0.8) {$\tA$};
        \node[blue!90,below] at (17,-0.8) {$\tA$};
        \node[orange!90,below] at (19,-0.8) {$\tB$};
        
        \filldraw[red] (22,0) circle (0.1);
        \filldraw[red] (24,0) circle (0.1);
        \filldraw[red] (26,0) circle (0.1);
        \filldraw[red] (28,0) circle (0.1);
        \filldraw[red] (30,0) circle (0.1);
        \draw[blue!90,very thick, ->] (20,-0.15) .. controls (20.5,-0.8) and (21.5,-0.8) .. (22,-0.15);
        \draw[blue!90,very thick, ->] (22,-0.15) .. controls (21.5,-0.8) and (20.5,-0.8) .. (20,-0.15);
        \draw[orange!90,very thick, ->] (22,-0.15) .. controls (22.5,-0.8) and (23.5,-0.8) .. (24,-0.15);
        \draw[orange!90,very thick, ->] (24,-0.15) .. controls (23.5,-0.8) and (22.5,-0.8) .. (22,-0.15);
        \draw[blue!90,very thick, ->] (24,-0.15) .. controls (24.5,-0.8) and (25.5,-0.8) .. (26,-0.15);
        \draw[blue!90,very thick, ->] (26,-0.15) .. controls (25.5,-0.8) and (24.5,-0.8) .. (24,-0.15);
        \draw[blue!90,very thick, ->] (26,-0.15) .. controls (26.5,-0.8) and (27.5,-0.8) .. (28,-0.15);
        \draw[blue!90,very thick, ->] (28,-0.15) .. controls (27.5,-0.8) and (26.5,-0.8) .. (26,-0.15);
        \draw[orange!90,very thick, ->] (28,-0.15) .. controls (28.5,-0.8) and (29.5,-0.8) .. (30,-0.15);
        \draw[orange!90,very thick, ->] (30,-0.15) .. controls (29.5,-0.8) and (28.5,-0.8) .. (28,-0.15);
        
        \node[blue!90,below] at (21,-0.8) {$\tA$};
        \node[orange!90,below] at (23,-0.8) {$\tB$};
        \node[blue!90,below] at (25,-0.8) {$\tA$};
        \node[blue!90,below] at (27,-0.8) {$\tA$};
        \node[orange!90,below] at (29,-0.8) {$\tB$};
    \end{tikzpicture}
    \caption{(a) Tight-binding model of a hybrid superconductor-quasicrystal Josephson junction, modeled by a Fibonacci chain with length $L$ and hoppings $\tA$ (blue) and $\tB$ (orange) at chemical potential $\mu$, attached via interface hopping $\tint$ to two superconducting leads with hopping $\tS$ (black), chemical potential $\muS$ and onsite spin-singlet $s$-wave superconducting order parameter $\Delta$.
    (b) Fibonacci hopping structure for the Fibonacci approximant $C_4$ with $F_4 = 5$ bonds. (c) Chain of multiple Fibonacci approximants, i.e.~a system with supercell $C_4$ repeated $N=3$ times.}
    \label{fig:sns_model}
\end{figure}
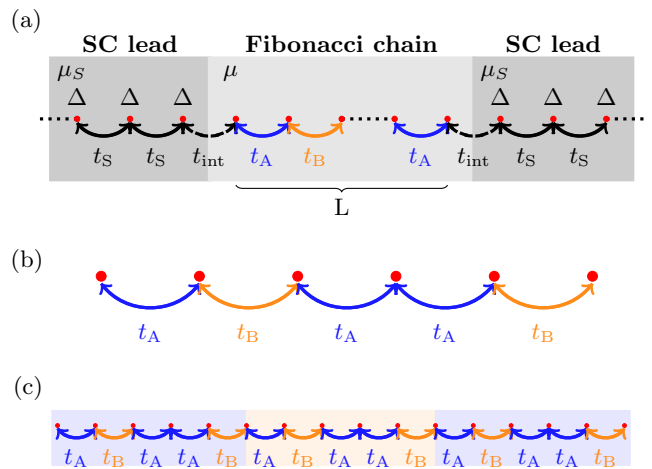

Motivated by the experimental timeliness and the predictions of enhanced superconductivity, we here study the influence of quasiperiodicity on one of the most technologically important superconducting phenomena, namely the Josephson effect~\cite{josephson.62,josephson.74,likharev.79,golubov.04}.
In particular, we study a non-superconducting Fibonacci chain of length $L$ attached to two crystalline superconducting (S) leads using microscopic tight-binding theory, see Fig.~\ref{fig:sns_model}(a).
Since the Fibonacci chain is either normal conducting (N) or insulating (I), depending on if the Fermi level is outside or inside the topological gaps, respectively, we effectively study both ballistic SNS and SIS Josephson junctions. 
Furthermore, we study the short ($\xi_0 \gg L$) to the long ($\xi_0 \ll L$) junction regime, where $\xi_0$ is the superconducting coherence length \cite{likharev.79}.
We also model repeated Fibonacci chain supercells, i.e.~so-called crystal approximants~\cite{goldman.93}, see Figs.~\ref{fig:sns_model}(b) and \ref{fig:sns_model}(c).
\newtext{Overall, we perform extensive model calculations in these systems to address whether quasiperiodicity significantly influences the Josephson effect. Specifically, we systematically study how each model parameter influences the energy spectrum, current-phase relation, and critical current, contrasting the quasiperiodic and crystalline scenarios at each step.}

We find that although the Josephson current-phase relation shows a conventional $2\pi$ sinusoidal or sawtooth profile, the low-energy Andreev bound states (ABS) are not conventional.
Specifically, we demonstrate that the ABS obtain a quasiperiodic probability density, also at perfect resonance.
Importantly, we also show how quasiperiodicity modifies the condition for zero-energy ABS, which generally generates the largest critical current~\cite{ishii.70}, associated with the sawtooth current-phase profile and perfect Andreev reflection (i.e.~zero normal reflection)~\cite{affleck.00}.
We are in fact able to obtain a set of simple functional forms for this zero-energy condition, depending on the model parameters and spatial realization of the Fibonacci chain.
We find that the system changes quasiperiodically between these forms with junction length.
Consequently, we find that the decay of the critical current with junction length is described by quasiperiodic oscillations, \newtext{on top of the traditional power law and exponential decays found in crystalline junctions~\cite{beenakker.92,affleck.00,nikolic.01,sonin.24}.
These quasiperiodic oscillations imply significant sample-to-sample fluctuations, unless the junction is created with atomic precision, possible with modern STM techniques~\cite{collins.17}.}
 
Surprisingly, while earlier studies have proposed that quasiperiodicity can enhance superconductivity and the proximity effect~\cite{fan.21,zhang.22,oliveira.23,sun.24,wang.24,rai.19,rai.20}, we do not find that it generally enhances the critical current, at least not when compared to an idealized ballistic crystalline junction with a zero-energy state.
However, beyond this idealized situation, we find that quasiperiodicity can cause a large enhancement of the critical current in junctions with poor transmission, due to it modifying the condition for zero-energy ABS.
Moreover, we show that by applying a gate voltage, the junction varies between SNS and SIS behavior in a fractal manner, as the Fermi level enters and exits the topological gaps of the quasicrystal energy spectrum.
Finally, we find that when the Fermi level is inside or close to the quasicrystal energy gaps, \newtext{the topological subgap states can carry the majority of the Josephson current, while their winding} leads to a similar winding in the critical current. We thus demonstrate how the Josephson current can probe the topological invariant in quasicrystals.

The rest of this work is organized as follows.
In Sec.~\ref{sec:model}, we describe our model and relevant properties of the Fibonacci chain.
In Sec.~\ref{sec:results:josephson_effect} we study \newtext{how quasiperiodicity influences} the ABS spectrum and current-phase relation.
In Sec.~\ref{sec:results:critical_current} we study \newtext{the important critical current, and how it depends on the quasiperiodic modulation and interface transparency, the Fibonacci approximant order, and junction length}.
In Sec.~\ref{sec:results:topological_invariant} we study the gate voltage dependence and also demonstrate how the critical current can measure the topological winding number.
The work is concluded in Sec.~\ref{sec:conclusions}.

\section{Model and background}
\label{sec:model}
In this work we investigate a quasiperiodic Josephson junction, shown schematically in Fig.~\ref{fig:sns_model}(a), by numerically simulating a non-superconducting quasiperiodic chain attached to two crystalline superconducting leads~\cite{sandberg.22}.
We model this system via the Hamiltonian
\begin{align}
    \label{eq:total_hamiltonian}
    H = \HQ + \HS + \HT, 
\end{align}
where $\HQ$ captures the quasiperiodic (Q) non-superconducting part arranged as a Fibonacci chain as described in Sec.~\ref{sec:model:fibonacci} and Sec.~\ref{sec:model:fibonacci_spectrum}, while $\HS$ and $\HT$ capture the superconducting (S) leads and interface tunneling (T), respectively, as described in Sec.~\ref{sec:model:superconductivity}.

\subsection{Fibonacci chain hopping model}
\label{sec:model:fibonacci}
In this subsection we define how to construct a Fibonacci chain, and our tight-binding model to study such a system, namely the Fibonacci hopping model~\cite{jagannathan.21}.

A Fibonacci chain is a 1D quasicrystal which can be constructed in a similar way to how Fibonacci numbers are generated~\cite{jagannathan.21}, but instead of numbers, a Fibonacci chain can be seen as a string of the letters \textit{A} and \textit{B}. A string is constructed recursively through concatenation, $C_n=C_{n-1}\oplus C_{n-2}$, where $C_0={\rm B}$, $C_1={\rm A}$, such that $C_2 = {\rm AB}$, $C_3= {\rm ABA}$, $C_4 = {\rm ABAAB}$ and so forth.
Here, $C_n$ is referred to as the \textit{n}:th approximant of the infinite Fibonacci chain~\cite{jagannathan.21}, with its length given by the corresponding Fibonacci number $F_n=F_{n-1}+F_{n-2}$ ($n\geq2$, $F_0=F_1=1$).
A more generalized Fibonacci chain can be constructed using the characteristic function~\cite{kraus.zilberberg.12}
\begin{align}
    \label{eq:characteristic_function}
    \chi_j=\sgn[\cos(2\pi j \tau^{-1} + \phi)-\cos(\pi \tau^{-1})],
\end{align}
where $j=1,2,3...$ is the letter index, $\tau=(1+\sqrt{5})/2$ is the golden ratio, with $\chi_j=-1$ giving the letter ${\rm A}$ and $\chi_j=1$ gives the letter ${\rm B}$.
Here, $\phi \in [0, 2\pi)$ is a phase factor referred to as the {\it phason angle}, which is related to the topology of the Fibonacci chain as discussed in Sec.~\ref{sec:model:fibonacci_spectrum}. 
We set constant $\phi = \pi\tau^{-1}$ unless otherwise specified, as this value ensures that terminating the characteristic function in Eq.~(\ref{eq:characteristic_function}) at the bond length $F_n$ reproduces the Fibonacci approximant $C_n$ given by the concatenation rule.
In contrast, varying $\phi$ from $0$ to $2\pi$ leads to successive letter changes called {\it phason flips} which generates the complete set of $F_n+1$ unique Fibonacci chain realizations of length $F_n$~\cite{jagannathan.21}.

In this work we model a Fibonacci chain using the Fibonacci hopping model via the Hamiltonian
\begin{align}
    \HQ= \sum_{\sigma,j} \mu c_{j\sigma}^\dagger c_{j\sigma} -\sum_{\sigma,\langle ij\rangle}\left(t_{\langle ij\rangle}c_{j\sigma}^\dagger c_{i\sigma}+{\rm H.c.}\right),
\end{align}
where $\mu$ is the chemical potential, e.g.~controlled by an external gate voltage, $c^\dagger_{j\sigma}$ ($c_{j\sigma}$) is the creation (annihilation) operator of electronic states at site $j$ with spin $\sigma$, ${\langle ij\rangle}$ denotes nearest-neighbor sites $i$ and $j$ in the Fibonacci chain with hopping $t_{\langle ij\rangle}$ taking one of the two values $\tA$ or $\tB$ following the Fibonacci chain, i.e.~substituting the letters ${\rm A} \mapsto \tA$ and ${\rm B} \mapsto \tB $ following Eq.~(\ref{eq:characteristic_function}).
From here on, $\tB$ is our natural unit of energy. We introduce the hopping ratio $\rho \equiv \tA/\tB$ where $\rho \neq 1$ ($\rho=1$) corresponds to a quasiperiodic (crystalline) system.
We note that the Fibonacci hopping model is closely related to the onsite Fibonacci model where an onsite potential instead varies quasiperiodically~\cite{kohmoto.83,ostlund.83,jagannathan.21}, with one main difference being that the Fibonacci number $F_n$ usually labels the number of bonds (sites) in the Fibonacci hopping (onsite) model.

We consider two quasiperiodic scenarios.
In the first, we consider Fibonacci chains with length $L\in[2,1000]a_0$ described by the characteristic function in Eq.~(\ref{eq:characteristic_function}) as depicted in Figs.~\ref{fig:sns_model}(a), where $a_0$ is the lattice spacing and our natural unit of length.
In the second scenario, we consider Fibonacci chains consisting of approximants $C_n$ repeated $N$ times as illustrated in Fig.~\ref{fig:sns_model}(c), thus with physical length $L=a_0N \times F_n$ (and $N\times F_n+1$ sites).
In particular, we study such repeated approximants up to several thousand sites, e.g.~$L=2584a_0$ for $F_{17}$ with $N=1$, or $N=500$ for low $F_n$, which is several orders of magnitude longer than both the microscopic length scale $a_0$ and the superconducting coherence length $\xi_0\approx17a_0$ (see Sec.~\ref{sec:model:superconductivity}).
Thus, these non-repeated and repeated scenarios essentially correspond to quasicrystals and approximant crystals~\cite{goldman.93} embedded across two superconducting leads separated by a distance $L$, respectively. 
We note that the most important features, such as topology and the major gap structure, remains unchanged with repetition~\cite{rai.21,kobialka.24} (see also Sec.~\ref{sec:model:fibonacci_spectrum}).
Furthermore, it is well-known that even such finite quasiperiodic Fibonacci approximants host the most relevant features of the Fibonacci quasicrystals~\cite{jagannathan.21}.

\subsection{Fibonacci chain spectrum and topology}
\label{sec:model:fibonacci_spectrum}
To provide a background for the relevant physics of the Fibonacci chain, we in this subsection discuss its spectrum and topology.

\begin{figure}[t!]
    \centering
    \includegraphics[width=\columnwidth]{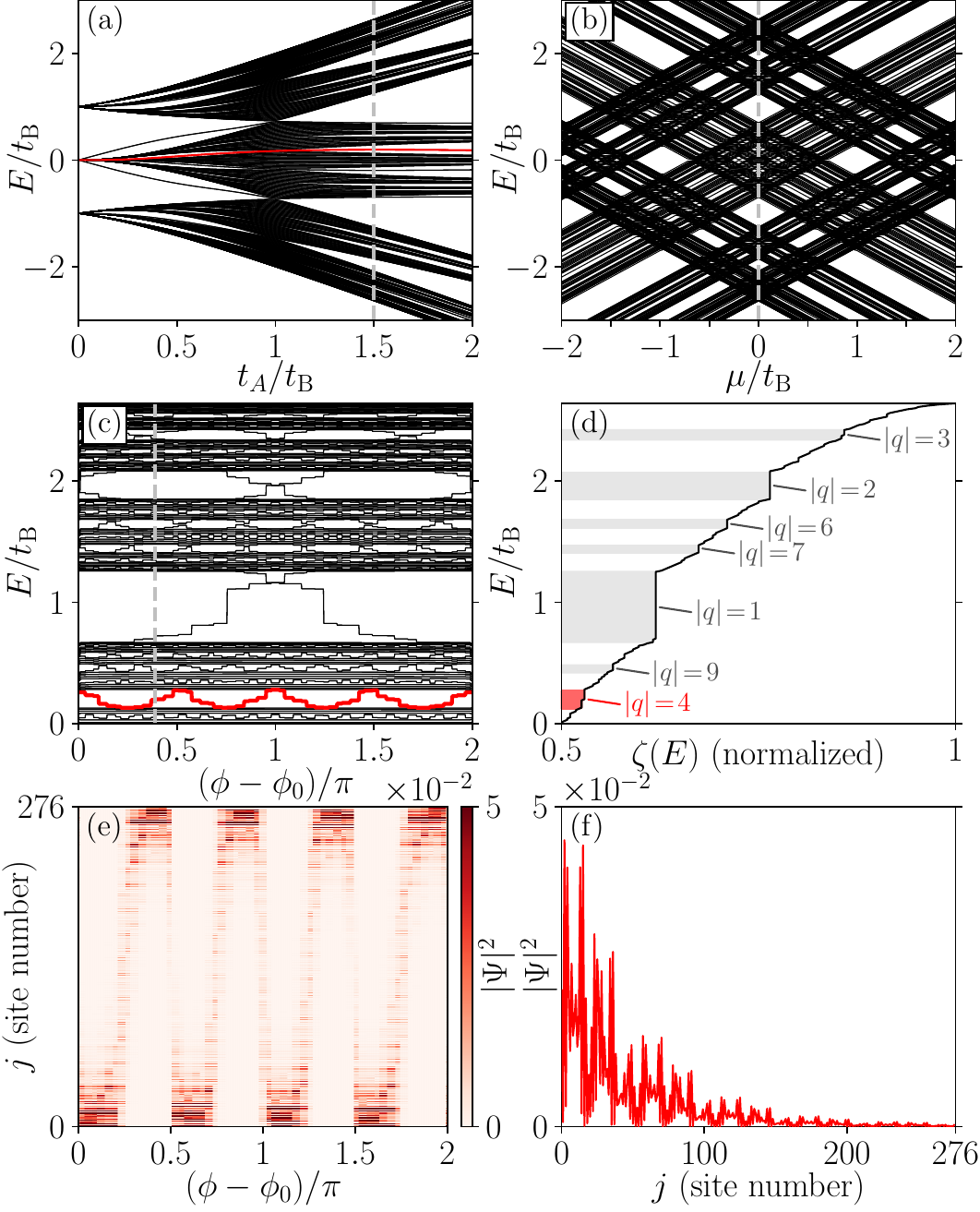}
    \caption{Energy spectrum versus hopping ratio $\rho$ (a), chemical potential $\mu$ (b), phason angle $\phi$ (c), for the Fibonacci approximant $C_9$ with $N=5$ repetitions and open boundary conditions. Vertical dashed lines: guides to the eye marking the constant parameter choice in every other figure. Here, $\phi_0=\pi\tau^{-1}(F_n+1)$ is a constant to symmetrize the spectrum around $\pi$~\cite{rai.19}.
    (d) Normalized number of energy levels $\zeta$ below energy $E$, with gap labels $q$ for the largest gaps (shaded), corresponding to the winding numbers in (c). (e) Probability density $|\Psi|^2$ as a function of $\phi$ on site $j$ for the $|q|=4$ the winding state [red energy level in (a),(c),(d)]. (f) $|\Psi|^2$ at fixed $\phi-\phi_0=1.5\pi$ in (e).}
    \label{fig:fibonacci_spectrum}
\end{figure}

One of the most remarkable features of the Fibonacci chain is the opening (closing) of topological gaps at $\rho \neq 1$ ($\rho=1$), as illustrated in Fig.~\ref{fig:fibonacci_spectrum}(a). We note that here $\mu=0$ corresponds to half-filling, see Fig.~\ref{fig:fibonacci_spectrum}(b).
Interestingly, at $\rho \neq 1$ there is a topological phase transition with the appearance of edge modes, see the red subgap state in Fig.~\ref{fig:fibonacci_spectrum}(a), in analogy with the dimerized Su-Schrieffer-Heeger (SSH) model~\cite{su.shrieffer.79}.
After all, a repeated Fibonacci approximant $C_n$ can in a sense be seen as a natural extension of the SSH model to include a richer sublattice structure, since the lowest-order non-trivial Fibonacci sublattice $C_2=\tA\tB$ is similar to the the dimerized SSH chain.
However, there are some important differences, like how the chain is usually terminated, and that the Fibonacci chain can host multiple topological gaps each with subgap states, see Fig.~\ref{fig:fibonacci_spectrum}(b).
Furthermore, the exact subgap energy of the states depend directly on the phason angle as illustrated in Fig.~\ref{fig:fibonacci_spectrum}(c).
Additionally, the Fibonacci chain might have edge modes for both $\rho<1$ and $\rho>1$ in contrast to the SSH chain, and there are three (two) bands in the limit $\rho\to0$ for the Fibonacci (SSH) chain, corresponding to the existence of three (two) kinds of nearest-neighbor hopping neighborhoods.

Each gap of the Fibonacci chain is related to an integer gap label $q$ according to a gap labeling theorem~\cite{bellissard.89,bellissard.92,mace.jagannathan.piechon.17} which states that for the Fibonacci approximant $C_n$, $q$ is related to the number of energy levels $\zeta$ below the corresponding gap via $\zeta = \operatorname{mod}\left[q(F_{n-1}),F_n\right]$~\cite{jagannathan.21}.
Figure~\ref{fig:fibonacci_spectrum}(d) visualizes the gap labeling theorem for the $C_9$ approximant, and we note that the gap size is usually inversely proportional to the gap label~\cite{mace.jagannathan.piechon.17,yamamoto.22}.

The gap label $q$ was recently shown to be equivalent to a Chern number~\cite{kraus.12}, which is easiest understood by first identifying the Fibonacci hopping model as topological equivalent with the 1D Aubry-Andr{\'e}-Harper (AAH) model~\cite{harper.55,aubry.andre.80,gordon.97} and the 2D integer quantum Hall system~\cite{kraus.zilberberg.12,flicker.wezel.15,kraus.16}.
Specifically, the Fibonacci chain has topological edge states which wind across the quasicrystal gaps labeled $q$, with corresponding winding number $q$.
However, the states wind as a function of the phason angle $\phi$, not momentum as in crystalline systems.
Specifically, as $\phi$ varies from $0$ to $2\pi$, the topological states wind inside the gap $|q|$ times with direction $\sgn(q)$ as shown in Fig.~\ref{fig:fibonacci_spectrum}(c).
Due to the discrete nature of the characteristic function in Eq.~(\ref{eq:characteristic_function}), there are $F_{n} + 1$ unique values (or phason flips) for the approximant $C_n$, which explains the step-like appearance of the energy levels in Fig.~\ref{fig:fibonacci_spectrum}(c).
Finally, we note that also the real-space localization of the winding state changes with $\phi$, such that the state winds back and forth across the chain $|q|$ times as shown in the probability density in Fig.~\ref{fig:fibonacci_spectrum}(e).
Recently, this was experimentally realized as topological pumping of the edge modes in phototonic~\cite{verbin.15} and polaritonic~\cite{goblot.20} systems.
Finally, Fig.~\ref{fig:fibonacci_spectrum}(f) shows a cut at fixed $\phi$, demonstrating that the spatial dependence of the edge modes is not extended or exponentially localized, but rather critical and furthermore with multifractal behavior~\cite{jagannathan.21}.
The goal of our work is to see how the interesting quasiperiodic properties mentioned above influence the Josephson effect, \newtext{by systematically investigating the model and parameter dependencies}.

\subsection{Superconductivity and Josephson junction}
\label{sec:model:superconductivity}
In this subsection, we describe our tight-binding model for the superconducting leads and interface. We also show how we compute the Josephson current and estimate the superconducting coherence length.

Each superconducting lead is modelled via the effective mean-field Hamiltonian
\begin{align}
    \nonumber
    \HS = & \sum_{\sigma,j}\muS c^\dagger_{j\sigma}c_{j\sigma} -\sum_{\sigma,\langle ij\rangle}\left(\tS c_{j\sigma}^\dagger c_{i\sigma} + {\rm H.c.}\right)\\
    \label{eq:superconducting_hamiltonian}
    & -\sum_{j}\left[ \Delta_{j}c^\dagger_{j\uparrow}c^\dagger_{j\downarrow} + {\rm H.c.} \right],
\end{align}
with chemical potential $\muS$ and hopping $\tS$, sites $i$ and $j$ in the respective superconducting lead, and where $\uparrow$ $(\downarrow)$ denotes spin up (down) states.
We introduce the notation $\rhoS\equiv\tS/\tB$, and for simplicity mainly focus on the scenario $\muS=0$ (half-filling).
We consider superconducting leads with $\NS=135$ sites (we verify that increasing this size does not yield any noticeable difference on our results), with on-site $s$-wave superconductivity via the mean-field order parameter $\Delta_j$.

We model the coupling between the superconducting leads and the Fibonacci chain via the Hamiltonian
\begin{align}
    \label{eq:tunneling_hamiltonian}
    \HT = -\sum_{\sigma,\langle ij\rangle} \tint c^\dagger_{j\sigma}c_{i\sigma} + {\rm H.c.},
\end{align}
where $\tint$ is the interface hopping between nearest neighbor sites ${\langle ij \rangle}$, one in the Fibonacci chain and the other in a superconducting lead
as illustrated in Fig.~\ref{fig:sns_model}(a).
We investigate different values of $\tint$, $\rho\equiv\tA/\tB$ and $\rhoS\equiv\tS/\tB$ as a simplified way to model different interface transparencies, and note that this can still be related to results obtained with e.g.~a scattering-matrix approach~\cite{affleck.00}.

We solve the resulting Hamiltonian $H = \HQ + \HS + \HT$ at zero temperature using the Bogoliubov-de Gennes method, \newtext{using uniform superconductivity in the leads $|\Delta_j| = \Delta_0 = 0.06\tB$. We verify that self-consistency, capturing the inverse proximity effect, does not qualitatively modify the results}~\footnote{\newtext{We diagonalize the Hamiltonian both with and without self-consistency using a homogeneous bulk magnitude $|\Delta_j|$. Our criterion for self-consistency is that from self-consistency iteration number $m$ to ${m+i}$, the global relative error $\epsilon_{\mathrm{G}} = \left \| \Delta_{m+1}-\Delta_{m} \right \|_2/\left \|\Delta_{m} \right \|_2 < 10^{-7}$, where we use a Polyak convergence accelerator~\cite{SuperConga:2023}. We generally find negligible influence of self-consistency apart from a small quantitative difference, which can be reproduced without self-consistency by simply adjusting $\Delta_0$.}}.
We fix the phases in the two superconducting leads using the phase difference $\Delta\theta \in [0,2\pi)$ as the relevant phase parameter.
A finite phase difference $\Delta\theta>0$ leads to a supercurrent through the Josephson junction, which we compute via the bond charge current from site $k$ to $j$ via the current operator~\cite{zhu.16}
\begin{align}
    \label{eq:current}
    I_{jk}=\frac{e}{i\hbar} \sum_{\nu,\sigma} \left[t_{jk}u_{j\sigma}^{\nu*}u_{k\sigma}^{\nu}f(E_\nu) - t_{jk}^*u_{k\sigma}^{\nu*}u_{j\sigma}^{\nu}f(E_\nu)\right],
\end{align}
with elementary charge $e$, reduced Planck constant $\hbar$, eigenvectors $u$ and $v$, Fermi-Dirac distribution $f(E_\nu)$, and $\nu$ labels the eigenstates.

The current contribution from individual energy levels $\nu$ are also obtained via the usual energy-phase dispersion at zero temperature $I_\nu(\Delta\theta) = -(2e/\hbar)(dE_\nu/d\Delta\theta)$~\cite{shumeiko.97}, and we verify that $I(\Delta\theta) = \sum_\nu I_\nu$ reproduces the same result as Eq.~(\ref{eq:current}).
The critical current $\Ic$ is defined as the maximum of the current-phase relation,
\begin{equation}
    \label{eq:critical_current}
    \Ic \equiv I(\Delta\thetac) \equiv \max \{I(\Delta\theta) : \Delta\theta\},    
\end{equation}
where $\Delta\thetac$ is the critical phase difference.

Next, we estimate the effective superconducting coherence length in our model system.
We consider the commonly used expression for the ballistic superconducting coherence length $\xi_0 = \hbar\vF/(\pi\Delta_0)$, with Fermi velocity $\vF$ computed from the normal-state dispersion in the leads (using a single-band model at half-filling).
We obtain $\xi_0 \approx 11 a_0$.
This value is slightly smaller but still the same order as obtained through fitting the exponential decay of the ABS in the main text (see Sec.~\ref{sec:results:josephson_effect}), where we find $\xi_0 \approx 17 a_0$.
We use these estimates as a representative length scale to quantify the normal junction length $L$, and for simplicity assume that $L$ is the same as the effective junction length~\cite{likharev.79}, such that a long (short) Josephson junction corresponds to $L \gg \xi_0$ ($L \ll \xi_0$). 
Thus, the short junction limit $L \ll \xi_0$ essentially corresponds to a quantum dot ($L \sim a_0$) in our system.
We note that the interesting physics comes from the unique quasiperiodic spatial dependence, while such a quantum dot has no spatial extent.
Still, for full transparency we consider junctions from single sites to thousands of sites, covering both the short and long junction regimes.

Finally, we note that recent studies of a superconductor-quasicrystal SN interface (i.e.~not an SNS or SIS Josephson junction as in our work) have shown that the proximity effect can be enhanced by the quasiperiodicity, due to the topological states in the quasicrystal~\cite{rai.19,rai.20}.
Our earlier calculations~\cite{sandberg.22} reproduce these results but in an SNS Josephson junction.
In the current work, we instead focus our attention on the ABS spectrum and Josephson current.
Furthermore, Refs.~\cite{rai.21} and~\cite{rai.20} have shown that moderate disorder or impurities do not significantly modify the most crucial properties of the Fibonacci chain, and we therefore leave such perturbations as an outlook.
Instead, our study focuses on quantifying the DC Josephson effect in ballistic weak links with a quasiperiodic normal region described by different Fibonacci chains with and without finite repetition $N$, for different values of the superconducting phase difference $\Delta\theta$, hopping ratios $\rho$ and $\rhoS$, interface hopping $\tint$, electrochemical potential $\mu$, and phason angle $\phi$.

\section{Spectrum and current-phase relation}
\label{sec:results:josephson_effect}
In order to highlight the influence of quasiperiodicity on the DC Josephson effect, we in this section study how the ABS spectrum ($E<|\Delta|$) and Josephson current depend on the superconducting phase difference $\Delta\theta$ for different hopping ratios $\rho$ and interface hoppings $\tint$, keeping other parameters fixed.

\begin{figure}[t!]
    \centering
    \includegraphics[width=\columnwidth]{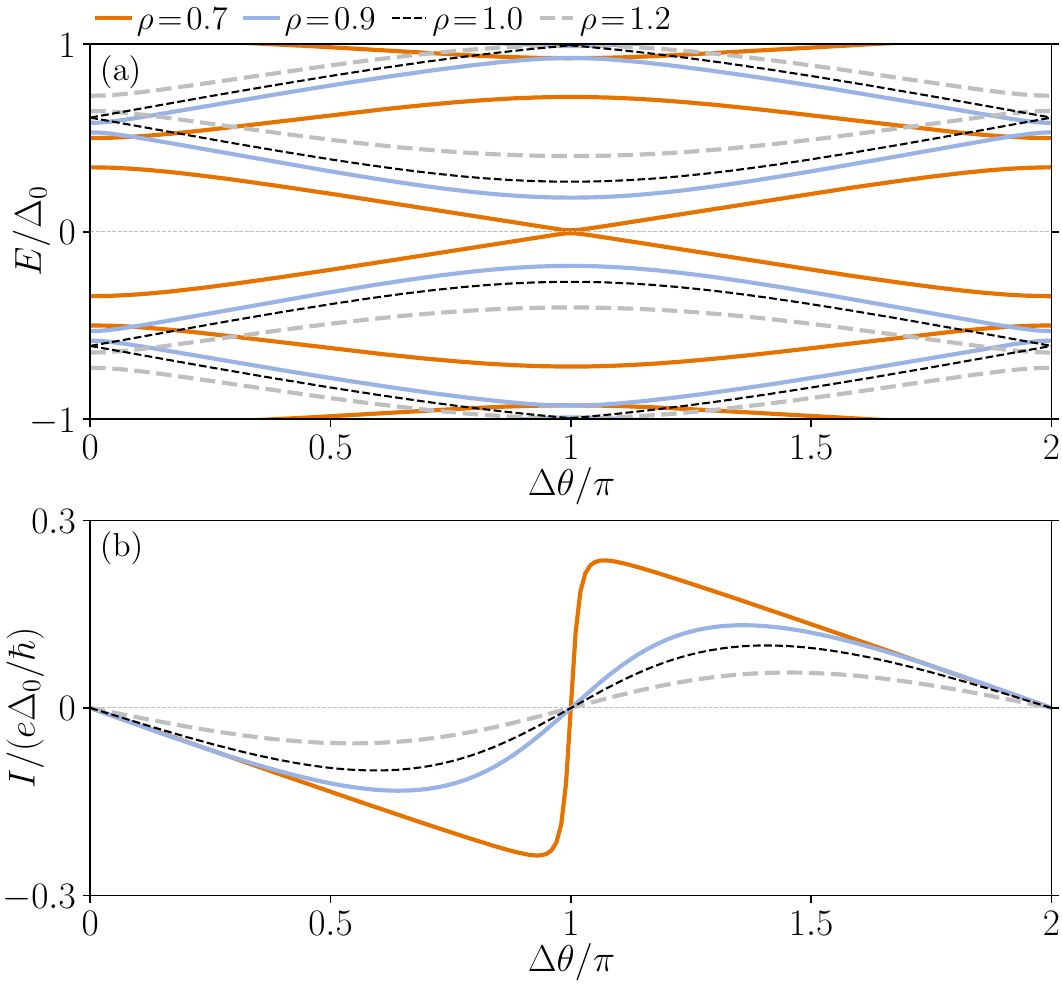}
    \caption{
    Energy eigenvalues $E$ (a) and supercurrent $I$ (b) as functions of the phase difference $\Delta\theta$ between two superconducting leads contacted by the Fibonacci approximant $C_9$ with $F_9+1 = 56$ sites. Parameters: $\tint = 0.7\tB$, $\rhoS=1$, $\mu=\muS=0$, $\Delta_0 = 0.06\tB$, where $\rho=1$ ($\rho \neq 1$) corresponds to a crystalline (quasiperiodic) junction.
    }
    \label{fig:basic_josephson_current:even}
\end{figure}

Figure~\ref{fig:basic_josephson_current:even} shows the low-energy spectrum (a) and Josephson current-phase relation (b) in a Josephson junction where the non-superconducting region is the Fibonacci approximant $C_9$ for different $\rho$ (line colors) at $\tint=0.7\tB$, $\rhoS=1$ and $\mu=0$.
We note that all energy levels in Fig.~\ref{fig:basic_josephson_current:even}(a) are subgap ABS and that the number of such states is directly related to e.g.~superconducting gap versus level spacing (or analogously the superconducting coherence length versus junction length)~\cite{bena.12}.
Notably, the crystalline junction ($\rho=1$) shows the usual ABS degeneracy at $\Delta\theta=0$ and $2\pi$, while the quasiperiodic junction ($\rho \neq 1$) shows no such degeneracy.
Generally, the degeneracy can break whenever there is an asymmetry between \newtext{left- and right-moving quasiparticle states, for instance related to an asymmetry between} the two leads~\cite{bena.12}.
Here, we find that the broken degeneracy \newtext{in the quasiperiodic junction is related to the $C_9 = \tA\tB\tA{\bm \tA} \ldots {\bm \tB}\tA\tB\tA$ approximant having} different local hopping neighborhood at each superconducting lead whenever $\rho \neq 1$.
We verify that the degeneracy is also broken in other asymmetric approximants e.g.~$C_{10} = \tA \ldots \tB$, while it is conserved in symmetric approximants e.g.~$C_3 = \tA\tB\tA$ regardless of $\rho$ (or number of repetitions $N$).
More generally, we also find conserved degeneracy when the characteristic function Eq.~(\ref{eq:characteristic_function}) is terminated at a point which conserves the symmetry between leads (e.g.~when the number of bonds are $F_n - 2$).
These results illustrate the important concept of how different Fibonacci chains can lead to both qualitatively and quantitatively different behavior, a reoccurring theme throughout our work.

Next, we focus on the Josephson current in Fig.~\ref{fig:basic_josephson_current:even}(b), which in this case is enhanced by quasiperiodicity for all $\rho<1$ \newtext{compared to the crystalline case ($\rho=1$)}.
We can explain this in terms of how the Andreev reflection is modified by the quasiperiodic hopping asymmetry $\rho\neq1$, together with the superconducting hopping ratio $\rhoS$ and interface hopping $\tint$, since they effectively model the transmission together.
To start, it is well-known that a crystalline junction shows maximal current at ideal junction transmission (i.e.~at perfect Andreev reflection and thus zero normal reflection), where the current-phase relation turns into a linear sawtooth profile with a discontinuity at $\Delta\theta = \pi$~\cite{ishii.70}.
At this phase difference, the lowest-energy ABS becomes an exact degenerate zero-energy state with perfect resonance (i.e.~wave-function matching)~\cite{affleck.00}, also associated with a Jackiw-Rebbi zero mode~\cite{jackiw.rebbi.76}.
But this perfect scenario is in a sense fine-tuned since any deviation from ideal transmission lifts the zero-energy degeneracy~\cite{affleck.00}.
It is reasonable to expect that realistic materials show a reduced transmission due to interface imperfections or other effects.
Specifically, Fig.~\ref{fig:basic_josephson_current:even} corresponds to a system with noticeably lower transmission for the crystalline junction $\rho=1$ (as modeled by $\tint=0.7\tB$).
For such reduced junction transmission, there is an increased occurrence of normal reflection instead of Andreev reflection, and the resonance is not perfect.
As a result, the ABS is shifted to finite energy, obtaining a smaller and non-linear phase dispersion, thus yielding a reduced current~\cite{furusaki.92} with the sinusoidal shape seen for $\rho=1$ in Fig.~\ref{fig:basic_josephson_current:even}(b).
The quasiperiodic junction, on the other hand, instead approaches large current and sawtooth profile for smaller values of 
$\rho$.
We note that this can also occur for larger $\rho$ depending on the other model parameters (which we show in Sec.~\ref{sec:results:critical_current}).
Thus, the quasiperiodic hopping asymmetry modifies the ABS spectrum and condition for an ABS at zero energy (i.e.~due to perfect Andreev reflection), the latter we denote $\tintStar(\rho,\rhoS)$ and which we fully quantify for different quasiperiodic junctions in Sec.~\ref{sec:results:critical_current}.

\begin{figure}[t!]
    \centering
    \includegraphics[width=\columnwidth]{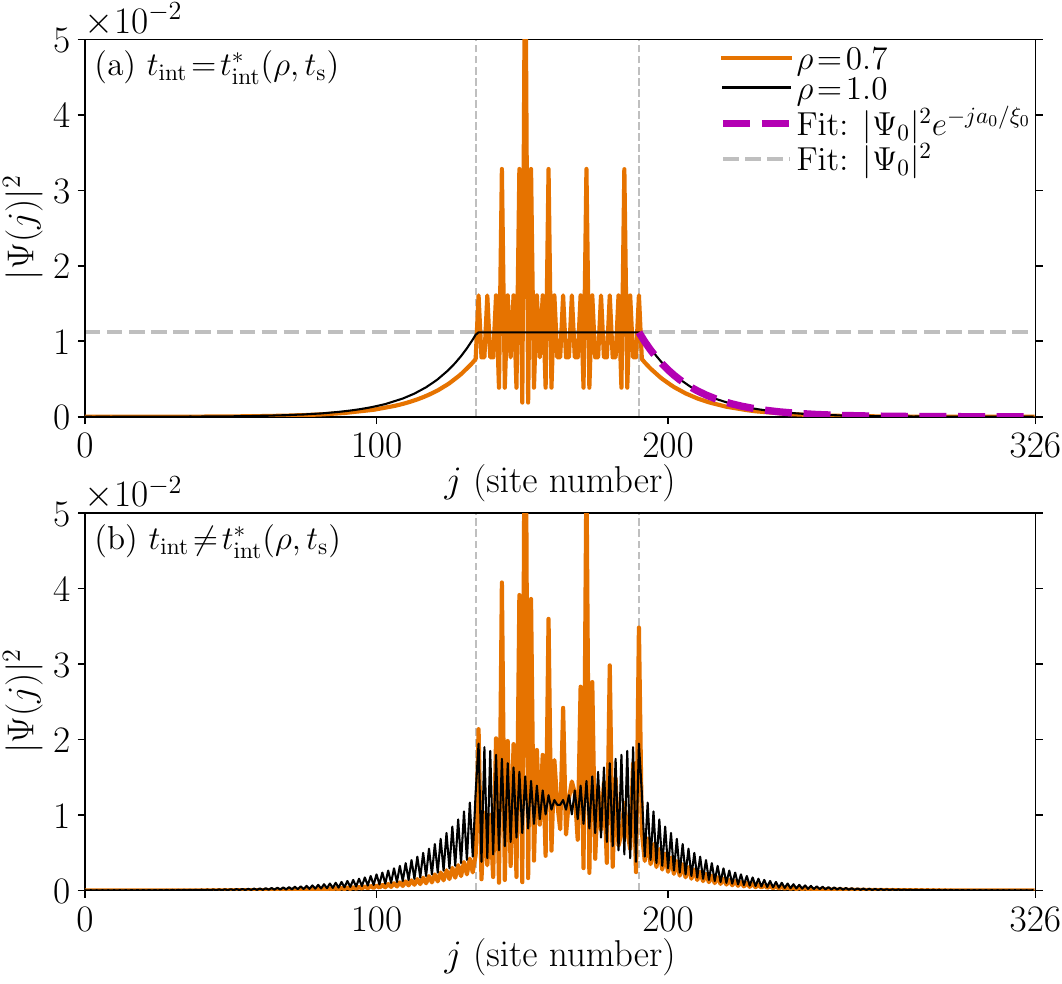}
    \caption{Probability density $|\Psi(j)|^2$ at site $j$ for lowest-energy ABS, at zero energy with perfect resonance $\tint=\tintStar(\rho,\rhoS)$ (a) and at finite energy off resonance $\tint=0.5\tintStar(\rho,\rhoS)$ (b). Vertical lines: guides to the eye marking the junction interfaces. Other lines: data (solid) or fits to the data (dashed) in the central region (gray) and leads (purple). Other parameters: same as Fig.~\ref{fig:basic_josephson_current:even}.
    }
    \label{fig:resonant_abs}
\end{figure}

To gain better insight into the lowest-energy ABS, we plot its probability density $|\Psi(j)|^2$ versus site $j$ in Fig.~\ref{fig:resonant_abs}, at zero energy (perfect resonance) $\tint = \tintStar(\rho,\rhoS)$ (a) and at finite energy (off perfect resonance) $\tint = 0.5\tintStar(\rho,\rhoS)$ (b).
The ABS shows an exponential localization in the superconducting lead for the crystalline junction, which we fit in (a) with $|\Psi^{\rm fit}(j)|^2 = |\Psi_0|^2 \exp(-ja_0/\xi_0)$, where $|\Psi_0|^2$ is the constant resonant value in the junction. We find an excellent fit for $\xi_0 \approx 17 a_0$. 
Interestingly, we see that quasiperiodicity causes the ABS to obtain similar quasiperiodic oscillations as critical states, but here superposed on top of the otherwise constant and resonant level $|\Psi_0|^2$. 
Off resonance in (b), the spatial dependence is superposed with microscopic oscillations due to wave-function mismatch at the interface, which grow with $|\tint-\tintStar|$.
We note that there is a broken symmetry in $|\Psi(j)|^2$ between the two leads for $\rho\neq1$, which is most visible in (b), and which becomes even more obvious for higher-energy modes (not shown).
We verify the above behavior for other approximants $C_n$ both with and without repetition $N$.

\newtext{Finally, we comment that the ABS degeneracy and current-phase relation studied above can be significantly altered in crystalline junctions by microscopic effects, e.g.~with a zero-energy state existing already in the normal state due to symmetry~\cite{bena.12}, which can result in a sawtooth current-phase profile for all interface transparencies.
In Appendix~\ref{app:even_odd} we investigate these effects in the presence of quasiperiodicity, and find that such a robust sawtooth profile is either maintained or broken depending on the quasiperiodic approximant structure.}

\section{Critical current}
\label{sec:results:critical_current}
Having established the influence of quasiperiodicity on the ABS spectrum and current-phase relation in Sec.~\ref{sec:results:josephson_effect}, we next look to its influence on observable features like the critical current.
In particular, we here quantify how the critical current $\Ic$ defined in Eq.~(\ref{eq:critical_current}) depends on the hopping parameters in different approximants $C_n$ with and without repetition, and in more general quasiperiodic junctions of length $L$ as modeled by the characteristic function in Eq.~(\ref{eq:characteristic_function}).
We further determine the functional form of the condition $\tintStar(\rho,\rhoS)$ for the emergence of a zero-energy ABS, which we use to analyze several emergent features.
\newtext{Finally, we compute the critical current as function of junction length and find that it exhibits quasiperiodic oscillations superposed on the usual decay found in crystalline junctions.
While some of the parameter dependencies considered in this section may not be tunable in solids (although they might be in certain atomically engineered chains or synthetic quasiperiodic metamaterials~\cite{kouwenhoven.90,chatterjee.22,kuzmanovski.23,splitthoff.24,zubchenko.24}), our theoretical model calculations serve to establish the fundamental influence of the quasiperiodic hopping modulation on the critical current.}

\subsection{Zero-energy ABS and critical current}
\label{sec:results:critical_current_conditions}
In order to determine the influence of quasiperiodicity on observable features, we start by studying how the lowest-energy ABS and the critical current vary with $\tint$ and $\rho$.
We then also vary $\rhoS$ to fully quantify the condition $\tintStar(\rho,\rhoS)$ for the emergence of zero-energy ABS.
We show that the condition $\tintStar(\rho,\rhoS)$ changes its functional form for different Fibonacci approximants and junction lengths, leading to qualitatively different behaviors for the critical current compared to crystalline junctions.

\begin{figure}[t!]
    \centering
    \includegraphics[width=\columnwidth]{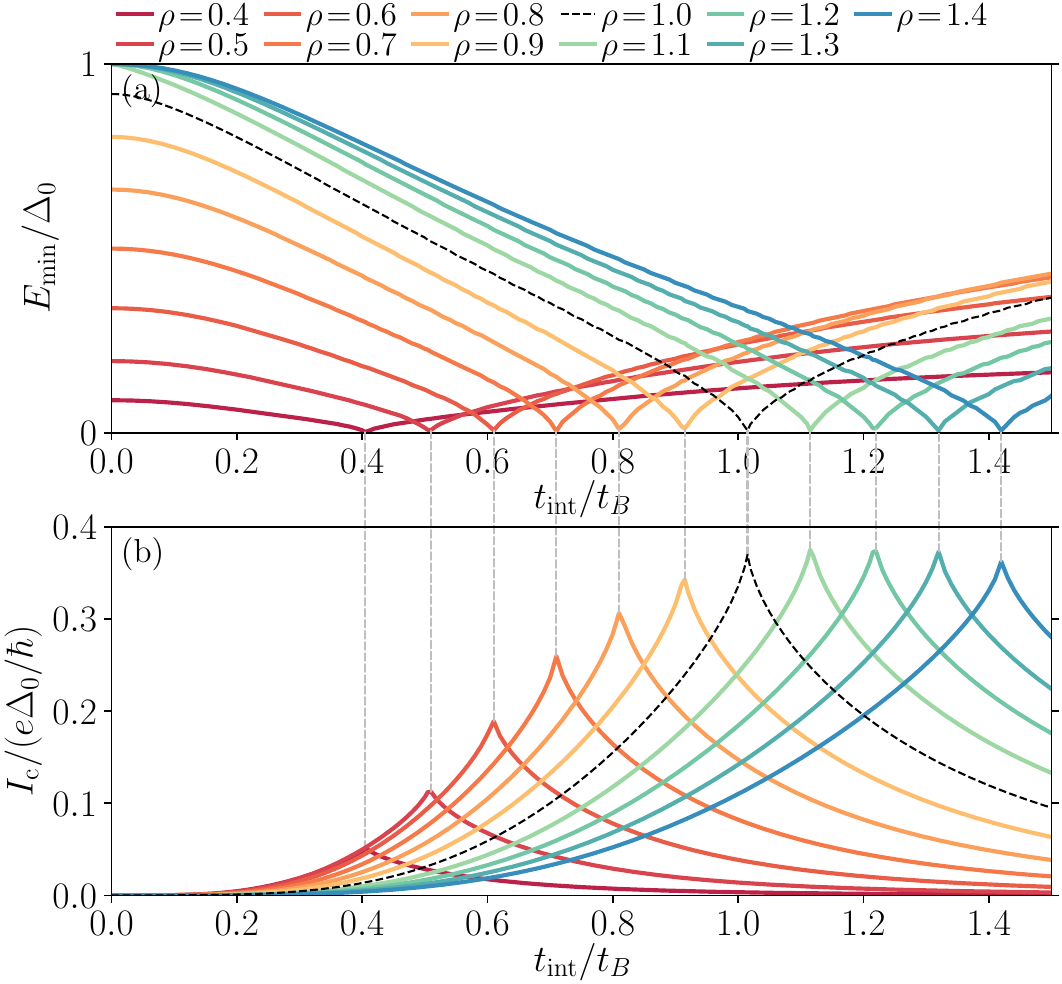}
    \caption{Energy of lowest-energy ABS ($\Emin$) (a) and critical current $\Ic$ (b) as functions of interface hopping $\tint$ for different hopping ratios $\rho$ (line colors), at the critical phase difference $\Delta\theta=\Delta\thetac$ for the Fibonacci approximant $C_9$. Parameters: $\mu=0$, $\rhoS=1$. Vertical dashes: guide to the eye indicating the overlapping extrema between (a) and (b).
    }
    \label{fig:critical_current_tint}
\end{figure}

We start by studying the dependence of the spectrum and critical current on $\tint$ and $\rho$.
In particular, Fig.~\ref{fig:critical_current_tint} shows the energy eigenvalue of the ABS closest to zero energy $\Emin$ (a) and the critical current $\Ic$ (b) as functions of $\tint$ for different $\rho$ (line colors), in a junction described by the Fibonacci approximant $C_9$.
Here, it is important to note that the regime $\tint > \tB$ may be as reasonable as $\tint < \tB$ since there are two hopping terms in the Fibonacci chain ($\tA$ and $\tB$) and another in the superconducting leads ($\tS$), which might be quite dissimilar.
In particular, we verify that the most important results in the following discussion are valid when e.g.~$\tB < \tint \leq \tS,\tA$.
In Fig.~\ref{fig:critical_current_tint}(a) $\Emin$ approaches zero at a specific $\tint = \tintStar(\rho,\rhoS)$ due to perfect Andreev reflection~\cite{affleck.00}, correlating with the maximal critical current in Fig.~\ref{fig:critical_current_tint}(b).
We find that the current is fully carried by the lowest-energy state at these extrema due to zero contribution from all other states, and that the current-phase relation becomes a sawtooth profile (not shown).
In contrast, away from the extrema in $\Emin$ and $\Ic$, the energy-phase slope reduces and the current-phase relation becomes sinusoidal, where other states also start contributing destructively.
Interestingly, Fig.~\ref{fig:critical_current_tint} illustrates that for $\tint < \tB$ ($\tint > \tB$), quasiperiodicity can significantly enhance the current for most $\rho < 1$ ($\rho > 1$) \newtext{as compared to the crystalline junction with $\rho=1$}, i.e.~without any parameter fine-tuning.

To understand the behavior in $\Emin$ and $\Ic$, we simultaneously vary $\rho$, $\rhoS$ and $\tint$ and find that $\tintStar(\rho,\rhoS)/\tB = \rho\sqrt{\rhoS}$ for the $C_9$ approximant, see Appendix~\ref{app:resonant_condition} for calculation details.
This is consistent with the positions of the extrema (e.g.~$\tintStar/\tB \approx 1$ for $\rho=\rhoS=1$) up to a small deviation which we attribute to corrections in powers of $|\Delta|/\tB$, i.e.~vanishing for $|\Delta| \ll \tB,\tA,\tS$~\cite{affleck.00}.
We proceed to quantify the condition $\tintStar(\rho,\rhoS)$ in Table~\ref{tab:resonant_ABS} for different Fibonacci approximants up to $C_{17}$ where the junction length $L=a_0F_n$ is more than two (three) orders of magnitude larger than the coherence length $\xi_0$ (atomic scale $a_0$).
We find that in junctions with even number of sites, $\tintStar(\rho,\rhoS)$ is described by one of three functional forms $\rho\sqrt{\rhoS}$, $\sqrt{\rho}\sqrt{\rhoS}$, $\sqrt{\rhoS}$, thus always scaling as $\propto \sqrt{\rhoS}$.
Thus, we find qualitatively different behavior from in crystalline junctions, where it is known that a normal region with a single hopping $\tA$ scales as $\tintStar(\tA,\tS) = \sqrt{\tA\tS}$ in the limit $|\Delta| \ll \tA,\tS$~\cite{affleck.00}.
Furthermore, in the junctions with odd number of sites (i.e.~every third approximant), there is already a zero-energy ABS in the normal state~\cite{bena.12}, and we here let $\tintStar(\rho,\rhoS)$ denote for which $\rho$ and $\rhoS$ the state remains at zero energy.
Thus, we find a zero-energy ABS $\forall \tint,\rhoS$ for $\rho=1$ (and $\forall\rho,\rhoS$ at $\tint\to0$) as described in Appendix~\ref{app:even_odd}.

\begin{table}[t!]
\begin{ruledtabular}
\begin{tabular}{l|l|l|l}
     $C_n$ & $F_n+1$ (\#sites) & bond structure & $\tintStar(\rho,\rhoS)/\tB$ \\
\hline
    $C_{0}$ & $2$ & $\tB$ & $\sqrt{\rhoS}$ \\
    $C_{1}$ & $2$ & $\tA$ & $\sqrt{\rho}\sqrt{\rhoS}$ \\
    $C_{2}$ & $3$ & $\tA\tB$ & $\rho=1, \forall\rhoS$ \\
    $C_{3}$ & $4$ & $\tA\tB\tA$ & $\rho\sqrt{\rhoS}$ \\
    $C_{4}$ & $6$ & $\tA\tB\tA\tA\tB$ & $\sqrt{\rho}\sqrt{\rhoS}$ \\
    $C_{5}$ & $9$ & $\tA\ldots\tA$ & $\rho=1, \forall\rhoS$ \\
    $C_{6}$ & $14$ & $\tA\ldots\tB$ & $\sqrt{\rhoS}$ \\
    $C_{7}$ & $22$ & $\tA\ldots\tA$ & $\sqrt{\rho}\sqrt{\rhoS}$ \\
    $C_{8}$ & $35$ & $\tA\ldots\tB$ & $\rho=1, \forall\rhoS$ \\
    $C_{9}$ & $56$ & $\tA\ldots\tA$ & $\rho\sqrt{\rhoS}$ \\
    $C_{10}$ & $90$ & $\tA\ldots\tB$ & $\sqrt{\rho}\sqrt{\rhoS}$ \\
    $C_{11}$ & $145$ & $\tA\ldots\tA$ & $\rho=1, \forall\rhoS$ \\
    $C_{12}$ & $234$ & $\tA\ldots\tB$ & $\sqrt{\rhoS}$ \\
    $C_{13}$ & $378$ & $\tA\ldots\tA$ & $\sqrt{\rho}\sqrt{\rhoS}$ \\
    $C_{14}$ & $611$ & $\tA\ldots\tB$ & $\rho=1, \forall\rhoS$ \\ 
    $C_{15}$ & $988$ & $\tA\ldots\tA$ & $\rho\sqrt{\rhoS}$ \\
    $C_{16}$ & $1598$ & $\tA\ldots\tB$ & $\sqrt{\rho}\sqrt{\rhoS}$ \\
    $C_{17}$ & $2585$ & $\tA\ldots\tA$ & $\rho=1, \forall\rhoS$ \\
\end{tabular}
\end{ruledtabular}
\caption{Condition $\tintStar(\rho,\rhoS)$ for emergent zero-energy ABS for different Fibonacci approximants $C_n$. Second column: the number of sites $F_n+1$ (i.e.~with physical length $L=a_0F_n$). 
Third column: the corresponding hopping structure in the Fibonacci chain, focusing on the first and last bond, i.e.~closest to the superconductor-Fibonacci chain interfaces. }
\label{tab:resonant_ABS}
\end{table}

Next, we note that $\tintStar(\rho,\rhoS)$ in Table~\ref{tab:resonant_ABS} varies cyclically in $n$ as we go to higher approximants $C_n$.
To better highlight this cycle, we write the condition for zero-energy ABS explicitly in terms of $\tA$ and $\tB$ starting with $C_3$ where $\tintStar(\tA,\tB) \propto \sqrt{\tA\tA}$, then $\sqrt{\tA\tB}$ in $C_4$, at $\tA=\tB$ in $C_5$, $\sqrt{\tB\tB}$ in $C_6$, $\sqrt{\tB\tA}$ in $C_7$, and finally at $\tA=\tB$ in $C_8$, after which the cycle repeats from $C_9$.
Thus the cycle corresponds to an equal occurrence of each of these scenarios, or, in other words, all the possible permutations with $\tA$ and $\tB$ under the square root.
Further in-depth analysis is left as an outlook.
Instead, we emphasize that the main point of the above analysis and results are to illustrate that different approximants have qualitatively different behavior for the zero-energy ABS (and thus the Andreev reflection) due to different hopping structures.
This naturally also leads to qualitatively different behaviors for the critical current, which we demonstrate below first for junctions with even number of sites then for odd number of sites.

\begin{figure}[t!]
    \centering
    \includegraphics[width=\columnwidth]{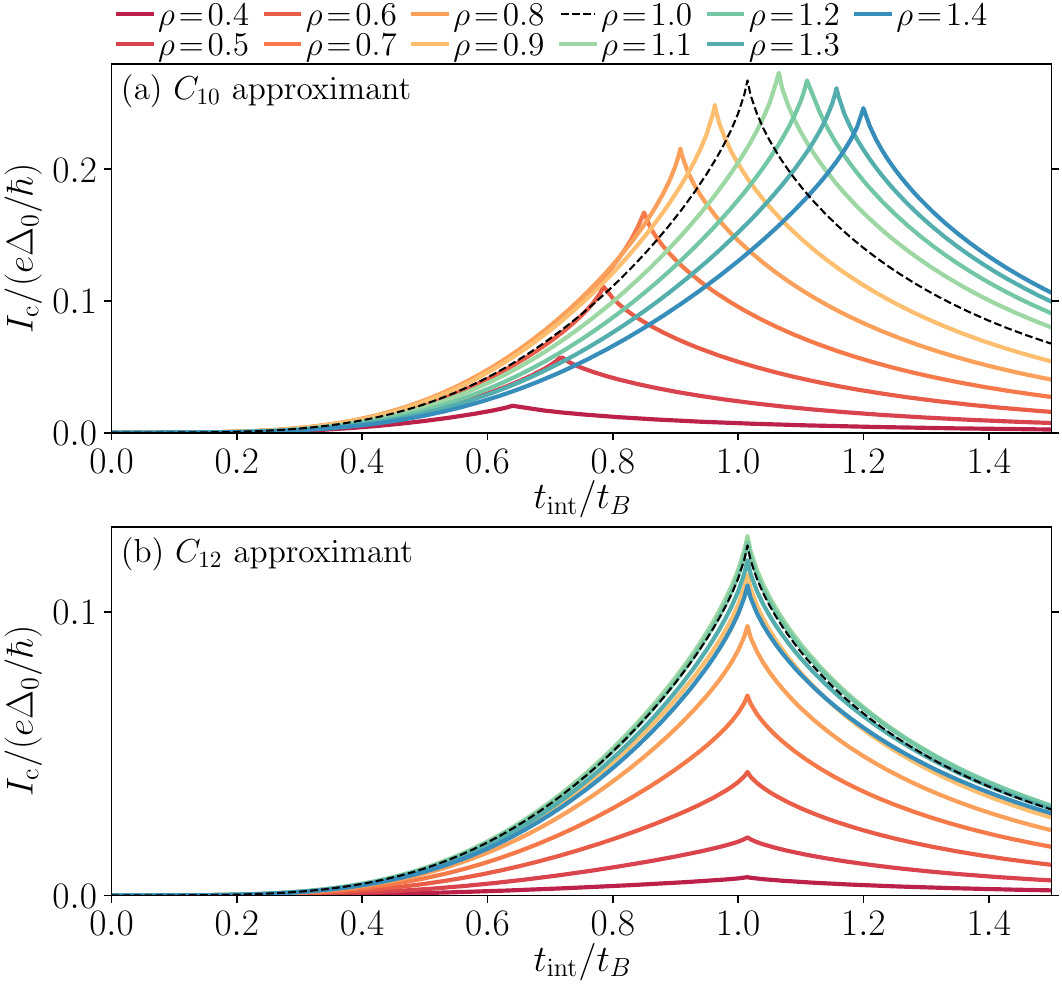}
    \caption{Same as Fig.~\ref{fig:critical_current_tint}(b) but for $C_{10}$ with $\tintStar(\rho,\rhoS) = \sqrt{\rho}\sqrt{\rhoS}$ (a) and $C_{12}$ with $\tintStar(\rho,\rhoS) = \sqrt{\rhoS}$ (b).
    }
    \label{fig:critical_current_tint_scenarios}
\end{figure}

Figure~\ref{fig:critical_current_tint_scenarios} shows the critical current as a function of $\tint$ through the $C_{10}$ junction with $\tintStar(\rho,\rhoS) = \sqrt{\rho}\sqrt{\rhoS}$ (a) and through the $C_{12}$ junction with $\tintStar(\rho,\rhoS) = \sqrt{\rhoS}$ (b).
These cases show smaller or no spread in $\Ic$ with $\rho$, respectively, as compared to $C_9$ with linear scaling $\rho\sqrt{\rhoS}$ in Fig.~\ref{fig:critical_current_tint}(b).
We also note that there is still a variation in the peak values for different $\rho$ in Fig.~\ref{fig:critical_current_tint_scenarios}(b), which we find to be caused by how $\rho$ changes the slope of the energy-phase dispersion.
Furthermore, we point out that quasiperiodicity can still influence the spread of the peaks in the $C_{12}$, since it still varies with $\tB$, but here we keep $\tB$ fixed, as the unit of energy.
Thus, allowing the hopping $\tB$ to vary in the model calculations can lead to $\tintStar(\tA,\tB,\tS)$ being proportional to e.g.~$\tB$ or $\sqrt{\tB}$.

\begin{figure}[t!]
    \centering
    \includegraphics[width=\columnwidth]{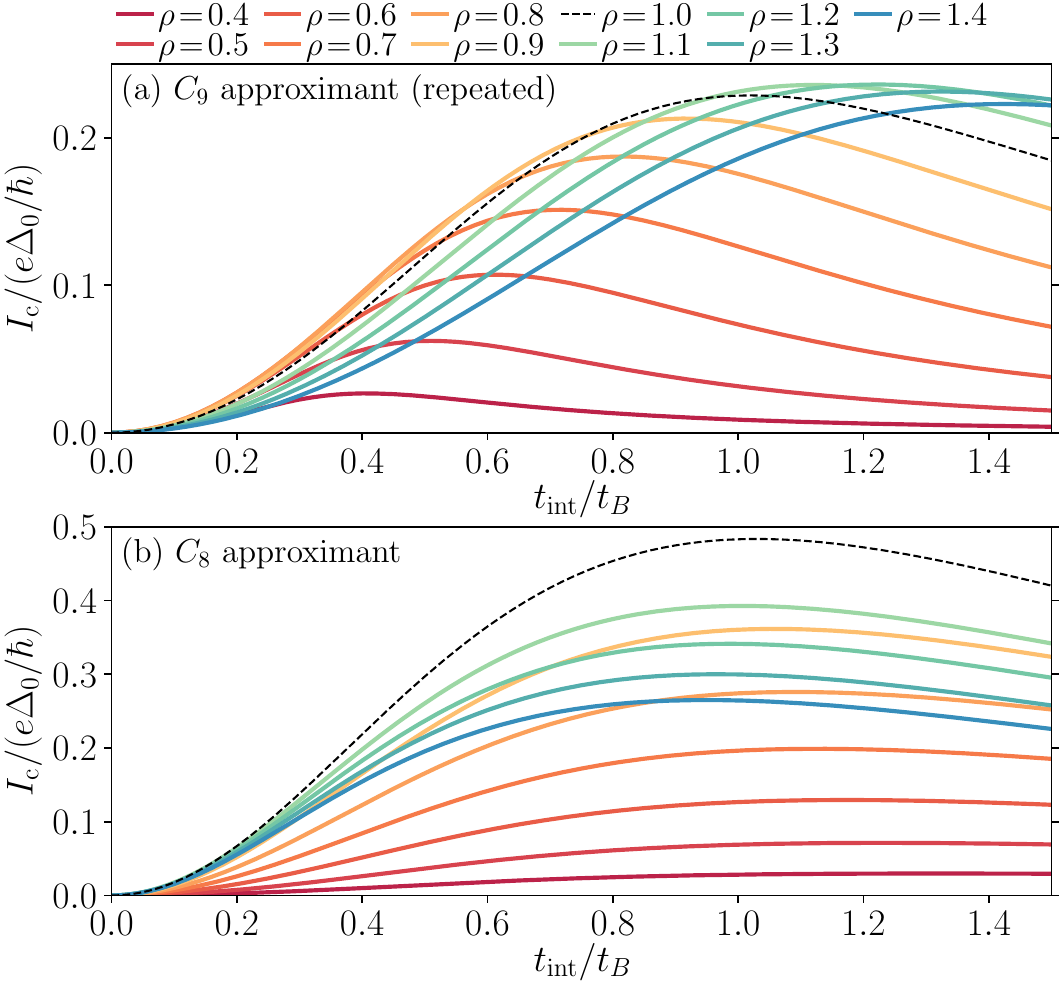}
    \caption{Same as Fig.~\ref{fig:critical_current_tint}(b) but for junctions with odd number of sites, for $C_{9}$ with $N=2$ repetitions (a) and $C_{8}$ without repetitions (b).
    }
    \label{fig:critical_current_tint_scenarios:odd}
\end{figure}

Next, Fig.~\ref{fig:critical_current_tint_scenarios:odd} shows $\Ic$ as function of $\tint$ in junctions with odd number of sites, corresponding to the approximant $C_9$ repeated $N=2$ times (a) and the approximant $C_8$ without repetitions (b).
Focusing first on the repeated $C_9$ approximant in Fig.~\ref{fig:critical_current_tint_scenarios:odd}(a), the critical current here shows a smooth variation with $\tint$ for every $\rho$. 
This is related to how varying $\tint$ and $\rho$ does not alter the functional form of the energy-phase dispersion of the lowest-energy ABS, and thus not the sawtooth current-phase relation, since there is always a zero-energy state (see Fig.~\ref{fig:basic_josephson_current:odd_repeated}). Instead, $\tint$ and $\rho$ only slightly alters the slope coefficient.
As a result, there is only a small difference in $\Ic$ between different $\rho$ close to the maxima in $\Ic$, i.e.~the effect of quasiperiodicty is small, while quasiperiodicity has a slightly more noticeable effect far from the maxima (i.e.~when normal reflection becomes significant).
This is in contrast to the $C_9$ approximant without repetition in Fig.~\ref{fig:critical_current_tint}(b), where the sharp peak is caused by a rapid decrease in phase dispersion due to changing from linear energy-phase slope to non-linear slope.
Comparing these scenarios with and without repetitions further, we note that the maximum in $\Ic$ occur at exactly the same $\tint$.
We explain this by that the two scenarios have the exact same local hopping structure close to each interface, resulting in the exact same wave-function matching criterion underlying the peak location.

Next, we note that for the $C_8$ approximant in Fig.~\ref{fig:critical_current_tint_scenarios:odd}(b), $\Ic$ also varies smoothly with $\tint$ at fixed $\rho$, again related to how $\tint$ influences the slope of the energy-phase dispersion.
In contrast to (a), however, the zero-energy degeneracy is broken for all $\rho\neq1$ causing a significant reduction of the critical current (see Fig.~\ref{fig:basic_josephson_current:odd}), such that the crystalline junction with $\rho=1$ always has a larger critical current.

Next, we briefly summarize how the condition for zero-energy ABS behaves in more generalized quasiperiodic junctions modeled by the characteristic function in Eq.~(\ref{eq:characteristic_function}), for different lengths $L \in [2,200]a_0$ (see Appendix~\ref{app:resonant_condition} for calculation details).
We find that junctions with odd number of sites follow the same behavior as either Fig.~\ref{fig:critical_current_tint_scenarios:odd}(a) or (b), i.e.~with a robust zero-energy mode $\forall\rho$, or only when $\rho=1$, respectively.
In junctions with even number of sites, we find that the condition $\tintStar(\rho,\rhoS)$ always scales as $\propto\sqrt{\rhoS}$, while the scaling in $\rho$ varies between different rational exponents, e.g.~$\rho^{3/2}$, $\rho$, $\sqrt{\rho}$, $1/\sqrt{\rho}$ and $1/\rho$.
Interestingly, we find that as $L$ increases, the system varies between the conditions in a quasiperiodic manner similar to how the hopping structure evolves with $L$.
This is consistent also with the approximants $C_n$, where the change in the functional form of $\tintStar(\rho,\rhoS)$ is cyclic in $n$, and therefore quasiperiodic in $L$ following the Fibonacci numbers, since $L = a_0F_n$.

Finally, we note that a deeper explanation for the exact analytic forms in $\tintStar(\rho,\rhoS)$ presented in this subsection is beyond our numeric calculations.
However, the quasiperiodic evolution indicates that the condition for perfect Andreev reflection relates non-trivially to the local quasiperiodic hopping structure close to the superconductor-Fibonacci chain interfaces.
We therefore propose studies based on other methods as an interesting outlook to shed additional light on this behavior, e.g.~using perturbation expansion or renormalization group theory~\cite{rai.21}.

\subsection{Critical current versus junction length}
\label{sec:results:size_scaling}
The previous Sec.~\ref{sec:results:critical_current_conditions} demonstrated a non-trivial behavior in the critical current for different approximants and thus different junction lengths.
Generally speaking, the critical current decays with junction length since the Josephson effect is a mesoscopic effect relying on phase coherence being mediated across the junction, via finite wave function overlap between the two superconducting leads~\cite{beenakker.92}.
In this subsection, we study how this decay is influenced by quasiperiodicity from $L \sim a_0 \ll \xi_0$ to $L = 1000a_0 \gg \xi_0$, thus varying from the short to long junction limit.
We start with the case of the repeated Fibonacci approximants $C_n$ with $L=a_0N\times F_N$ \newtext{(i.e.~relevant for the crystal approximants~\cite{goldman.93})}, followed by the non-repeated Fibonacci chain with more general $L$ modeled by the characteristic function Eq.~(\ref{eq:characteristic_function}) \newtext{(i.e.~more relevant for the Fibonacci quasicrystals~\cite{jagannathan.21})}.
\newtext{Although modern experimental techniques allows the fabrication of materials with atomic precision~\cite{eigler.90,stroscio.91,gomes.12,polini.13,drost.17,khajetoorians.19,schneider.20,huda.20,kuster.22,freeney.22}, the main purpose of this section is to theoretically establish how quasiperiodicity influences the overall trend $\Ic(L)$.} 

\begin{figure}[t!]
    \centering
    \includegraphics[width=\columnwidth]{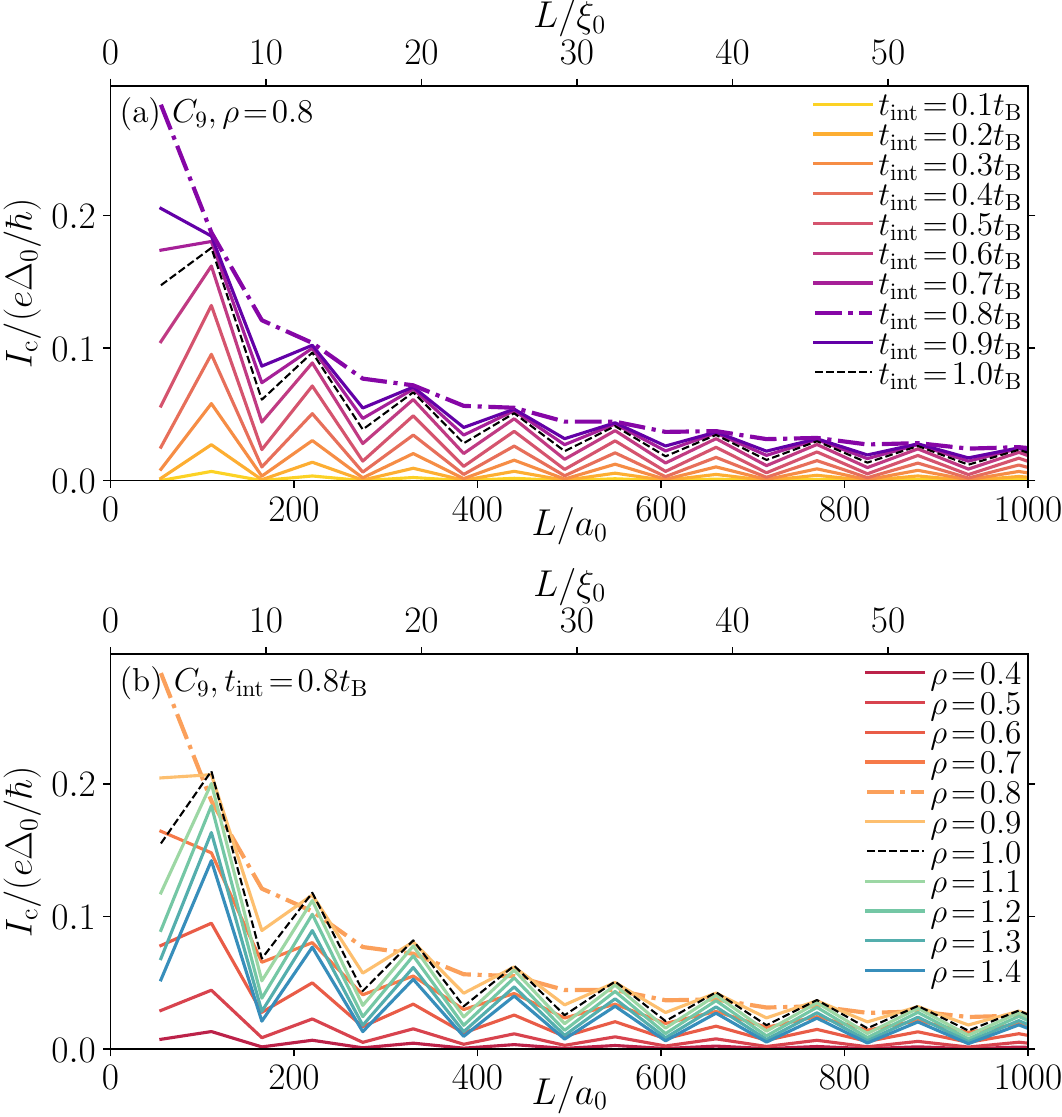}
    \caption{Critical current $\Ic$ as function of junction length $L = a_0N\times F_9$ at fixed $\rho=0.8$ (a) and fixed $\tint=0.8\tB$ (b). Here $\xi_0 \approx 17a_0$, and the junction is the Fibonacci approximant $C_9$ repeated $N$ times.
    }
    \label{fig:critical_current_repetition:C9}
\end{figure}

\begin{figure}[t!]
    \centering
    \includegraphics[width=\columnwidth]{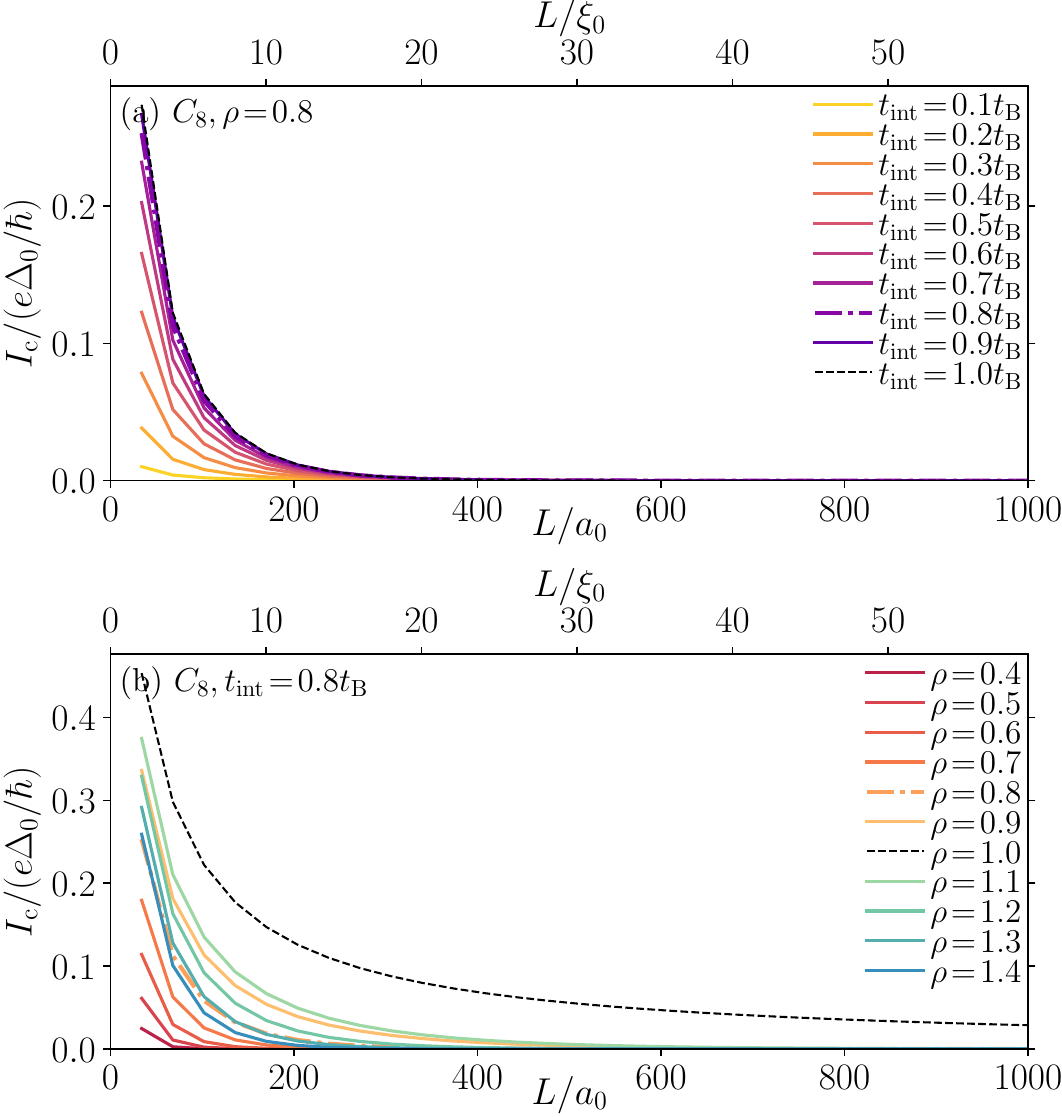}
    \caption{Same as Fig.~\ref{fig:critical_current_repetition:C9} but for $C_8$ approximant, which always has an odd number of sites regardless of repetition.
    }
    \label{fig:critical_current_repetition:C8}
\end{figure}

Beginning with a junction corresponding to a repeated Fibonacci approximant, Fig.~\ref{fig:critical_current_repetition:C9} shows $\Ic$ as a function of $L$ for the approximant $C_9$ at fixed $\rho=0.8$ (a) and at fixed $\tint=0.8\tB$ (b).
The critical current shows an overall decrease with $L$ as expected, but is superposed with a staggered behavior due to the microscopic even-odd effect described in Appendix~\ref{app:even_odd}.
Specifically, odd number of sites (even repetitions) leads to a robust zero-energy state which significantly increases the critical current, while for even number of sites there is only a zero-energy state at perfect Andreev reflection~\cite{affleck.00}, i.e.~here at $\tintStar(\rho,\rhoS)=\rho\sqrt{\rhoS}$.
Thus, the scenario $\tint/\tB=\rho=0.8$ at $\rhoS=1$ in Fig.~\ref{fig:critical_current_repetition:C9}(a) signifies the maximum possible critical current, yielding an envelope (purple dash-dotted line) of all other curves.
In (b) the curve for $\tint/\tB=\rho=0.8$ is only the maximum at even number of sites since the system with odd number of sites generally favors larger $\rho > \tint/\tB$ [see Fig.~\ref{fig:critical_current_tint_scenarios:odd}(a)].
We verify that other Fibonacci approximants follow a similar staggered behavior with an envelope curve given by the corresponding $\tintStar(\rho,\rhoS)$ expression given by Table~\ref{tab:resonant_ABS}.
The exception to the behavior is the Fibonacci approximants which always have an odd number of sites (i.e.~every third approximant), which we demonstrate in Fig.~\ref{fig:critical_current_repetition:C8} for the approximant $C_8$.
This approximant thus lacks the even-odd staggering, only illustrating the overall decay of $\Ic$ with increased $L$.
The current is maximal for $\rho=1$ and $\tint=\tB$ where there is a degenerate zero-energy ABS, while any other $\rho\neq1$ splits the ABS to finite energies thus reducing the current (see Fig.~\ref{fig:basic_josephson_current:odd}).
\newtext{Interestingly, by comparing the curves in Fig.~\ref{fig:critical_current_repetition:C8}(b) we find for $\rho\neq1$ a rapid decay that is approximately exponential and corresponds to the contribution from the finite-energy modes. For $\rho=1$, this is superposed with a long $\propto 1/L$ tail due to the zero-energy ABS, which is a well-known behavior in crystalline junctions~\cite{beenakker.92,affleck.00,nikolic.01,sonin.24}}.

\begin{figure}[t!]
    \centering
    \includegraphics[width=\columnwidth]{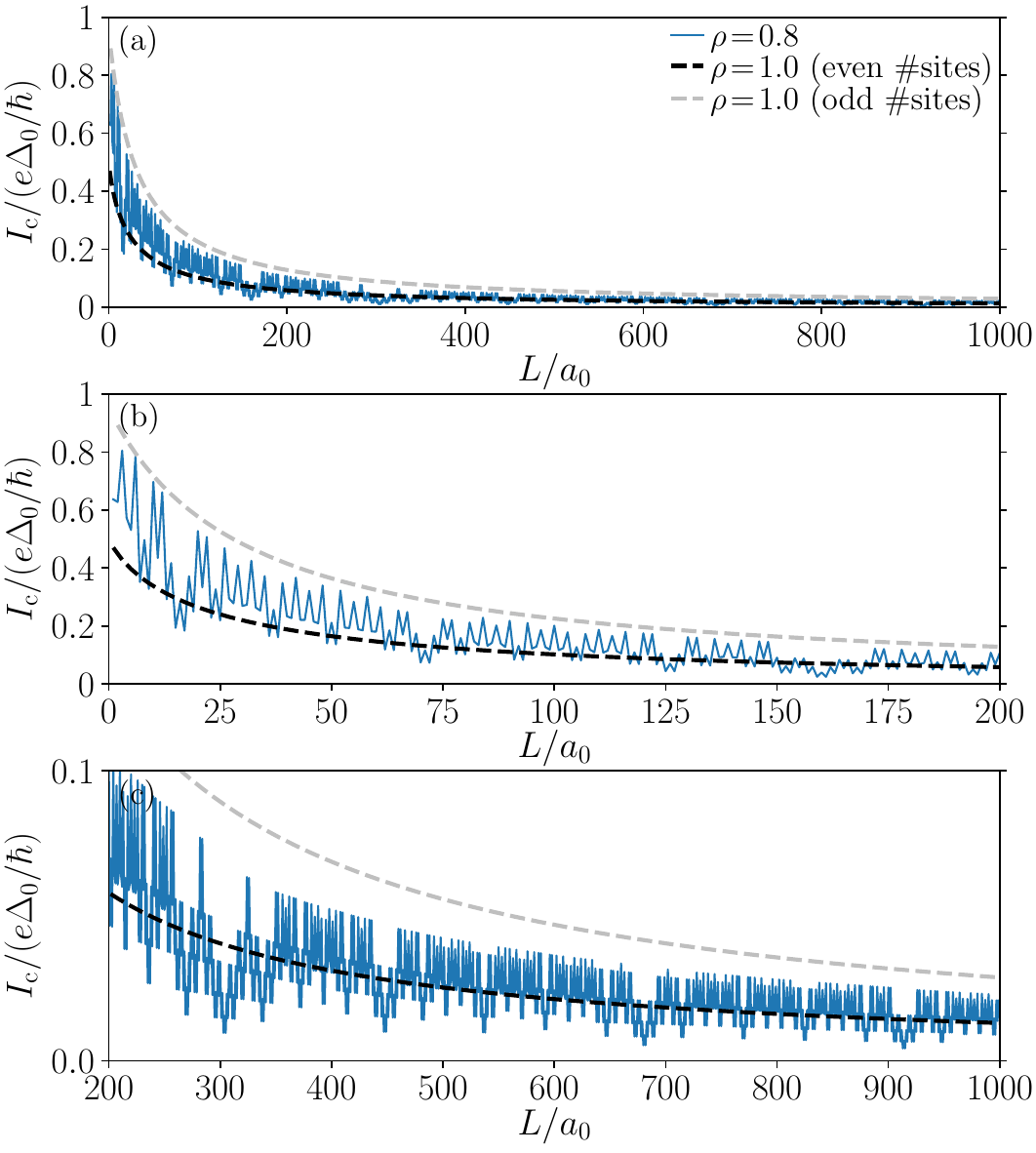}
    \caption{(a) Critical current $\Ic$ as a function of junction length $L$ for a Fibonacci chain (solid) modeled by the characteristic function Eq.~(\ref{eq:characteristic_function}) at $\rho=0.8$, $\tint=0.8\tB$ and $\rhoS=1$.
    Dashed black (gray) line: same but for a crystalline junction $\rho=1$ with even (odd) number of sites. Panels (b) and (c) show zooms of (a) in different regions.
    }
    \label{fig:critical_current_length}
\end{figure}

Next, we study the critical current decay in the more general quasiperiodic junctions without repetitions and of arbitrary length $L$, by adding one site at a time following the characteristic function in Eq.~(\ref{eq:characteristic_function}).
We note that the crystalline junction in this case obtains an extreme oscillation due to the even-odd effect.
To improve visibility, we therefore plot the crystalline results for even and odd number of sites separately.
We also note that each Fibonacci approximant $C_n$ is represented exactly once, i.e.~when $L = a_0F_n$, and we verify that this reproduces the result above.
Figure~\ref{fig:critical_current_length}(a) shows $\Ic$ as function of $L$ for a crystalline (quasiperiodic) junction as a dashed (solid) line, where Figs.~\ref{fig:critical_current_length}(b) and (c) are zooms of (a) at $L<200a_0$ and $L>200a_0$, respectively.
We begin by focusing on the crystalline results $\rho=1$, where even (odd) number of sites is shown as a black (gray) dashed line, \newtext{again showing the well-known monotonic decay~\cite{beenakker.92,affleck.00,nikolic.01,sonin.24}}.
The junction with odd number of sites has a significantly higher critical current since it is an idealized scenario with an exact zero-energy state as $\Delta\theta\to\pi$.
The junction with even number of sites approaches this curve as $\tint\to\tintStar(\rho,\rhoS)$, due to also obtaining a zero-energy state.

\begin{figure}[t!]
    \centering
    \includegraphics[width=\columnwidth]{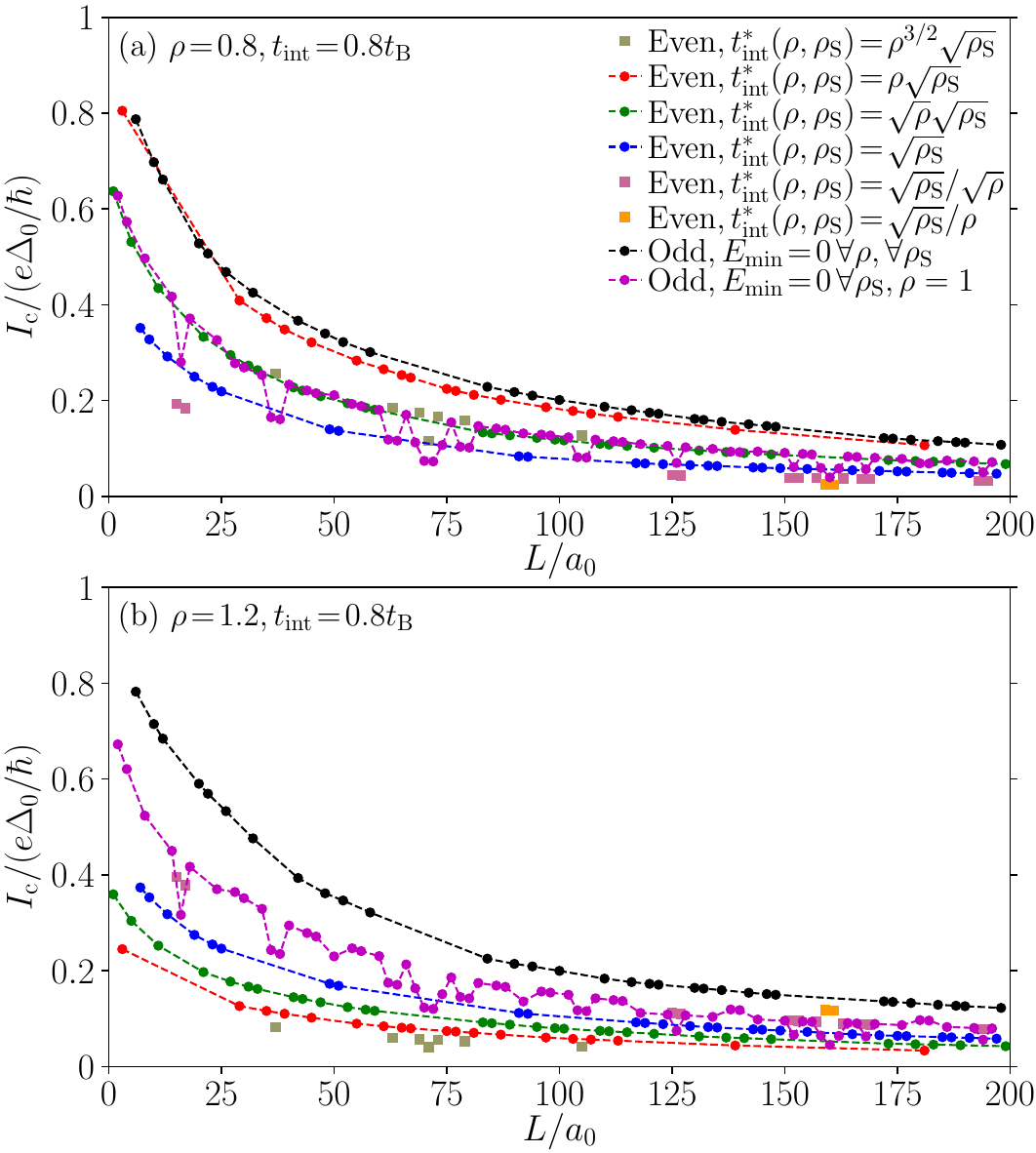}
    \caption{Same as Fig.~\ref{fig:critical_current_length}(b) but for the quasiperiodic junction at $\rho=0.8$, $\tint=0.8\tB$ (a), and at $\rho=1.2$, $\tint=0.8\tB$ (b).
    Data points (markers) are grouped by even or odd number of sites, and by the condition for zero-energy ABS, as given by Table~\ref{tab:resonant_ABS:characteristic_function} in Appendix~\ref{app:resonant_condition}. Dashes: guide to the eye showing the trend $\Ic(L)$ for the data points within the most populated groups.
    }
    \label{fig:critical_current_length:conditions}
\end{figure}

In the quasiperiodic junction $\rho\neq1$, Fig.~\ref{fig:critical_current_length} also illustrates an overall decay in $\Ic$ with $L$, but the microscopic oscillations are much less trivial, going beyond just an even-odd effect.
We find that these oscillations are not random or erratic, however, but rather correlate with how the junction varies between different expressions $\tintStar(\rho,\rhoS)$ with $L$.
We illustrate this Fig.~\ref{fig:critical_current_length:conditions} by plotting $\Ic$ as a function of $L$ with data points colored according to their functional form of $\tintStar(\rho,\rhoS)$ (as given by Table~\ref{tab:resonant_ABS:characteristic_function} in  Appendix~\ref{app:resonant_condition}).
The figure thus shows that all data points with $\tintStar(\rho,\rhoS)$ being e.g.~$\rho\sqrt{\rhoS}$, $\sqrt{\rho}\sqrt{\rhoS}$ or $\sqrt{\rhoS}$ each follow their respective monotonic decay as indicated by the dashed lines, except for the purple curve (see discussion further below).
\newtext{We have fitted these monotonic decays and find that they are similar to the well-known exponential and power law decays in the crystalline junctions, depending on system and parameter regimes~\cite{beenakker.92,affleck.00,nikolic.01,sonin.24}.}
The data points without dashed lines follow a similar trend, but these points are fewer and farther between, and we therefore omit plotting their lines for visibility reasons.
Comparing the curves for the different hopping parameters in (a) and (b), the lines are shifted with respect to each other since $\Ic$ is maximal at $\tintStar$, e.g.~the red line has higher $\Ic$ than the blue one in (a) since $\tint$ is closer to $\rho\sqrt{\rhoS}$ than $\sqrt{\rhoS}$, and vice versa in (b).
Next, we comment on the spatial shape of the microscopic oscillations.
Interestingly, the junction changes functional form of $\tintStar(\rho,\rhoS)$ with $L$ in a quasiperiodic manner as discussed in Sec.~\ref{sec:results:critical_current_conditions} and Appendix~\ref{app:resonant_condition}, such that the critical current $\Ic(L)$ therefore also varies in a quasiperiodic way.
This conclusion is based on studying the first 200 values of $L$ in the characteristic function, and the first 18 approximants $C_n$.
The exception to this behavior are the purple data points which instead show oscillations around a monotonic decay.
These data points correspond to when quasiperiodicity breaks the zero-energy degeneracy, and we attribute the oscillations to variations in the size of the finite-energy shifts.
We propose that studying the Andreev reflection amplitudes (i.e.~with a scattering-matrix approach) might yield additional insight into these oscillations, but leave it as a future outlook.

Finally, we briefly comment on how the overall magnitude of $\Ic$ compares between the quasiperiodic and crystalline junctions, when varying $\tint \in [0.1,1.0]\tB$, $\rho\in[0.1,1.4]$ and $L \in [2,1000]a_0$.
If taking the idealized scenario for odd number of sites into account, we find that the crystalline junction quite generally shows a higher critical current \newtext{than the quasiperiodic junction} except in a few cases, since the crystalline junction is ideal with perfectly ballistic extended states and exactly degenerate ABS.
However, if this idealized scenario with odd number of sites is not taken into account (in either the quasiperiodic or crystalline junctions), then the quasiperiodic junction can host a significantly higher critical current \newtext{than the crystalline junction} by several factors, especially at low $\tint$.
\newtext{Overall, we find microscopic quasiperiodic oscillations present in $\Ic(L)$, indicating significant sample-to-sample fluctuations in an experimental setup, while beyond these oscillations the current shows the same overall decay as in the crystalline scenario. }

\section{Fractal gap structure and topological invariant}
\label{sec:results:topological_invariant}
So far, we have considered a system at half filling through fixed chemical potential $\mu=0$ and with fixed phason angle $\phi=\pi\tau^{-1}$, at which the system behaves as a hybrid SNS Josephson junction.
\newtext{The purpose of this section is to investigate how the Josephson effect is influenced by quasiperiodicity when varying $\mu$, i.e.~varying an applied gate voltage in an experimental setup, as well as by the topological gaps and winding states of the Fibonacci chain.}
\newtext{In particular, in Sec.~\ref{sec:results:sns_to_sis} we demonstrate that varying the gate voltage leads to a controlled SNS to SIS transition as the Fermi level enters the gaps of the Fibonacci chain.}
Then, we show in Sec.~\ref{sec:results:phason_angle} how the winding of the subgap states in the different gaps lead to distinct oscillations in the critical current $\Ic$, such that each winding number (i.e.~gap label) can be determined, thus effectively measuring the topological invariant.
This occurs when the Fermi level is located inside or close to the respective gap, thus \newtext{accessible} by tuning the applied gate voltage.

\subsection{Gate-voltage tuning: fractal SNS to SIS transitions}
\label{sec:results:sns_to_sis}
\newtext{So far, most of our work has centered on systematic model calculations to establish a fundamental theoretical understanding of the Josephson effect in quasiperiodic junctions.
While some of the model parameters investigated may not be tunable in situ in an experiment once the device has been fabricated, our calculations still provide important fundamental understanding of the influence of the quasiperiodic modulation.
Here, we study the influence of the chemical potential $\mu$, as envisioned by the application of a gate voltage to the non-superconducting part of the junction.
Such a gate voltage is highly accessible in experiment, and our following results predict that it should lead to tunable and directly observable fractal oscillations between SNS and SIS behavior.}

\begin{figure}[t!]
    \centering
    \includegraphics[width=\columnwidth]{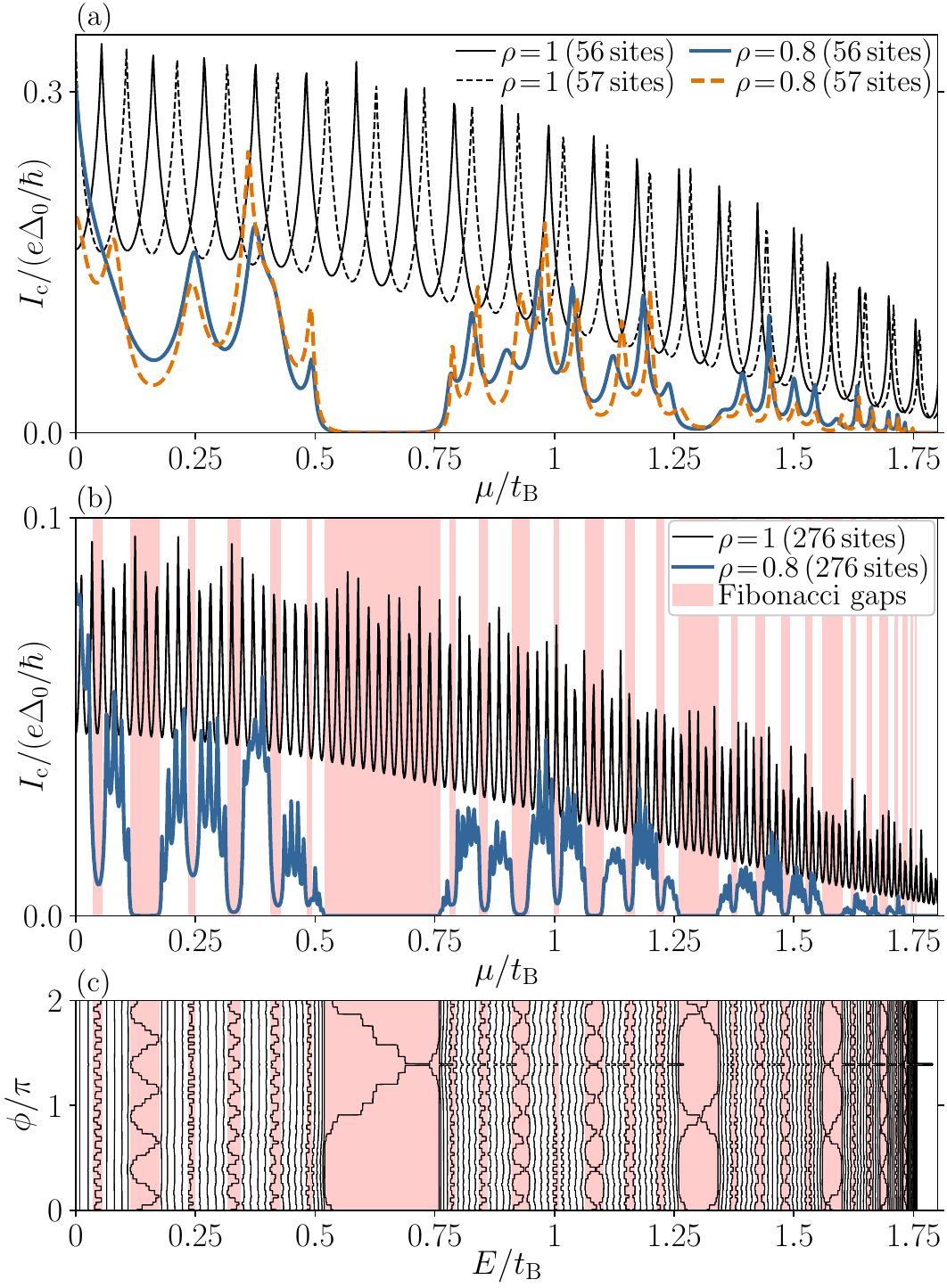}
    \caption{(a) Critical current versus chemical potential $\mu$ in junctions with $56$ sites (solid), i.e.~the $C_9$ approximant, and $57$ sites (dashed), for both quasiperiodic $\rho=0.8$ (thick) and crystalline hoppings $\rho=1$ (thin). Here, $\rhoS=1$ and $\tint=0.81\tB\approx\tintStar(\rho,\rhoS)$, while $\mu=0$ corresponds to half-filling. The peaks in $\Ic$ correspond to the discrete level spacing.
    (b) Same but for the $C_9$ approximant repeated $N=5$ times. Shaded regions show the location of topological gaps of the repeated Fibonacci chain when isolated ($\tint=0$, $\Delta=0$).
    (c) Phason angle-energy spectrum of the isolated Fibonacci chain in (b) illustrating the winding in each corresponding gap.
    }
    \label{fig:critical_current_mu}
\end{figure}

Figure~\ref{fig:critical_current_mu}(a) shows the critical current through a junction with even (odd) number of sites $F_n+1=56$ ($57$) as solid (dashed) lines, for both a quasiperiodic (thick lines) and crystalline (thin lines) junction.
Figure~\ref{fig:critical_current_mu}(b) shows the same but for the $C_9$ approximant repeated $N=5$ times ($276$ sites) leading to an overall lower current, and where the Fibonacci gaps are shown as shaded regions [not shown in (a) for visibility reasons].
We note that these are the corresponding gaps occurring at finite energy at $\mu=0$, i.e.~a gap around $E=E_1$ at $\mu=0$ occurs around $E=0$ at $\mu=E_1$ [see Fig.~\ref{fig:fibonacci_spectrum}(b)].
To show this correspondence and the winding number $q$ in each gap, we plot in Fig.~\ref{fig:critical_current_mu}(c) the phason angle $\phi$ versus energy spectrum.

Focusing first on the crystalline scenario $\rho=1$, we note that the sharp peaks in $\Ic$ in Figs.~\ref{fig:critical_current_mu}(a) and (b) are due to the idealized sharp and discrete energy levels at zero temperature, i.e.~occurring whenever an energy level crosses zero energy as function of $\mu$.
The peak repetition is thus given by the level spacing, and the even-odd effect is explicitly seen by the staggering of the peak position between the junction with even and odd number of sites, also related to the degeneracy in the ABS spectrum~\cite{bena.12}.
In a less idealized system, we expect the peaks to be broadened such that the variation of $\Ic$ with $\mu$ is much smoother.
Apart from these sharp peaks, the critical current shows an overall reduction from half-filling ($\mu=0$) towards the band edges ($|\mu|\approx2\tB$) beyond which the junction becomes insulating and the current vanishes.
The critical current is symmetric around $\mu=0$ and we therefore only show results for $\mu>0$.

Next, we focus on the quasiperiodic scenario $\rho\neq1$ and note the same features with peaks and overall larger current close to half-filling (but lower bandwidth $|\mu|<1.8\tB$).
Importantly, however, the quasiperiodic gap structure is directly probed by a sudden drop in current as the Fermi level enters the quasiperiodic gaps $\DeltaQ$.
We find that the size of these gaps relative to the proximitized superconducting gap $|\Delta|$ determines the junction behavior.
In particular, inside the larger gaps ($\DeltaQ>|\Delta|$) we find SIS behavior with the current becoming orders of magnitude smaller than outside the gaps where there is instead SNS behavior.
Inside the smaller gaps ($\DeltaQ<|\Delta|$) the behavior strongly depends on the junction length.
For instance, in the shorter junction in Fig.~\ref{fig:critical_current_mu}(a) the current is still the same order of magnitude as outside the gaps due to pronounced ABS and supergap contributions, and thus of SNS type, but the gaps still cause a reduction or even complete absence of some peaks that are otherwise present in the crystalline junction.
In contrast, in the longer junction in Fig.~\ref{fig:critical_current_mu}(b), the current is more strongly suppressed in the small gaps, showing more SIS-type behavior.
Thus, the gap ratio $\DeltaQ/|\Delta|$ together with the Fibonacci chain length can qualitatively alter the junction properties, and the SNS to SIS transitions in $\mu$ are fractal since the Fibonacci spectrum itself is fractal~\cite{jagannathan.21}. 
We note that the gap ratio was recently shown to also be important for both emergent topological superconductivity~\cite{kobialka.24} and intrinsic superconductivity~\cite{wang.24}.

Finally, we briefly comment on the overall magnitude of $\Ic$ in quasiperiodic versus crystalline junctions when varying $\tint \in [0.1,1.0]\tB$, $\rho\in[0.1,1.4]$ and $\mu$ between the band edges.
Similar to Sec.~\ref{sec:results:size_scaling}, we find that the crystalline junction in most cases has a higher critical current due to the idealized perfect ballistic model with an exact zero-energy state.
Only in a few parts of the parameter space does the quasiperiodic junction provide a significantly higher $\Ic$ \newtext{than the crystalline junction}, usually when $\tint$ is closer to the quasiperiodically modified expression for $\tintStar(\rho,\rhoS)$, or when $\tint$ is small and $\rho$ large.

\subsection{Phason angle dependence and topological invariant from critical current}
\label{sec:results:phason_angle}
In the previous Sec.~\ref{sec:results:sns_to_sis} we investigated the behavior of $\mu$ on $\Ic$, demonstrating that the junction can change between SNS and SIS behavior whenever the Fermi level enter the largest topological gaps of the Fibonacci chain.
In this section, we show how the critical current can measure the winding number in each of these gaps.
\newtext{We begin by studying the influence of the phason angle $\phi$, in order to better understand and quantify how phason modes influence the Josephson physics. Similar to phonons, phason modes propagate through quasiperiodic materials where they induce phason flips~\cite{jagannathan.21}, and we demonstrate that they have quantifiable influence on the critical current.}
We then describe the principle idea behind connecting the winding number to the critical current, followed by our results and proof-of-principle.
\newtext{Beyond providing a theoretical understanding of the influence of phason modes, these results could be measured e.g.~by using STM techniques~\cite{drost.17,huda.20,collins.17} to fabricate an ensemble of junctions each with a different phason angle, or exciting phason modes, or realizing an effective Fibonacci hopping model in a metamaterial~\cite{kouwenhoven.90,chatterjee.22,kuzmanovski.23,splitthoff.24,zubchenko.24} with quantum dots where phason flips are induced by in situ tuning the the coupling between the quantum dots.}

As shown in Fig.~\ref{fig:critical_current_mu}(c), the subgap states in the Fibonacci chain wind across the gap with the phason angle $\phi$~\cite{jagannathan.21} (see also discussion in Sec.~\ref{sec:model:fibonacci_spectrum}).
Our idea to measure the winding is based on that the closer these states wind to the Fermi level, the stronger their contribution is to the current-phase relation, and thus the critical current.
Specifically, by tuning the chemical potential close to a gap with winding number $|q|$, the variation with $\phi$ in the spectrum comes predominantly from the winding state, since all other states in comparison show no significant winding (per definition) and therefore have a constant contribution as function of $\phi$.
Furthermore, the winding state is at its closest and furthest from the Fermi level exactly $|q|$ times.
The conjecture is therefore that this should produce $|q|$ periodic oscillations in the current.
We note that this should furthermore even be observable in a small region outside the gap.

\begin{figure}[t!]
    \centering
    \includegraphics[width=\columnwidth]{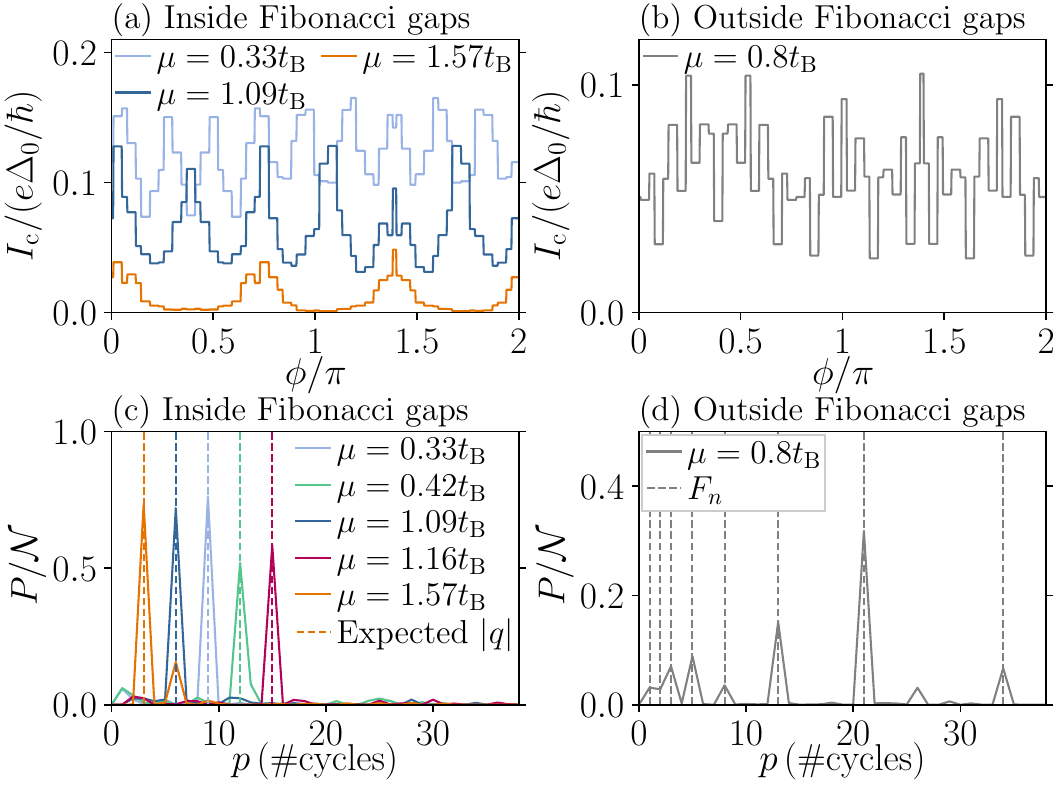}
    \caption{(a) Critical current $\Ic$ as a function of phason angle $\phi$ in a $C_9$ junction (no repetition) with $\tint=\tB$, $\rho=0.8$, $\rhoS=1$, when the chemical potential $\mu$ lies inside a gap of the Fibonacci chain, and (b) outside the gaps.
    (c) Power spectrum $P$ of $\Ic$ (solid) of (a) \newtext{scaled to show the integer number of periodic cycles $p$ (as $\phi$ varies from $0$ to $2\pi$)}, thus corresponding to the winding number $|q|$ of each gap, $p=|q|$ (dashed). (d) Power spectrum of (b) and Fibonacci numbers $F_n$ (dashed). $\mathcal{N}$ is a normalization constant.
    }
    \label{fig:critical_current_phason_angle}
\end{figure}

To verify the conjecture that the winding numbers $|q|$ appear in the critical current, we plot in Fig.~\ref{fig:critical_current_phason_angle}(a) $\Ic$ as a function of $\phi$ at several $\mu$ (line colors) inside the topological gaps, clearly illustrating that $\Ic$ \newtext{oscillates periodically $p$ times as $\phi$ varies from $0$ to $2\pi$}.
In contrast, Fig.~\ref{fig:critical_current_phason_angle}(b) illustrates that when $\mu$ is between the gaps there is no single distinct oscillation frequency [see Fig.~\ref{fig:critical_current_mu}(c) for gap locations].
To more clearly highlight the periodic oscillations in $\Ic$, we compute the power spectrum $P$ via the discrete Fourier transform $\mathcal{F}$ with respect to $\phi$
\begin{equation}
P(\mu) = \left|\mathcal{F}_\phi[\Ic(\phi,\mu)]\right|^2,
\end{equation}
which we normalize with $\mathcal{N} = \sum_p P(\mu)$.
We also subtract the mean critical current to avoid the trivial peak at $p=0$ in the power spectrum, but refrain from further signal processing operations (e.g.~windowing).
Figure~\ref{fig:critical_current_phason_angle}(c) shows the power spectrum \newtext{as a function of the number of completed periodic cycles $p$} when $\mu$ is inside the gaps (solid), illustrating a single well-defined peak corresponding to the expected winding number, $p=|q|$ (dashed vertical lines).
Barely visible are additional peaks mainly at integer multiples of the main peak frequency, e.g.~spurious peaks due to the discrete Fourier transform.
In contrast, Fig.~\ref{fig:critical_current_phason_angle}(d) shows that there is no single well-defined peak in the power spectrum when $\mu$ is outside the gaps, but instead a number of small peaks of roughly equal size.
Coincidentally, we note that these peaks occur at the Fibonacci numbers $F_n \in \{1,2,3,5,8,13,\ldots\}$, with additional smaller peaks at integer multiples of these ``Fibonacci peaks'' due to the discrete Fourier transform.
We consistently find these Fibonacci peaks at other values of $\mu$ and for other junctions (see further below), but have no clear explanation for their appearance, other than that the Fibonacci numbers and golden ratio have a tendency to appear throughout various quantities in the Fibonacci chain~\cite{rai.19,jagannathan.21}.
Next, we note that when the Fermi level lies in the middle of the largest gaps ($|q|=1,2,3$), $\Ic$ vanishes when the state lies at the gap edge, while $\Ic$ is finite when the winding state is inside the gap, see for instance $\mu=1.57\tB$ for the $|q|=3$ gap in Fig.~\ref{fig:critical_current_phason_angle}(a).
These scenarios correspond to when either no state or only the winding state is close to the Fermi level, respectively, thus demonstrating that the topological winding states can fully carry the critical current.

\newtext{Finally, the above results and conclusions hold more generally for other parameter ranges and approximants, as we demonstrate and discuss in Appendix~\ref{app:phason_angle}.}

\section{Concluding remarks}
\label{sec:conclusions}
We study the influence of quasiperiodicity on the DC Josephson effect, by considering a ballistic hybrid superconductor-quasicrystal-superconductor Josephson junction.
We consider the Fibonacci chain as a quasicrystal model system, from the short junction limit with just a few atomic sites, to the long junction limit with several thousand sites.
Furthermore, we study Fibonacci chains both without and with repetition, thus essentially modelling both quasicrystals and their approximants~\cite{goldman.93}, embedded across two superconducting leads.
As potential experimental realizations of these models, we propose a quasiperiodic engineered atomic chain~\cite{yan.19} or metamaterial~\cite{kouwenhoven.90,chatterjee.22,kuzmanovski.23,splitthoff.24,zubchenko.24},
or moir{\'e} structure~\cite{lubin.12,joon-ahn.18,mahmood.21,uri.23}, embedded across two bulk superconductors.
Alternatively, we propose a 3D generalization with superconductors sandwiching a Fibonacci superlattice~\cite{merlin.85,todd.86,bajema.merlin.87,cohn.88,zhu.97}, consisting of 2D periodic lattices stacked according to the Fibonacci sequence along the third dimension~\cite{merlin.85,todd.86,bajema.merlin.87,cohn.88,zhu.97} along which a Josephson current is applied.
It was recently shown that superconductivity in the 1D Fibonacci chain extend to such a 3D scenario~\cite{wang.24}.

We study the Josephson effect via the low-energy ABS spectrum, current-phase relation, and critical current.
We exhaustively investigate how these depend on the superconducting phase difference, quasiperiodic degrees of freedom, hopping parameters, chemical potential, and junction length.
We find that although the current-phase relation is still $2\pi$ periodic, with either a sinusoidal or sawtooth profile, quasiperiodicity leads to ABS with quasiperiodic oscillations in their probability density, also at perfect resonance.
Importantly, we find that quasiperiodicity qualitatively modifies the condition for emergent zero-energy ABS and maximal critical current.
We demonstrate that this condition changes between a few simple functional forms in a quasiperiodic manner as the junction length increases.
Consequently, we find that the critical current shows quasiperiodic oscillations as a function of junction length, on top of the monotonic decay also found in crystalline junctions~\cite{beenakker.92,affleck.00,nikolic.01,sonin.24}.
Based on these results we conclude that there might be large sample-to-sample fluctuations between different quasiperiodic junctions, depending on the microscopic details.

Surprisingly, despite proposals for \newtext{quasiperiodicity enhancing} superconductivity and the proximity effect~\cite{fan.21,zhang.22,oliveira.23,sun.24,wang.24,kobialka.24,rai.19,rai.20}, we find that the critical current is not generally enhanced \newtext{by quasiperiodicity}, especially compared to a  crystalline junction in the perfect ballistic limit and with a zero-energy state.
However, beyond this idealized scenario, we find that quasiperiodicity can significantly enhance the critical current, especially at reduced coupling to the superconducting leads.

We find that by varying the chemical potential in the junction, the junction changes between SNS and SIS behavior in a fractal manner, due to the intrinsic fractal energy spectrum and topological gap structure of the Fibonacci chain~\cite{jagannathan.21}.
\newtext{These predictions are directly accessible in experiments, e.g.~via an applied gate voltage.}
Each topological gap hosts a topological subgap state that winds as a function of the phason angle $\phi$, both across the gap and in real space across the Fibonacci chain, with winding number $|q|$ given by a gap labeling theorem~\cite{bellissard.89,bellissard.92,mace.jagannathan.piechon.17,yamamoto.22}.
We find that when the chemical potential is tuned through the gaps, \newtext{these topological subgap states can fully carry the current and that their} winding leads to the critical current oscillating with the winding number, while outside the gaps the critical current instead oscillates according to the Fibonacci numbers.
We therefore demonstrate how the critical current can in principle measure the topological invariant in the Fibonacci chain.
In summary, these results show how Josephson junctions can be used to probe the intricate physics of quasiperiodic systems, including their interplay with ordered states such as superconductivity.

As an outlook, there are many open questions such as the influence of quasiperiodicity on the AC Josephson effect, different heterostructures and sample realizations, the influence of defects and disorder, finite temperature, \newtext{self-consistent phase gradients}, as well as different strengths or symmetries of the superconducting order parameter.
It would also be interesting to find a connection with the topological invariant without relying on the phason angle~\cite{rai.21}.
\newtext{Another interesting topic is the connection, if any, between the quasiperiodic critical localization with the current and its decay characteristics}.
Furthermore, other theoretical models like the scattering matrix approach, or analytic calculations using perturbation theory and renormalization group theory~\cite{rai.21}, might shine additional light on the rich physics of these systems.
Beyond quasicrystals, fractal lattices are another example of interesting aperiodic structures to study the interplay of proximitized phase-coherent phenomena~\cite{kempkes.19,canyellas.23,iliasov.24}.
We further note that the supercurrent-magnetic field relation was recently studied in the transverse direction through a single layer of such a fractal~\cite{amundsen.24}, i.e.~no current was transmitted along fractal degree of freedom itself.
Studying the Josephson effect and currents propagating along a fractal would thus also be an interesting outlook.

\section{Acknowledgements}
We thank R.~Arouca, L.~Baldo, J.~Cayao, P.~Dutta, A.~Kobiałka, and T.~L{\"othman} for valuable discussions. We acknowledge financial support from the Swedish Research Council (Vetenskapsr{\aa}det Grant No.~2022-03963, and the Knut and Alice Wallenberg Foundation through the Wallenberg Academy Fellows program, KAW 2019.0309. A.S.~acknowledges partial funding by the Knut and Alice Wallenberg Foundation under grant No.~2017.0157. O.A.A~acknowledges funding from NanoLund, the Swedish Research Council (Grant Agreement No. 2020-03412) and the European Research Council (ERC) under the European Union's Horizon 2020 research and innovation programme under the Grant Agreement No. 856526. The computations were enabled by resources provided by the National Academic Infrastructure for Supercomputing in Sweden (NAISS) and the Swedish National Infrastructure for Computing (SNIC) at UPPMAX and PDC, partially funded by the Swedish Research Council through grant agreements No.~2022-06725 and No.~2018-05973. 

\appendix

\section{Microscopic even-odd effect}
\label{app:even_odd}
\newtext{In this Appendix, we comment on the strong influence of microscopic even-odd effects on the spectrum, specifically in a junction with even or odd number of energy levels, supplementing Secs.~\ref{sec:results:josephson_effect}--\ref{sec:results:topological_invariant} in the main text.
The scenario also illustrates what happens in a system where there is already a zero-energy state before the onset of superconductivity.
Depending on the actual physical realization of the system, this might either be an idealized or a crucial effect.}

\begin{figure}[t!]
    \centering
    \includegraphics[width=\columnwidth]{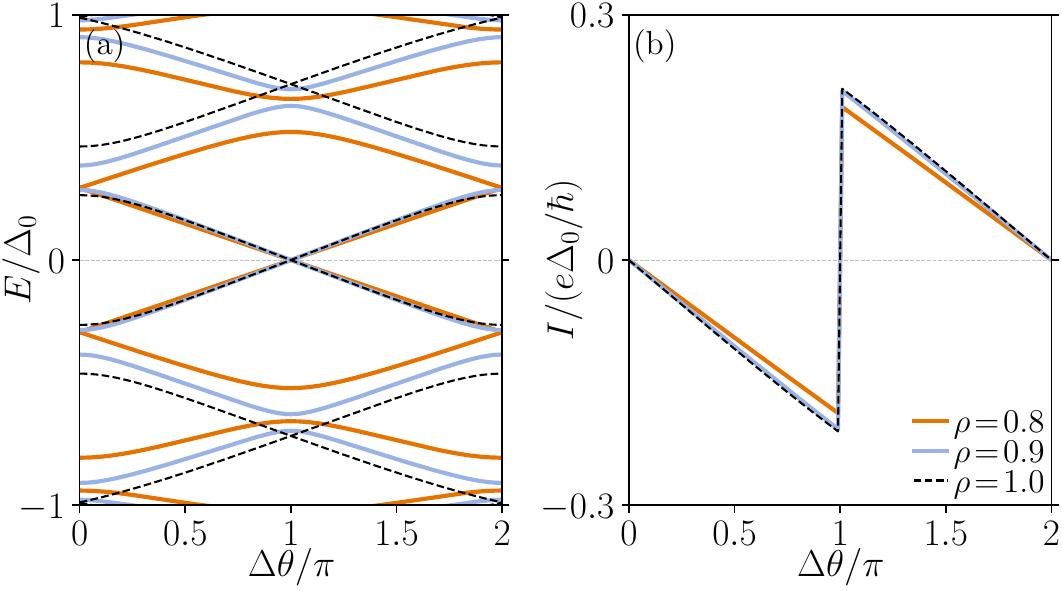}
    \caption{Same as Fig.~\ref{fig:basic_josephson_current:even} but where the Fibonacci approximant $C_9$ is repeated $N=2$ times such that the number of sites is odd $2 \times F_9 + 1 = 111$. The zero-energy state in (a) is robust for all $\forall\rho,\rhoS,\tint$.
    }
    \label{fig:basic_josephson_current:odd_repeated}
\end{figure}

In Sec.~\ref{sec:results:josephson_effect} we showed how the ABS in the crystalline junction ($\rho=1$) are degenerate at $\Delta\theta=0$ and $2\pi$, and how the degeneracy is lifted for all $\rho \neq 1$ when the quasiperiodic junction breaks the symmetry between the leads.
The Fibonacci approximant $C_9$ studied in that case has an even number of sites $F_9+1 = 56$, while the number of sites instead becomes odd when repeating the approximant an even number of times, e.g.~$2 \times F_9 + 1 = 111$, or in every third approximant e.g.~$C_8$ with $F_8+1=35$ sites. 
\newtext{At half-filling, nearest-neighbor hopping endows the chain with chiral symmetry which imposes that for every state with positive energy there exists another state with equal but negative energy. Hence, 
the} normal-state spectrum with an odd number of sites (and thus energy levels) must have an energy level at exactly zero energy due to chiral symmetry, and the ABS spectrum becomes qualitatively modified with the ABS degeneracy points shifting to $\Delta\theta=\pi$ instead~\cite{bena.12}.
We verify the modified degeneracy in quasiperiodic junctions in Fig.~\ref{fig:basic_josephson_current:odd_repeated}, showing the energy-phase spectrum (a) and current-phase relation (b) for a junction consisting of a repeated $C_9$ approximant (i.e.~otherwise equivalent to Fig.~\ref{fig:basic_josephson_current:even}).
In Fig.~\ref{fig:basic_josephson_current:odd_repeated}(a) quasiperiodicity breaks the degeneracy at $\Delta\theta=\pi$ at all higher energies due to the broken symmetry between the leads, but the degeneracy at zero energy is maintained leading to a fully linear phase dispersion with only small variations in slope with $\rho$ at low energy.
Consequently, since this zero-energy state carries most of the current, $\rho$ therefore has negligible influence on the current, causing a similar sawtooth profile in all cases shown in Fig.~\ref{fig:basic_josephson_current:odd_repeated}(b).

\begin{figure}[t!]
    \centering
    \includegraphics[width=\columnwidth]{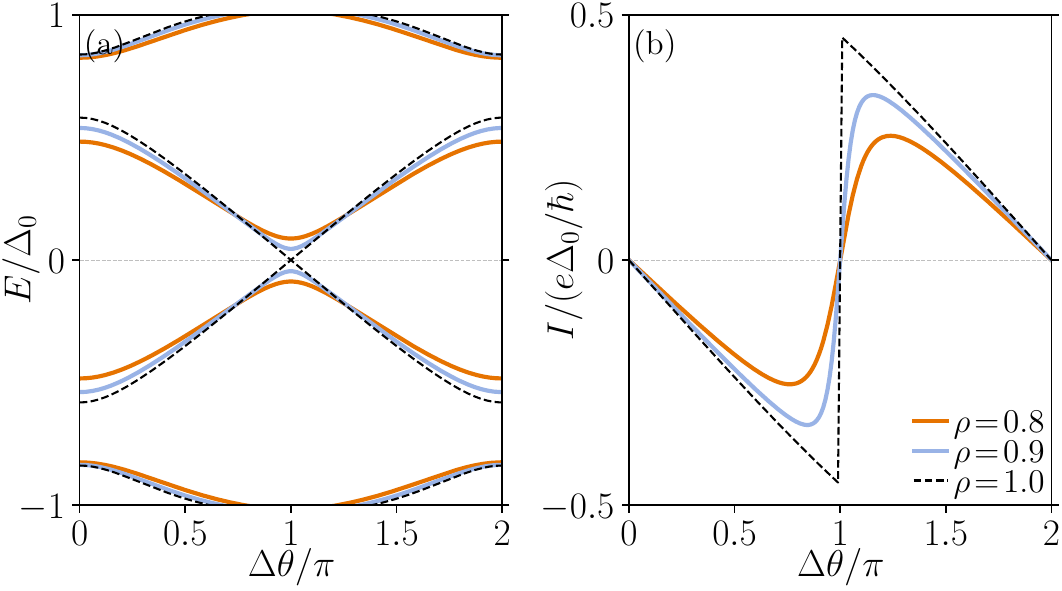}
    \caption{Same as Fig.~\ref{fig:basic_josephson_current:odd_repeated} but for $C_{8}$ without repetition, with $F_{8} + 1 = 35$ sites. The zero-energy state in (a) is not robust as it splits for any $\rho \neq 1$ as soon as $\tint \neq 0$.
    }
    \label{fig:basic_josephson_current:odd}
\end{figure}

Next, we find that at the critical phase difference $\Delta\theta=\Delta\thetac$, the current is completely carried by the lowest-energy state, where the current contribution from all other states are either exactly zero or pair-wise cancelling.
Away from this phase, the reduction in current comes primarily from the variation in the energy-phase dispersion from a linear slope at $\Delta\theta=\pi$ to zero slope at $\Delta\theta=0$ or $2\pi$, and secondarily from a destructive contribution of higher-energy states.
This behavior is qualitatively different from the junction with even number of sites described in Sec.~\ref{sec:results:josephson_effect}, where the energy-phase dispersion changes also its functional form away from $\Delta\thetac$.
Importantly, the zero-energy state in Fig.~\ref{fig:basic_josephson_current:odd_repeated}(a) is robust against variations in the hopping parameters.
We find that this is related to a preserved degeneracy where e.g.~the probability density of the zero-energy state is exactly the same in both leads (not shown), in contrast to the higher-energy states which break the degeneracy.

The situation changes qualitatively in a non-repeated approximant with odd number of states, e.g.~every third approximant $C_5$, $C_8$, $C_{11}$, and so on.
Here the zero-energy degeneracy lifts for all $\rho \neq 1$ as soon as $\tint \neq 0$ as illustrated in Fig.~\ref{fig:basic_josephson_current:odd} for the $C_{8}$ approximant.
The broken zero-energy degeneracy in Fig.~\ref{fig:basic_josephson_current:odd}(a) leads to a smoother energy-phase dispersion and thus a softening of the sawtooth profile in the current in Fig.~\ref{fig:basic_josephson_current:odd}(b) for $\rho\neq1$.
Thus the broken degeneracy for $\rho\neq1$ significantly reduces the current, such that the crystalline junction with $\rho=1$ always has a larger current.
We find that the broken zero-energy degeneracy is caused by a broken degeneracy between the leads, which we see in e.g.~the probability density of the zero-energy state (not shown), which apart from the critical behavior shows an overall decrease from one lead to the other.
Finally, we note that as $\tint \to 0$ the zero-energy degeneracy is recovered for all $\rho$, but the current vanishes accordingly due to the transparency approaching zero.

\section{Condition for zero-energy ABS}
\label{app:resonant_condition}
In this Appendix, we demonstrate how we obtain the condition $\tintStar(\rho,\rhoS)$ for the emergence of zero-energy ABS, first in the Fibonacci approximants $C_n$ with lengths $L=a_0F_n$ presented in Table~\ref{tab:resonant_ABS}, then for more general Fibonacci chains of arbitrary discrete lengths $L$ modeled by the characteristic function Eq.~(\ref{eq:characteristic_function}), thus supplementing the discussion in Sec.~\ref{sec:results:critical_current_conditions} and Sec.~\ref{sec:results:size_scaling}.
Specifically, we vary simultaneously the hopping parameters $\tint$, $\rho\equiv\tA/\tB$ and $\rhoS\equiv\tS/\tB$ and look for when the lowest-energy ABS reaches zero energy non-trivially (i.e.~ignoring trivial cases such as $\rho\to0$ or $\tint\to0$).
We parametrize the phase space with zero-energy ABS as $\tint=\tintStar(\rho,\rhoS)$, where we find a sawtooth current-phase relation and the largest critical current.
In a Fibonacci chain with even number of sites, the zero-energy ABS at $\tintStar(\rho,\rhoS)$ occurs due to perfect Andreev reflection (zero normal reflection)~\cite{affleck.00}, also with a perfect wave function matching at the interface as evidenced by the probability density of the wave function.
In a Fibonacci chain with odd number of sites, there is already a zero-energy state in the normal state~\cite{bena.12} (see Appendix~\ref{app:even_odd}), and $\tintStar(\rho,\rhoS)$ corresponds to the parameter space where this state remains at zero energy.

\begin{figure}[t!]
    \centering
    \includegraphics[width=\columnwidth]{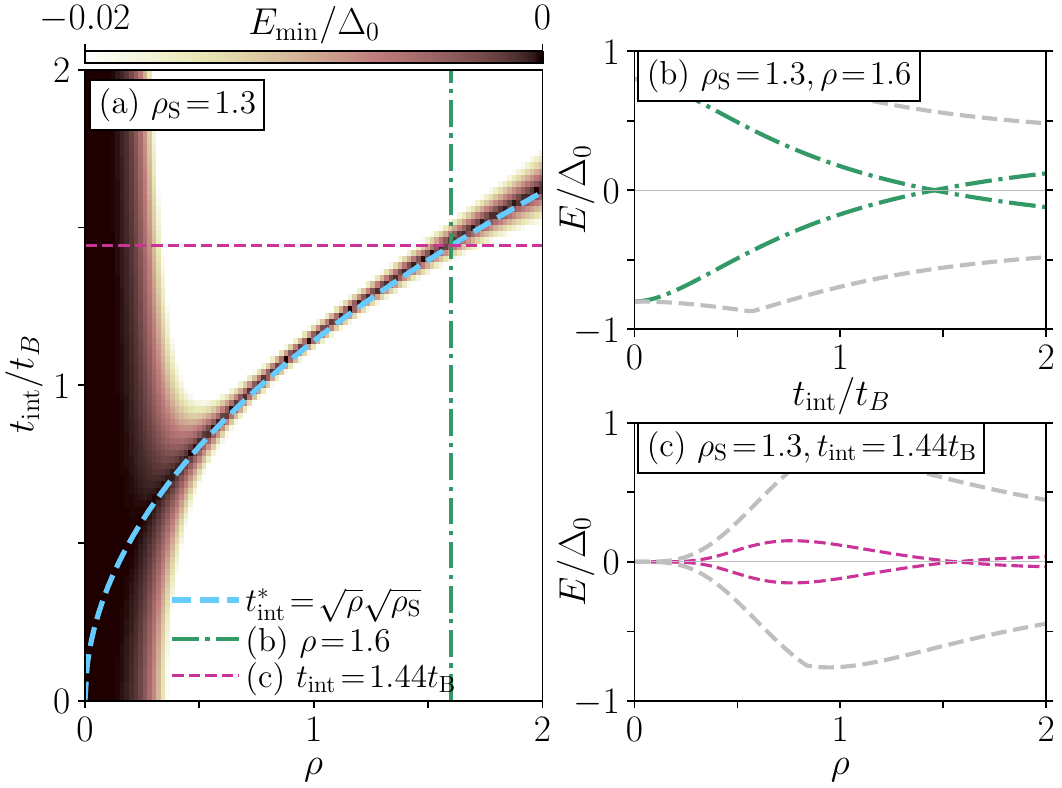}
    \caption{(a) Heatmap showing the energy of the lowest-energy ABS ($\Emin$) as function of hopping ratio $\rho$ and interface hopping $\tint$, in the approximant $C_{10}$ at fixed $\rhoS=1.3$, superconducting phase difference $\Delta\theta=\pi$ and $\mu=0$. Cyan dashed line shows that the energy is zero at $\tint=\tintStar(\rho,\rhoS)=\sqrt{\rho}\sqrt{\rhoS}$. (b),(c) Line cuts at fixed $\rho$ and $\tint$ as indicated by the vertical and horizontal lines in (a), respectively. Gray dashed lines show the energy levels of the next-lowest ABS.}
    \label{fig:tint_star_fit:c10}
\end{figure}
\begin{figure}[t!]
    \centering
    \includegraphics[width=\columnwidth]{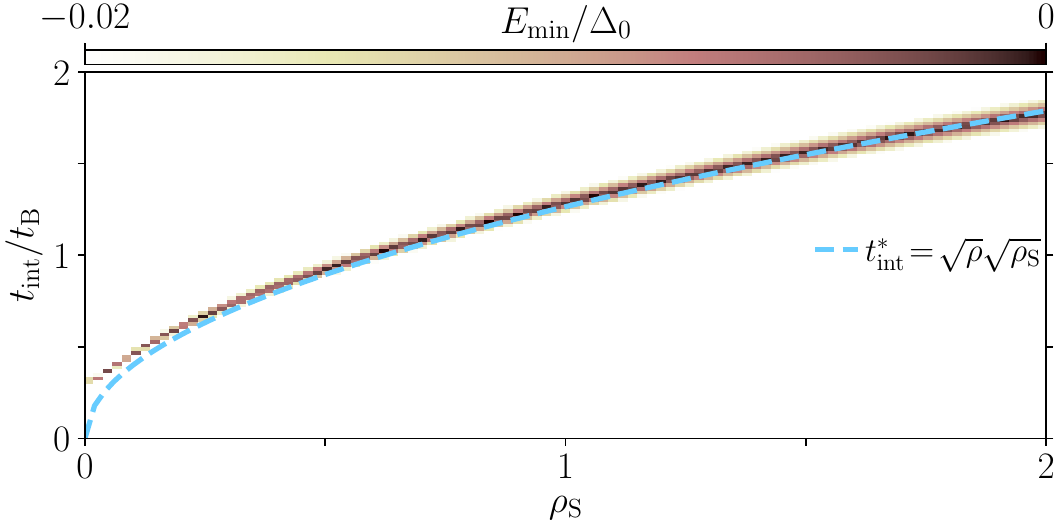}
    \caption{Same as Fig.~\ref{fig:tint_star_fit:c10}(a) but showing $\Emin$ as function of hopping ratio $\rhoS$ and interface hopping $\tint$ at fixed $\rho=1.6$.}
    \label{fig:tint_star_fit:t_S}
\end{figure}
\begin{figure}[t!]
    \centering
    \includegraphics[width=\columnwidth]{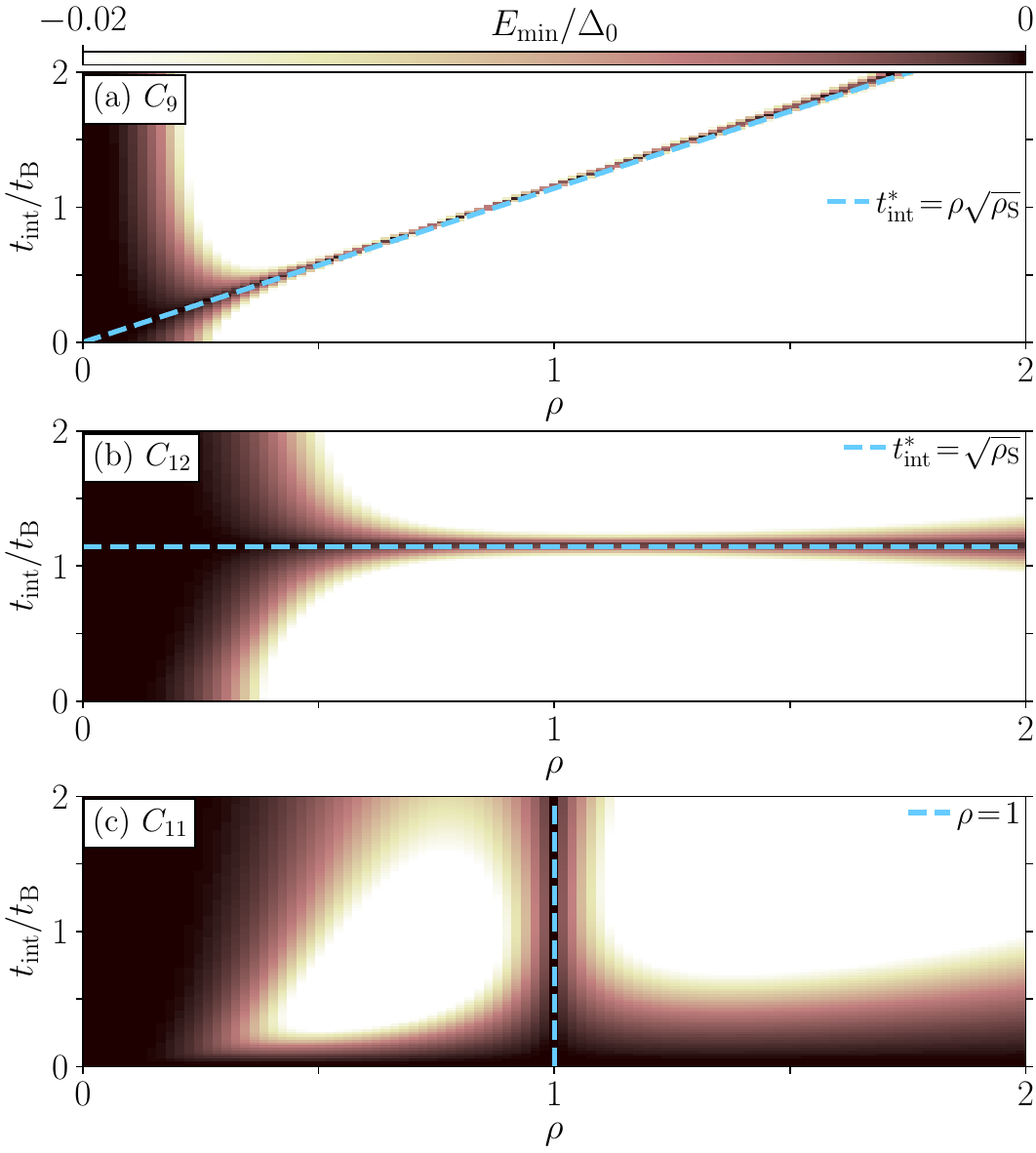}
    \caption{Same as Fig.~\ref{fig:tint_star_fit:c10}(a) but for $C_{9}$ (a), $C_{12}$ (b), and $C_{11}$ (c).}
    \label{fig:tint_star_fit:collage}
\end{figure}

We start by demonstrating the emergence of zero-energy ABS and the procedure for obtaining $\tintStar(\rho,\rhoS)$ in a Josephson junction where the non-superconducting region consists of the $C_{10}$ approximant, followed by the other approximants.
Figure~\ref{fig:tint_star_fit:c10}(a) shows a heatmap of the energy for the lowest-energy ABS ($\Emin$) as a function of $\tint$ and $\rho$ in the $C_{10}$ approximant (without repetitions), at fixed $\rhoS=1.3$ and phase difference $\Delta\theta=\pi$.
The ABS energy is zero along $\tint=\tintStar(\rho,\rhoS) = \sqrt{\rho}\sqrt{\rhoS}$ (cyan dashed lines), due to perfect Andreev reflection~\cite{affleck.00}, also associated with the appearance of a Jackiw-Rebbi zero mode~\cite{jackiw.rebbi.76}.
In contrast, the zero-energy state occurring for all $\tint$ at $\rho\to0$ corresponds to the trivial scenario where the entire spectrum reduces to three energy levels (see Sec.~\ref{sec:model:fibonacci}), and is thus of no interest here.
Figures~\ref{fig:tint_star_fit:c10}(b) and (c) are obtained from line cuts at fixed $\rho$ and $\tint$ along the vertical and horizontal lines in (a), respectively.
These plots illustrate more clearly the shape of the spectrum around zero energy, also showing the energy level of the next-lowest ABS (gray dashed lines).
We obtain the full functional form $\tintStar(\rho,\rhoS)$ by varying also $\rhoS$ which produces an equally good overlap as in Fig.~\ref{fig:tint_star_fit:c10}(a) for each $\rhoS$, which is further illustrated in Fig.~\ref{fig:tint_star_fit:t_S} showing $\Emin$ as a function of $\tint$ and $\rhoS$ at fixed $\rho=1.6$ (which we verify also for other values of $\rho$).
In other words, all plots here are projections of our higher-dimensional spectra $E(\tint,\rho,\rhoS)$.

Next we study different Fibonacci approximants using the same methodology as above, and find that all approximants with even number of sites follow the same scaling $\sqrt{\rhoS}$ in $\tintStar(\rho,\rhoS)$, while the scaling in $\rho$ varies between three different functional forms.
Beyond the one shown for the $C_{10}$ approximant, we find $\tintStar(\rho,\rhoS)=\rho\sqrt{\rhoS}$ for e.g.~$C_9$ in Fig.~\ref{fig:tint_star_fit:collage}(a), and $\tintStar(\rho,\rhoS)=\sqrt{\rhoS}$ for e.g.~$C_{12}$ in Fig.~\ref{fig:tint_star_fit:collage}(b).
The situation changes qualitatively in the approximants with odd number of sites, (i.e.~every third approximant $C_2,C_5,C_8,\ldots$), as shown for $C_{11}$ in Fig.~\ref{fig:tint_star_fit:collage}(c).
Here, there is already a zero-energy state in the normal state, related to the microscopic even-odd effect discussed in Appendix~\ref{app:even_odd}, which changes the degeneracy points in the ABS spectrum~\cite{bena.12}.
As a result, the phase-space of zero-energy ABS $\tintStar(\rho,\rhoS)$ technically becomes a manifold instead of an analytic function, where we find zero-energy ABS $\forall \tint,\rhoS$ at $\rho=1$, or alternatively for $\forall \rho,\rhoS$ at $\tint\to0$.
However, we are not interested in the latter scenario since it describes an uncoupled system.
We find that the underlying reason for the broken zero-energy degeneracy, and thus the shift to finite energy, is related to the Fibonacci chain breaking the symmetry between the leads, together with the coupling to the superconducting leads ($\tint \neq 0$).
By repeating the above procedure for every approximant from $C_0$ to $C_{17}$, we obtain the results presented in Table~\ref{tab:resonant_ABS} in Sec.~\ref{sec:results:critical_current_conditions}.

Next, we comment on the small numeric deviations between $\Emin=0$ and the $\tintStar(\rho,\rhoS)$ seen in the above figures.
Specifically, we attribute these deviations to small corrections in powers of $\Delta_0/\tB$ similar to in the crystalline case~\cite{affleck.00}, i.e.~disappearing in the limit $\Delta_0 \ll \tA,\tS,\tB$.

\begin{figure}[t!]
    \centering
    \includegraphics[width=\columnwidth]{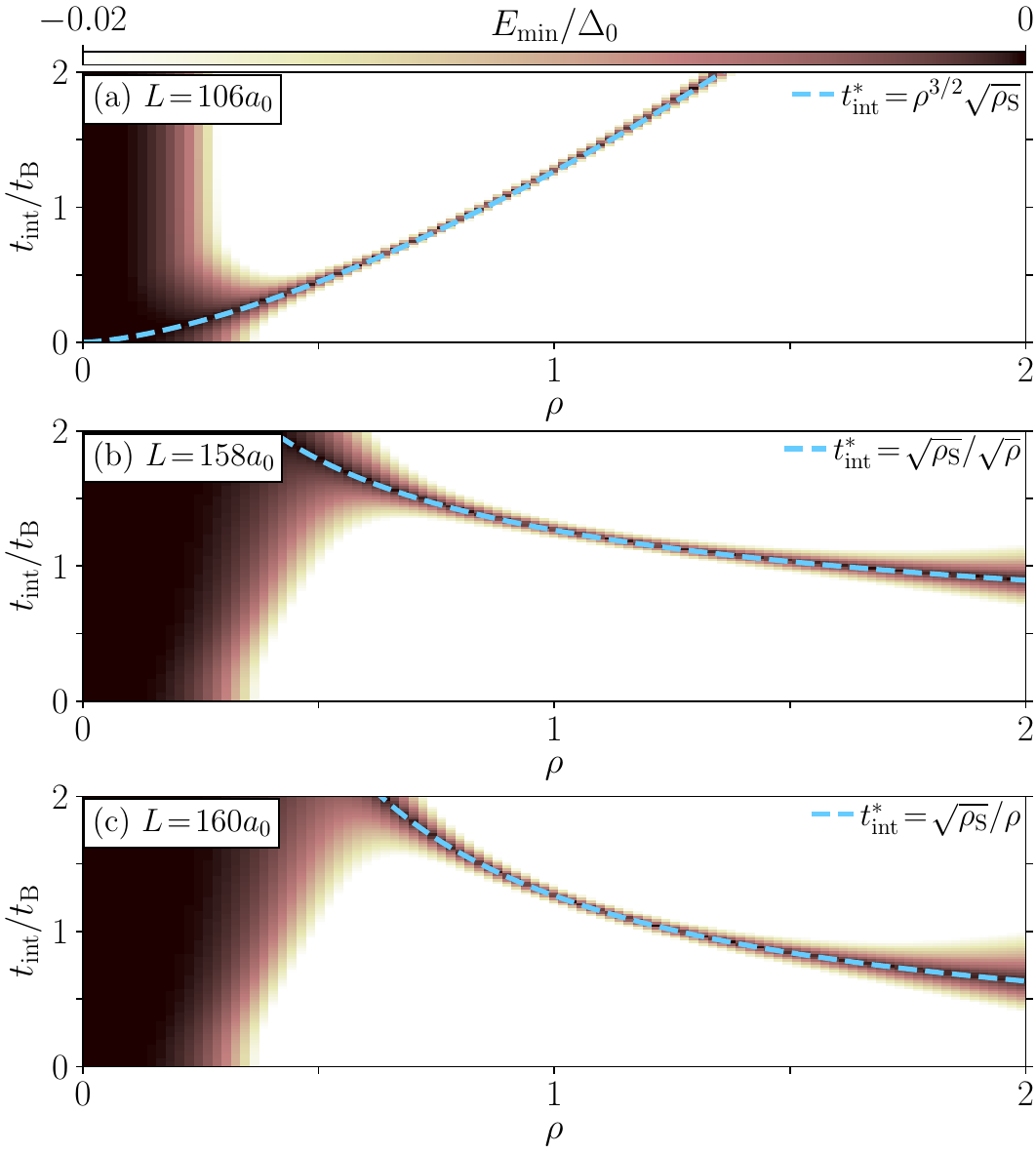}
    \caption{Same as Fig.~\ref{fig:tint_star_fit:c10}(a) but for a Fibonacci chain following the characteristic function Eq.~(\ref{eq:characteristic_function}) with $\rhoS=1.6$ and lengths $L=106a_0$ (a), $L=158a_0$ (b), and $L=160a_0$ (c).}
    \label{fig:tint_star_fit:collage:charfunc}
\end{figure}

\begin{table*}[t!]
\begin{ruledtabular}
\begin{tabular}{|ll|l|ll|l|ll|l|ll|l|ll|}
     \#sites & $\tintStar(\rho,\rhoS)/\tB$ & &
     \#sites & $\tintStar(\rho,\rhoS)/\tB$ & & 
     \#sites & $\tintStar(\rho,\rhoS)/\tB$ & & 
     \#sites & $\tintStar(\rho,\rhoS)/\tB$ & & 
     \#sites & $\tintStar(\rho,\rhoS)/\tB$ \\
\hline
 & & & $41$ & $\rho=1, \forall\rhoS$ & & $81$ & $\rho=1, \forall\rhoS$ & & $121$ & $\forall\rhoS, \forall\rho$ & & $161$ & $\rho=1, \forall\rhoS$ \\
$2$ & $\sqrt{\rho}\sqrt{\rhoS}$ & & $42$ & $\sqrt{\rho}\sqrt{\rhoS}$ & & $82$ & $\rho\sqrt{\rhoS}$ & & $122$ & $\sqrt{\rho}\sqrt{\rhoS}$ & & $162$ & $\sqrt{\rhoS}/\rho$ \\
$3$ & $\rho=1, \forall\rhoS$ & & $43$ & $\forall\rhoS, \forall\rho$ & & $83$ & $\rho=1, \forall\rhoS$ & & $123$ & $\forall\rhoS, \forall\rho$ & & $163$ & $\rho=1, \forall\rhoS$ \\
$4$ & $\rho\sqrt{\rhoS}$ & & $44$ & $\sqrt{\rho}\sqrt{\rhoS}$ & & $84$ & $\sqrt{\rho}\sqrt{\rhoS}$ & & $124$ & $\sqrt{\rhoS}$ & & $164$ & $\sqrt{\rhoS}/\sqrt{\rho}$ \\
$5$ & $\rho=1, \forall\rhoS$ & & $45$ & $\rho=1, \forall\rhoS$ & & $85$ & $\forall\rhoS, \forall\rho$ & & $125$ & $\rho=1, \forall\rhoS$ & & $165$ & $\rho=1, \forall\rhoS$ \\
$6$ & $\sqrt{\rho}\sqrt{\rhoS}$ & & $46$ & $\rho\sqrt{\rhoS}$ & & $86$ & $\sqrt{\rho}\sqrt{\rhoS}$ & & $126$ & $\sqrt{\rhoS}/\sqrt{\rho}$ & & $166$ & $\sqrt{\rhoS}$ \\
$7$ & $\forall\rhoS, \forall\rho$ & & $47$ & $\rho=1, \forall\rhoS$ & & $87$ & $\rho=1, \forall\rhoS$ & & $127$ & $\rho=1, \forall\rhoS$ & & $167$ & $\rho=1, \forall\rhoS$ \\
$8$ & $\sqrt{\rhoS}$ & & $48$ & $\sqrt{\rho}\sqrt{\rhoS}$ & & $88$ & $\rho\sqrt{\rhoS}$ & & $128$ & $\sqrt{\rhoS}/\sqrt{\rho}$ & & $168$ & $\sqrt{\rhoS}/\sqrt{\rho}$ \\
$9$ & $\rho=1, \forall\rhoS$ & & $49$ & $\forall\rhoS, \forall\rho$ & & $89$ & $\rho=1, \forall\rhoS$ & & $129$ & $\rho=1, \forall\rhoS$ & & $169$ & $\rho=1, \forall\rhoS$ \\
$10$ & $\sqrt{\rhoS}$ & & $50$ & $\sqrt{\rhoS}$ & & $90$ & $\sqrt{\rho}\sqrt{\rhoS}$ & & $130$ & $\sqrt{\rhoS}$ & & $170$ & $\sqrt{\rhoS}/\sqrt{\rho}$ \\
$11$ & $\forall\rhoS, \forall\rho$ & & $51$ & $\rho=1, \forall\rhoS$ & & $91$ & $\forall\rhoS, \forall\rho$ & & $131$ & $\forall\rhoS, \forall\rho$ & & $171$ & $\rho=1, \forall\rhoS$ \\
$12$ & $\sqrt{\rho}\sqrt{\rhoS}$ & & $52$ & $\sqrt{\rhoS}$ & & $92$ & $\sqrt{\rhoS}$ & & $132$ & $\sqrt{\rho}\sqrt{\rhoS}$ & & $172$ & $\sqrt{\rhoS}$ \\
$13$ & $\forall\rhoS, \forall\rho$ & & $53$ & $\forall\rhoS, \forall\rho$ & & $93$ & $\rho=1, \forall\rhoS$ & & $133$ & $\forall\rhoS, \forall\rho$ & & $173$ & $\forall\rhoS, \forall\rho$ \\
$14$ & $\sqrt{\rhoS}$ & & $54$ & $\sqrt{\rho}\sqrt{\rhoS}$ & & $94$ & $\sqrt{\rhoS}$ & & $134$ & $\sqrt{\rhoS}$ & & $174$ & $\sqrt{\rho}\sqrt{\rhoS}$ \\
$15$ & $\rho=1, \forall\rhoS$ & & $55$ & $\rho=1, \forall\rhoS$ & & $95$ & $\forall\rhoS, \forall\rho$ & & $135$ & $\rho=1, \forall\rhoS$ & & $175$ & $\forall\rhoS, \forall\rho$ \\
$16$ & $\sqrt{\rhoS}/\sqrt{\rho}$ & & $56$ & $\rho\sqrt{\rhoS}$ & & $96$ & $\sqrt{\rho}\sqrt{\rhoS}$ & & $136$ & $\sqrt{\rhoS}$ & & $176$ & $\sqrt{\rhoS}$ \\
$17$ & $\rho=1, \forall\rhoS$ & & $57$ & $\rho=1, \forall\rhoS$ & & $97$ & $\rho=1, \forall\rhoS$ & & $137$ & $\forall\rhoS, \forall\rho$ & & $177$ & $\rho=1, \forall\rhoS$ \\
$18$ & $\sqrt{\rhoS}/\sqrt{\rho}$ & & $58$ & $\sqrt{\rho}\sqrt{\rhoS}$ & & $98$ & $\rho\sqrt{\rhoS}$ & & $138$ & $\sqrt{\rho}\sqrt{\rhoS}$ & & $178$ & $\sqrt{\rhoS}$ \\
$19$ & $\rho=1, \forall\rhoS$ & & $59$ & $\forall\rhoS, \forall\rho$ & & $99$ & $\rho=1, \forall\rhoS$ & & $139$ & $\rho=1, \forall\rhoS$ & & $179$ & $\forall\rhoS, \forall\rho$ \\
$20$ & $\sqrt{\rhoS}$ & & $60$ & $\sqrt{\rho}\sqrt{\rhoS}$ & & $100$ & $\sqrt{\rho}\sqrt{\rhoS}$ & & $140$ & $\rho\sqrt{\rhoS}$ & & $180$ & $\sqrt{\rho}\sqrt{\rhoS}$ \\
$21$ & $\forall\rhoS, \forall\rho$ & & $61$ & $\rho=1, \forall\rhoS$ & & $101$ & $\forall\rhoS, \forall\rho$ & & $141$ & $\rho=1, \forall\rhoS$ & & $181$ & $\rho=1, \forall\rhoS$ \\
$22$ & $\sqrt{\rho}\sqrt{\rhoS}$ & & $62$ & $\rho\sqrt{\rhoS}$ & & $102$ & $\sqrt{\rho}\sqrt{\rhoS}$ & & $142$ & $\sqrt{\rho}\sqrt{\rhoS}$ & & $182$ & $\rho\sqrt{\rhoS}$ \\
$23$ & $\forall\rhoS, \forall\rho$ & & $63$ & $\rho=1, \forall\rhoS$ & & $103$ & $\rho=1, \forall\rhoS$ & & $143$ & $\forall\rhoS, \forall\rho$ & & $183$ & $\rho=1, \forall\rhoS$ \\
$24$ & $\sqrt{\rhoS}$ & & $64$ & $\rho^{3/2}\sqrt{\rhoS}$ & & $104$ & $\rho\sqrt{\rhoS}$ & & $144$ & $\sqrt{\rhoS}$ & & $184$ & $\sqrt{\rho}\sqrt{\rhoS}$ \\
$25$ & $\rho=1, \forall\rhoS$ & & $65$ & $\rho=1, \forall\rhoS$ & & $105$ & $\rho=1, \forall\rhoS$ & & $145$ & $\rho=1, \forall\rhoS$ & & $185$ & $\forall\rhoS, \forall\rho$ \\
$26$ & $\sqrt{\rhoS}$ & & $66$ & $\rho\sqrt{\rhoS}$ & & $106$ & $\rho^{3/2}\sqrt{\rhoS}$ & & $146$ & $\sqrt{\rhoS}$ & & $186$ & $\sqrt{\rhoS}$ \\
$27$ & $\forall\rhoS, \forall\rho$ & & $67$ & $\rho=1, \forall\rhoS$ & & $107$ & $\rho=1, \forall\rhoS$ & & $147$ & $\forall\rhoS, \forall\rho$ & & $187$ & $\rho=1, \forall\rhoS$ \\
$28$ & $\sqrt{\rho}\sqrt{\rhoS}$ & & $68$ & $\rho\sqrt{\rhoS}$ & & $108$ & $\rho\sqrt{\rhoS}$ & & $148$ & $\sqrt{\rho}\sqrt{\rhoS}$ & & $188$ & $\sqrt{\rhoS}$ \\
$29$ & $\rho=1, \forall\rhoS$ & & $69$ & $\rho=1, \forall\rhoS$ & & $109$ & $\rho=1, \forall\rhoS$ & & $149$ & $\forall\rhoS, \forall\rho$ & & $189$ & $\forall\rhoS, \forall\rho$ \\
$30$ & $\rho\sqrt{\rhoS}$ & & $70$ & $\rho^{3/2}\sqrt{\rhoS}$ & & $110$ & $\sqrt{\rho}\sqrt{\rhoS}$ & & $150$ & $\sqrt{\rhoS}$ & & $190$ & $\sqrt{\rho}\sqrt{\rhoS}$ \\
$31$ & $\rho=1, \forall\rhoS$ & & $71$ & $\rho=1, \forall\rhoS$ & & $111$ & $\forall\rhoS, \forall\rho$ & & $151$ & $\rho=1, \forall\rhoS$ & & $191$ & $\forall\rhoS, \forall\rho$ \\
$32$ & $\sqrt{\rho}\sqrt{\rhoS}$ & & $72$ & $\rho^{3/2}\sqrt{\rhoS}$ & & $112$ & $\sqrt{\rho}\sqrt{\rhoS}$ & & $152$ & $\sqrt{\rhoS}/\sqrt{\rho}$ & & $192$ & $\sqrt{\rhoS}$ \\
$33$ & $\forall\rhoS, \forall\rho$ & & $73$ & $\rho=1, \forall\rhoS$ & & $113$ & $\rho=1, \forall\rhoS$ & & $153$ & $\rho=1, \forall\rhoS$ & & $193$ & $\rho=1, \forall\rhoS$ \\
$34$ & $\sqrt{\rho}\sqrt{\rhoS}$ & & $74$ & $\rho^{3/2}\sqrt{\rhoS}$ & & $114$ & $\rho\sqrt{\rhoS}$ & & $154$ & $\sqrt{\rhoS}/\sqrt{\rho}$ & & $194$ & $\sqrt{\rhoS}/\sqrt{\rho}$ \\
$35$ & $\rho=1, \forall\rhoS$ & & $75$ & $\rho=1, \forall\rhoS$ & & $115$ & $\rho=1, \forall\rhoS$ & & $155$ & $\rho=1, \forall\rhoS$ & & $195$ & $\rho=1, \forall\rhoS$ \\
$36$ & $\rho\sqrt{\rhoS}$ & & $76$ & $\rho\sqrt{\rhoS}$ & & $116$ & $\sqrt{\rho}\sqrt{\rhoS}$ & & $156$ & $\sqrt{\rhoS}$ & & $196$ & $\sqrt{\rhoS}/\sqrt{\rho}$ \\
$37$ & $\rho=1, \forall\rhoS$ & & $77$ & $\rho=1, \forall\rhoS$ & & $117$ & $\forall\rhoS, \forall\rho$ & & $157$ & $\rho=1, \forall\rhoS$ & & $197$ & $\rho=1, \forall\rhoS$ \\
$38$ & $\rho^{3/2}\sqrt{\rhoS}$ & & $78$ & $\rho\sqrt{\rhoS}$ & & $118$ & $\sqrt{\rhoS}$ & & $158$ & $\sqrt{\rhoS}/\sqrt{\rho}$ & & $198$ & $\sqrt{\rhoS}$ \\
$39$ & $\rho=1, \forall\rhoS$ & & $79$ & $\rho=1, \forall\rhoS$ & & $119$ & $\rho=1, \forall\rhoS$ & & $159$ & $\rho=1, \forall\rhoS$ & & $199$ & $\forall\rhoS, \forall\rho$ \\
$40$ & $\rho\sqrt{\rhoS}$ & & $80$ & $\rho^{3/2}\sqrt{\rhoS}$ & & $120$ & $\sqrt{\rhoS}$ & & $160$ & $\sqrt{\rhoS}/\rho$ & & $200$ & $\sqrt{\rho}\sqrt{\rhoS}$ \\
\end{tabular}
\end{ruledtabular}
\caption{Condition $\tintStar(\rho,\rhoS)$ for zero-energy ABS, in a Fibonacci chain with physical length $L$ (i.e.~$L/a_0+1$ number of sites) as modeled by the characteristic function in Eq.~(\ref{eq:characteristic_function}).
Here, junctions with odd number of sites already hosts a zero-energy state in the normal state (see Appendix~\ref{app:even_odd}), and the notation of style $\forall\rhoS$ and $\rho=1$ denotes for which values the ABS remains at zero energy $\forall\tint$. We note that when the length is $L=a_0F_n$, we obtain the Fibonacci approximants $C_n$ consistent with Table~\ref{tab:resonant_ABS}. }
\label{tab:resonant_ABS:characteristic_function}
\end{table*}

Next, we follow the same methodology to obtain $\tintStar(\rho,\rhoS)$ for Fibonacci chains of arbitrary discrete length $L$ as modeled by the characteristic function Eq.~(\ref{eq:characteristic_function}).
We find that these Fibonacci chains alternate between the same functional forms as the Fibonacci approximants $C_n$ (we verify that when $L=a_0F_n$ we obtain the same results as in Table~\ref{tab:resonant_ABS}), but also a few additional ones as shown in Table~\ref{tab:resonant_ABS:characteristic_function} for all lengths between $L \in [2,200]a_0$.
Specifically, we find that junctions with an even number of sites can also have $\tintStar(\rho,\rhoS) = \rho^{3/2}\sqrt{\rhoS}$, $\tintStar(\rho,\rhoS) = \sqrt{\rhoS}/\sqrt{\rho}$ and $\tintStar(\rho,\rhoS) = \sqrt{\rhoS}/\rho$, as shown in Fig.~\ref{fig:tint_star_fit:collage:charfunc}.
Thus, for such junctions with even number of sites the scaling in $\rhoS$ thus always goes as $\sqrt{\rhoS}$, and the scaling exponent in $\rho$ is always a rational number.
We do not rule out that there might be additional scalings for other lengths $L>200a_0$, but note that the new scalings occur for very few $L$ in comparison to the ones found for the approximants $C_n$.
We find that the junctions with odd number of sites either show a zero-energy state for the same parameter space as the approximants ($\forall \tint,\rhoS$ at $\rho=1$), or host a zero-energy state that is also degenerate regardless of quasiperiodicity (i.e~$\forall \tint,\rhoS,\rho$).

Finally, we comment on the variations in $\tintStar(\rho,\rhoS)$ and $\Ic$ with junction length $L$.
In particular, we find symmetric variations in both of these quantities repeating across multiple length scales, from a few sites to hundreds of sites.
For instance, $\tintStar(\rho,\rhoS)$ varies symmetrically for increasing and decreasing length around $72\pm70$ \#sites and around $161\pm40$ \#sites (see Table~\ref{tab:resonant_ABS:characteristic_function}), but also with similar local symmetry around other points within these intervals, e.g.~$=38\pm16$ \#sites.
The symmetry can also be seen by e.g.~selecting any two closest lengths with condition $\forall\rhoS,\forall\rho$ (such as $91$ and $95$ \#sites), where the symmetry can extend further beyond the points depending on which two points are chosen.
We note that the microscopic oscillations in $\Ic(L)$ show a similar symmetry, e.g.~across $L=(681\pm303)a_0$ in Fig.~\ref{fig:critical_current_length}(c), and also at multiple shorter length scales within this span.
These variations thus imply self-similarity and scale invariance.
This is further supported by noting that $\tintStar(\rho,\rhoS)$ alternates cyclically with $n$ in the Fibonacci approximants $C_n$, and thus quasiperiodically in $L = a_0F_n$ according to the Fibonacci number, i.e.~quasiperiodic and self-similar variation in $\tintStar(\rho,\rhoS)$ with $L$.
Thus, based on our data for $L\in[1,2584]a_0$ in the approximants $C_0$ to $C_{17}$ and $L\in[2,1000]a_0$ for Fibonacci chains modeled by the characteristic function, we find scale-invariant and quasiperiodic variations in both $\tintStar(\rho,\rhoS)$ and $\Ic(L)$.

\begin{figure*}[t!]
    \centering
    \includegraphics[width=\textwidth]{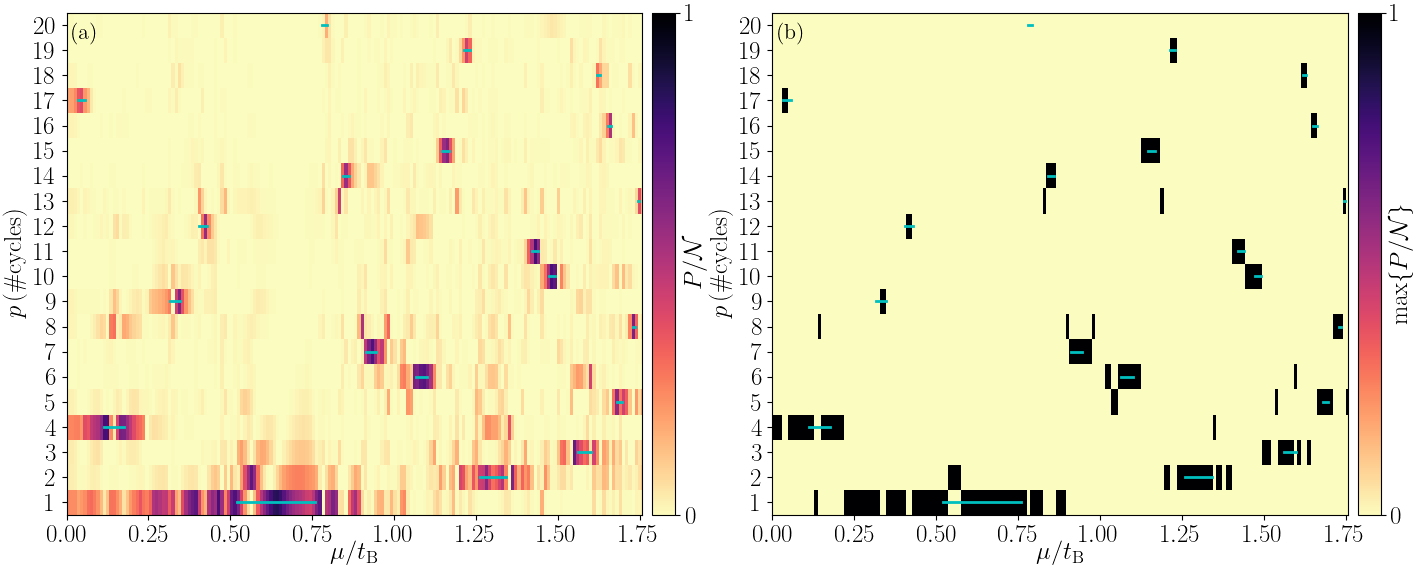}
    \caption{(a) Power spectrum $P$ of the critical current $\Ic$ as a function of chemical potential $\mu$ and \newtext{number of periodic cycles} $p$ in the $C_9$ junction (no repetition) at $\rho=0.8$, $\rhoS=1$, $\tint=0.5\tB$, with normalization $\mathcal{N}$. (b) Maximum of the power spectrum at each $\mu$. Cyan lines: winding numbers $|q|=p$, with length corresponding to each gap width.
    }
    \label{fig:critical_current_phason_angle:heatmap}
\end{figure*}

\section{Topological invariant from critical current: additional results}
\label{app:phason_angle}

\newtext{Section~\ref{sec:results:phason_angle} demonstrated the proof-of-principle that the critical current can be related to the topological invariant of the Fibonacci chain via the winding of the subgap states.
Specifically, Fig.~\ref{fig:critical_current_phason_angle} illustrated how the winding states cause distinct $p=|q|$ number of periodic oscillations in the critical current. In this Appendix, we discuss and show how this holds for other parameter values and approximants.}

In Fig.~\ref{fig:critical_current_phason_angle:heatmap}(a) we plot the corresponding full power spectrum between half-filling and the band edge of Figs.~\ref{fig:critical_current_phason_angle}(c) and Figs.~\ref{fig:critical_current_phason_angle}(d), illustrating the same three kinds of peaks found in the latter figures, namely the main peaks from winding numbers $|q|$ when $\mu$ is inside the gaps (cyan lines), the Fibonacci peaks outside the gaps, and additional spurious peaks due to the discrete Fourier transform.
This is more clearly illustrated in Fig.~\ref{fig:critical_current_phason_angle:heatmap}(b) where we plot $\mathrm{max}\{P,p\}$, i.e.~only the largest peak at each $\mu$.
All gaps with winding numbers $|q|<20$ are clearly distinguishable.
Furthermore, the signature of the largest gaps extends slightly beyond the gap itself.
This is one reason why the $|q|=20$ gap is not visible in Fig.~\ref{fig:critical_current_phason_angle:heatmap}(b) since it lies too close to the largest gap $|q|=1$, although it is clearly visible in Fig.~\ref{fig:critical_current_phason_angle:heatmap}(a) as the largest peak for $p=20$.
Apart from the Fibonacci peaks appearing outside the gaps, the only additional peak found are those correspond to the \newtext{frequency doubling}, e.g.~when the Fermi level lies inside the gaps with $|q| \leq 4$.
We expect that a significantly better result can be obtained by applying refined signal processing operations, and that peak fitting can yield a better quantification of the uncertainty and spurious peaks.
Still, our results show that even our simple approach can be used to infer the topological invariant and winding number from the critical current.

Finally, for the sake of completeness, we comment on the above procedure for other parameter values and Fibonacci chains.
We generally find that close to $\tint=\tintStar(\rho,\rhoS)$, the spurious peaks due to the discrete Fourier transform (e.g.~frequency doubling) become smaller, while the Fibonacci peaks are quite pronounced.
Further from $\tint=\tintStar(\rho,\rhoS)$, the situation is reversed, i.e.~with larger influence of the spurious peaks, and smaller influence of the Fibonacci peaks.
Importantly, the peaks due to the winding numbers $p=|q|$ of the topological gaps are clearly distinguishable in both cases, where in addition these peaks stretch further outside the gaps further from $\tint=\tintStar(\rho,\rhoS)$.
Additionally, we find that by comparing the results for different parameter values, we can more reliably capture the winding numbers in the smallest gaps.
We find a similar conclusion also for $\rho=0.5$, $\rho=1.2$ and $\rho=1.5$.
We also find that the procedure works with and without repetition of the Fibonacci approximant, as well as in longer Fibonacci approximants since these show the same major gap structure and winding numbers.
The instances when the procedure works less well are e.g.~in too short approximants due less clear winding with $\phi$, or when the hopping ratio becomes either very small or very large, or very far from $\tint=\tintStar(\rho,\rhoS)$, e.g.~as $\tint\to0$, since the spurious peaks become so large that it can be difficult to reliably identify the winding numbers.

\bibliography{cite.bib}

\begin{thebibliography}{167}%
\makeatletter
\providecommand \@ifxundefined [1]{%
 \@ifx{#1\undefined}
}%
\providecommand \@ifnum [1]{%
 \ifnum #1\expandafter \@firstoftwo
 \else \expandafter \@secondoftwo
 \fi
}%
\providecommand \@ifx [1]{%
 \ifx #1\expandafter \@firstoftwo
 \else \expandafter \@secondoftwo
 \fi
}%
\providecommand \natexlab [1]{#1}%
\providecommand \enquote  [1]{``#1''}%
\providecommand \bibnamefont  [1]{#1}%
\providecommand \bibfnamefont [1]{#1}%
\providecommand \citenamefont [1]{#1}%
\providecommand \href@noop [0]{\@secondoftwo}%
\providecommand \href [0]{\begingroup \@sanitize@url \@href}%
\providecommand \@href[1]{\@@startlink{#1}\@@href}%
\providecommand \@@href[1]{\endgroup#1\@@endlink}%
\providecommand \@sanitize@url [0]{\catcode `\\12\catcode `\$12\catcode
  `\&12\catcode `\#12\catcode `\^12\catcode `\_12\catcode `\%12\relax}%
\providecommand \@@startlink[1]{}%
\providecommand \@@endlink[0]{}%
\providecommand \url  [0]{\begingroup\@sanitize@url \@url }%
\providecommand \@url [1]{\endgroup\@href {#1}{\urlprefix }}%
\providecommand \urlprefix  [0]{URL }%
\providecommand \Eprint [0]{\href }%
\providecommand \doibase [0]{https://doi.org/}%
\providecommand \selectlanguage [0]{\@gobble}%
\providecommand \bibinfo  [0]{\@secondoftwo}%
\providecommand \bibfield  [0]{\@secondoftwo}%
\providecommand \translation [1]{[#1]}%
\providecommand \BibitemOpen [0]{}%
\providecommand \bibitemStop [0]{}%
\providecommand \bibitemNoStop [0]{.\EOS\space}%
\providecommand \EOS [0]{\spacefactor3000\relax}%
\providecommand \BibitemShut  [1]{\csname bibitem#1\endcsname}%
\let\auto@bib@innerbib\@empty
\bibitem [{\citenamefont {Shechtman}\ \emph {et~al.}(1984)\citenamefont
  {Shechtman}, \citenamefont {Blech}, \citenamefont {Gratias},\ and\
  \citenamefont {Cahn}}]{shechtman.84}%
  \BibitemOpen
  \bibfield  {author} {\bibinfo {author} {\bibfnamefont {D.}~\bibnamefont
  {Shechtman}}, \bibinfo {author} {\bibfnamefont {I.}~\bibnamefont {Blech}},
  \bibinfo {author} {\bibfnamefont {D.}~\bibnamefont {Gratias}},\ and\ \bibinfo
  {author} {\bibfnamefont {J.~W.}\ \bibnamefont {Cahn}},\ }\bibfield  {title}
  {\bibinfo {title} {Metallic phase with long-range orientational order and no
  translational symmetry},\ }\href
  {https://doi.org/10.1103/PhysRevLett.53.1951} {\bibfield  {journal} {\bibinfo
   {journal} {Phys. Rev. Lett.}\ }\textbf {\bibinfo {volume} {53}},\ \bibinfo
  {pages} {1951} (\bibinfo {year} {1984})}\BibitemShut {NoStop}%
\bibitem [{\citenamefont {Levine}\ and\ \citenamefont
  {Steinhardt}(1984)}]{levine.steinhardt.84}%
  \BibitemOpen
  \bibfield  {author} {\bibinfo {author} {\bibfnamefont {D.}~\bibnamefont
  {Levine}}\ and\ \bibinfo {author} {\bibfnamefont {P.~J.}\ \bibnamefont
  {Steinhardt}},\ }\bibfield  {title} {\bibinfo {title} {Quasicrystals: A new
  class of ordered structures},\ }\href
  {https://doi.org/10.1103/PhysRevLett.53.2477} {\bibfield  {journal} {\bibinfo
   {journal} {Phys. Rev. Lett.}\ }\textbf {\bibinfo {volume} {53}},\ \bibinfo
  {pages} {2477} (\bibinfo {year} {1984})}\BibitemShut {NoStop}%
\bibitem [{\citenamefont {Levine}\ and\ \citenamefont
  {Steinhardt}(1986)}]{levine.steinhardt.86}%
  \BibitemOpen
  \bibfield  {author} {\bibinfo {author} {\bibfnamefont {D.}~\bibnamefont
  {Levine}}\ and\ \bibinfo {author} {\bibfnamefont {P.~J.}\ \bibnamefont
  {Steinhardt}},\ }\bibfield  {title} {\bibinfo {title} {{Quasicrystals. I.
  Definition and structure}},\ }\href {https://doi.org/10.1103/PhysRevB.34.596}
  {\bibfield  {journal} {\bibinfo  {journal} {Phys. Rev. B}\ }\textbf {\bibinfo
  {volume} {34}},\ \bibinfo {pages} {596} (\bibinfo {year} {1986})}\BibitemShut
  {NoStop}%
\bibitem [{\citenamefont {Socolar}\ and\ \citenamefont
  {Steinhardt}(1986)}]{socolar.steinhardt.86}%
  \BibitemOpen
  \bibfield  {author} {\bibinfo {author} {\bibfnamefont {J.~E.~S.}\
  \bibnamefont {Socolar}}\ and\ \bibinfo {author} {\bibfnamefont {P.~J.}\
  \bibnamefont {Steinhardt}},\ }\bibfield  {title} {\bibinfo {title}
  {Quasicrystals. ii. unit-cell configurations},\ }\href
  {https://doi.org/10.1103/PhysRevB.34.617} {\bibfield  {journal} {\bibinfo
  {journal} {Phys. Rev. B}\ }\textbf {\bibinfo {volume} {34}},\ \bibinfo
  {pages} {617} (\bibinfo {year} {1986})}\BibitemShut {NoStop}%
\bibitem [{\citenamefont {Stadnik}(1998)}]{stadnik.98}%
  \BibitemOpen
  \bibfield  {author} {\bibinfo {author} {\bibfnamefont {Z.~M.}\ \bibnamefont
  {Stadnik}},\ }\href@noop {} {\emph {\bibinfo {title} {Physical properties of
  quasicrystals}}},\ Vol.\ \bibinfo {volume} {126}\ (\bibinfo  {publisher}
  {Springer Science \& Business Media},\ \bibinfo {year} {1998})\BibitemShut
  {NoStop}%
\bibitem [{\citenamefont {Maciá}(2005)}]{macia.06}%
  \BibitemOpen
  \bibfield  {author} {\bibinfo {author} {\bibfnamefont {E.}~\bibnamefont
  {Maciá}},\ }\bibfield  {title} {\bibinfo {title} {The role of aperiodic
  order in science and technology},\ }\href
  {https://doi.org/10.1088/0034-4885/69/2/R03} {\bibfield  {journal} {\bibinfo
  {journal} {Rep. Prog. Phys.}\ }\textbf {\bibinfo {volume} {69}},\ \bibinfo
  {pages} {397} (\bibinfo {year} {2005})}\BibitemShut {NoStop}%
\bibitem [{\citenamefont {Kempkes}\ \emph {et~al.}(2019)\citenamefont
  {Kempkes}, \citenamefont {Slot}, \citenamefont {Freeney}, \citenamefont
  {Zevenhuizen}, \citenamefont {Vanmaekelbergh}, \citenamefont {Swart},\ and\
  \citenamefont {Smith}}]{kempkes.19}%
  \BibitemOpen
  \bibfield  {author} {\bibinfo {author} {\bibfnamefont {S.~N.}\ \bibnamefont
  {Kempkes}}, \bibinfo {author} {\bibfnamefont {M.~R.}\ \bibnamefont {Slot}},
  \bibinfo {author} {\bibfnamefont {S.~E.}\ \bibnamefont {Freeney}}, \bibinfo
  {author} {\bibfnamefont {S.~J.~M.}\ \bibnamefont {Zevenhuizen}}, \bibinfo
  {author} {\bibfnamefont {D.}~\bibnamefont {Vanmaekelbergh}}, \bibinfo
  {author} {\bibfnamefont {I.}~\bibnamefont {Swart}},\ and\ \bibinfo {author}
  {\bibfnamefont {C.~M.}\ \bibnamefont {Smith}},\ }\bibfield  {title} {\bibinfo
  {title} {Design and characterization of electrons in a fractal geometry},\
  }\href {https://doi.org/10.1038/s41567-018-0328-0} {\bibfield  {journal}
  {\bibinfo  {journal} {Nat. Phys.}\ }\textbf {\bibinfo {volume} {15}},\
  \bibinfo {pages} {127} (\bibinfo {year} {2019})}\BibitemShut {NoStop}%
\bibitem [{\citenamefont {Canyellas}\ \emph {et~al.}(2023)\citenamefont
  {Canyellas}, \citenamefont {Liu}, \citenamefont {Arouca}, \citenamefont
  {Eek}, \citenamefont {Wang}, \citenamefont {Yin}, \citenamefont {Guan},
  \citenamefont {Li}, \citenamefont {Wang}, \citenamefont {Zheng},
  \citenamefont {Liu}, \citenamefont {Jia},\ and\ \citenamefont
  {Smith}}]{canyellas.23}%
  \BibitemOpen
  \bibfield  {author} {\bibinfo {author} {\bibfnamefont {R.}~\bibnamefont
  {Canyellas}}, \bibinfo {author} {\bibfnamefont {C.}~\bibnamefont {Liu}},
  \bibinfo {author} {\bibfnamefont {R.}~\bibnamefont {Arouca}}, \bibinfo
  {author} {\bibfnamefont {L.}~\bibnamefont {Eek}}, \bibinfo {author}
  {\bibfnamefont {G.}~\bibnamefont {Wang}}, \bibinfo {author} {\bibfnamefont
  {Y.}~\bibnamefont {Yin}}, \bibinfo {author} {\bibfnamefont {D.}~\bibnamefont
  {Guan}}, \bibinfo {author} {\bibfnamefont {Y.}~\bibnamefont {Li}}, \bibinfo
  {author} {\bibfnamefont {S.}~\bibnamefont {Wang}}, \bibinfo {author}
  {\bibfnamefont {H.}~\bibnamefont {Zheng}}, \bibinfo {author} {\bibfnamefont
  {C.}~\bibnamefont {Liu}}, \bibinfo {author} {\bibfnamefont {J.}~\bibnamefont
  {Jia}},\ and\ \bibinfo {author} {\bibfnamefont {C.~M.}\ \bibnamefont
  {Smith}},\ }\href@noop {} {\bibinfo {title} {{Topological edge and corner
  states in Bi fractals on InSb}}} (\bibinfo {year} {2023}),\ \Eprint
  {https://arxiv.org/abs/2309.09860} {arXiv:2309.09860 [cond-mat.mes-hall]}
  \BibitemShut {NoStop}%
\bibitem [{\citenamefont {Iliasov}\ \emph {et~al.}(2024)\citenamefont
  {Iliasov}, \citenamefont {Katsnelson},\ and\ \citenamefont
  {Bagrov}}]{iliasov.24}%
  \BibitemOpen
  \bibfield  {author} {\bibinfo {author} {\bibfnamefont {A.~A.}\ \bibnamefont
  {Iliasov}}, \bibinfo {author} {\bibfnamefont {M.~I.}\ \bibnamefont
  {Katsnelson}},\ and\ \bibinfo {author} {\bibfnamefont {A.~A.}\ \bibnamefont
  {Bagrov}},\ }\href@noop {} {\bibinfo {title} {Strong enhancement of
  superconductivity on finitely ramified fractal lattices}} (\bibinfo {year}
  {2024}),\ \Eprint {https://arxiv.org/abs/2310.11497} {arXiv:2310.11497
  [cond-mat.supr-con]} \BibitemShut {NoStop}%
\bibitem [{\citenamefont {Deguchi}\ \emph {et~al.}(2015)\citenamefont
  {Deguchi}, \citenamefont {Nakayama}, \citenamefont {Matsukawa}, \citenamefont
  {Imura}, \citenamefont {Tanaka}, \citenamefont {Ishimasa},\ and\
  \citenamefont {Sato}}]{deguchi.15}%
  \BibitemOpen
  \bibfield  {author} {\bibinfo {author} {\bibfnamefont {K.}~\bibnamefont
  {Deguchi}}, \bibinfo {author} {\bibfnamefont {M.}~\bibnamefont {Nakayama}},
  \bibinfo {author} {\bibfnamefont {S.}~\bibnamefont {Matsukawa}}, \bibinfo
  {author} {\bibfnamefont {K.}~\bibnamefont {Imura}}, \bibinfo {author}
  {\bibfnamefont {K.}~\bibnamefont {Tanaka}}, \bibinfo {author} {\bibfnamefont
  {T.}~\bibnamefont {Ishimasa}},\ and\ \bibinfo {author} {\bibfnamefont
  {N.~K.}\ \bibnamefont {Sato}},\ }\bibfield  {title} {\bibinfo {title}
  {{Superconductivity of Au–Ge–Yb Approximants with Tsai-Type Clusters}},\
  }\href {https://doi.org/10.7566/JPSJ.84.023705} {\bibfield  {journal}
  {\bibinfo  {journal} {J. Phys. Soc. Jpn.}\ }\textbf {\bibinfo {volume}
  {84}},\ \bibinfo {pages} {023705} (\bibinfo {year} {2015})}\BibitemShut
  {NoStop}%
\bibitem [{\citenamefont {Fulga}\ \emph {et~al.}(2016)\citenamefont {Fulga},
  \citenamefont {Pikulin},\ and\ \citenamefont {Loring}}]{fulga.16}%
  \BibitemOpen
  \bibfield  {author} {\bibinfo {author} {\bibfnamefont {I.~C.}\ \bibnamefont
  {Fulga}}, \bibinfo {author} {\bibfnamefont {D.~I.}\ \bibnamefont {Pikulin}},\
  and\ \bibinfo {author} {\bibfnamefont {T.~A.}\ \bibnamefont {Loring}},\
  }\bibfield  {title} {\bibinfo {title} {Aperiodic weak topological
  superconductors},\ }\href {https://doi.org/10.1103/PhysRevLett.116.257002}
  {\bibfield  {journal} {\bibinfo  {journal} {Phys. Rev. Lett.}\ }\textbf
  {\bibinfo {volume} {116}},\ \bibinfo {pages} {257002} (\bibinfo {year}
  {2016})}\BibitemShut {NoStop}%
\bibitem [{\citenamefont {Sakai}\ \emph {et~al.}(2017)\citenamefont {Sakai},
  \citenamefont {Takemori}, \citenamefont {Koga},\ and\ \citenamefont
  {Arita}}]{sakai.17}%
  \BibitemOpen
  \bibfield  {author} {\bibinfo {author} {\bibfnamefont {S.}~\bibnamefont
  {Sakai}}, \bibinfo {author} {\bibfnamefont {N.}~\bibnamefont {Takemori}},
  \bibinfo {author} {\bibfnamefont {A.}~\bibnamefont {Koga}},\ and\ \bibinfo
  {author} {\bibfnamefont {R.}~\bibnamefont {Arita}},\ }\bibfield  {title}
  {\bibinfo {title} {Superconductivity on a quasiperiodic lattice:
  Extended-to-localized crossover of cooper pairs},\ }\href
  {https://doi.org/10.1103/PhysRevB.95.024509} {\bibfield  {journal} {\bibinfo
  {journal} {Phys. Rev. B}\ }\textbf {\bibinfo {volume} {95}},\ \bibinfo
  {pages} {024509} (\bibinfo {year} {2017})}\BibitemShut {NoStop}%
\bibitem [{\citenamefont {Kamiya}\ \emph {et~al.}(2018)\citenamefont {Kamiya},
  \citenamefont {Takeuchi}, \citenamefont {Kabeya}, \citenamefont {Wada},
  \citenamefont {Ishimasa}, \citenamefont {Ochiai}, \citenamefont {Deguchi},
  \citenamefont {Imura},\ and\ \citenamefont {Sato}}]{kamiya.18}%
  \BibitemOpen
  \bibfield  {author} {\bibinfo {author} {\bibfnamefont {K.}~\bibnamefont
  {Kamiya}}, \bibinfo {author} {\bibfnamefont {T.}~\bibnamefont {Takeuchi}},
  \bibinfo {author} {\bibfnamefont {N.}~\bibnamefont {Kabeya}}, \bibinfo
  {author} {\bibfnamefont {N.}~\bibnamefont {Wada}}, \bibinfo {author}
  {\bibfnamefont {T.}~\bibnamefont {Ishimasa}}, \bibinfo {author}
  {\bibfnamefont {A.}~\bibnamefont {Ochiai}}, \bibinfo {author} {\bibfnamefont
  {K.}~\bibnamefont {Deguchi}}, \bibinfo {author} {\bibfnamefont
  {K.}~\bibnamefont {Imura}},\ and\ \bibinfo {author} {\bibfnamefont {N.~K.}\
  \bibnamefont {Sato}},\ }\bibfield  {title} {\bibinfo {title} {Discovery of
  superconductivity in quasicrystal},\ }\href
  {https://doi.org/10.1038/s41467-017-02667-x} {\bibfield  {journal} {\bibinfo
  {journal} {Nat. Commun.}\ }\textbf {\bibinfo {volume} {9}},\ \bibinfo {pages}
  {154} (\bibinfo {year} {2018})}\BibitemShut {NoStop}%
\bibitem [{\citenamefont {Sakai}\ and\ \citenamefont {Arita}(2019)}]{sakai.19}%
  \BibitemOpen
  \bibfield  {author} {\bibinfo {author} {\bibfnamefont {S.}~\bibnamefont
  {Sakai}}\ and\ \bibinfo {author} {\bibfnamefont {R.}~\bibnamefont {Arita}},\
  }\bibfield  {title} {\bibinfo {title} {Exotic pairing state in
  quasicrystalline superconductors under a magnetic field},\ }\href
  {https://doi.org/10.1103/PhysRevResearch.1.022002} {\bibfield  {journal}
  {\bibinfo  {journal} {Phys. Rev. Research}\ }\textbf {\bibinfo {volume}
  {1}},\ \bibinfo {pages} {022002(R)} (\bibinfo {year} {2019})}\BibitemShut
  {NoStop}%
\bibitem [{\citenamefont {Ara\'ujo}\ and\ \citenamefont
  {Andrade}(2019)}]{araujo.19}%
  \BibitemOpen
  \bibfield  {author} {\bibinfo {author} {\bibfnamefont {R.~N.}\ \bibnamefont
  {Ara\'ujo}}\ and\ \bibinfo {author} {\bibfnamefont {E.~C.}\ \bibnamefont
  {Andrade}},\ }\bibfield  {title} {\bibinfo {title} {Conventional
  superconductivity in quasicrystals},\ }\href
  {https://doi.org/10.1103/PhysRevB.100.014510} {\bibfield  {journal} {\bibinfo
   {journal} {Phys. Rev. B}\ }\textbf {\bibinfo {volume} {100}},\ \bibinfo
  {pages} {014510} (\bibinfo {year} {2019})}\BibitemShut {NoStop}%
\bibitem [{\citenamefont {Shiino}\ \emph {et~al.}(2021)\citenamefont {Shiino},
  \citenamefont {Gebresenbut}, \citenamefont {Denoel}, \citenamefont {Mathieu},
  \citenamefont {H\"aussermann},\ and\ \citenamefont {Rydh}}]{shiino.21}%
  \BibitemOpen
  \bibfield  {author} {\bibinfo {author} {\bibfnamefont {T.}~\bibnamefont
  {Shiino}}, \bibinfo {author} {\bibfnamefont {G.~H.}\ \bibnamefont
  {Gebresenbut}}, \bibinfo {author} {\bibfnamefont {F.}~\bibnamefont {Denoel}},
  \bibinfo {author} {\bibfnamefont {R.}~\bibnamefont {Mathieu}}, \bibinfo
  {author} {\bibfnamefont {U.}~\bibnamefont {H\"aussermann}},\ and\ \bibinfo
  {author} {\bibfnamefont {A.}~\bibnamefont {Rydh}},\ }\bibfield  {title}
  {\bibinfo {title} {{Superconductivity at 1 K in Y-Au-Si quasicrystal
  approximants}},\ }\href {https://doi.org/10.1103/PhysRevB.103.054510}
  {\bibfield  {journal} {\bibinfo  {journal} {Phys. Rev. B}\ }\textbf {\bibinfo
  {volume} {103}},\ \bibinfo {pages} {054510} (\bibinfo {year}
  {2021})}\BibitemShut {NoStop}%
\bibitem [{\citenamefont {Jagannathan}(2021)}]{jagannathan.21}%
  \BibitemOpen
  \bibfield  {author} {\bibinfo {author} {\bibfnamefont {A.}~\bibnamefont
  {Jagannathan}},\ }\bibfield  {title} {\bibinfo {title} {{The Fibonacci
  quasicrystal: Case study of hidden dimensions and multifractality}},\ }\href
  {https://doi.org/10.1103/RevModPhys.93.045001} {\bibfield  {journal}
  {\bibinfo  {journal} {Rev. Mod. Phys.}\ }\textbf {\bibinfo {volume} {93}},\
  \bibinfo {pages} {045001} (\bibinfo {year} {2021})}\BibitemShut {NoStop}%
\bibitem [{\citenamefont {Nagai}\ \emph {et~al.}(2024)\citenamefont {Nagai},
  \citenamefont {Iwasaki}, \citenamefont {Kitahara}, \citenamefont {Takagiwa},
  \citenamefont {Kimura},\ and\ \citenamefont {Shiga}}]{nagai.24}%
  \BibitemOpen
  \bibfield  {author} {\bibinfo {author} {\bibfnamefont {Y.}~\bibnamefont
  {Nagai}}, \bibinfo {author} {\bibfnamefont {Y.}~\bibnamefont {Iwasaki}},
  \bibinfo {author} {\bibfnamefont {K.}~\bibnamefont {Kitahara}}, \bibinfo
  {author} {\bibfnamefont {Y.}~\bibnamefont {Takagiwa}}, \bibinfo {author}
  {\bibfnamefont {K.}~\bibnamefont {Kimura}},\ and\ \bibinfo {author}
  {\bibfnamefont {M.}~\bibnamefont {Shiga}},\ }\bibfield  {title} {\bibinfo
  {title} {High-temperature atomic diffusion and specific heat in
  quasicrystals},\ }\href {https://doi.org/10.1103/PhysRevLett.132.196301}
  {\bibfield  {journal} {\bibinfo  {journal} {Phys. Rev. Lett.}\ }\textbf
  {\bibinfo {volume} {132}},\ \bibinfo {pages} {196301} (\bibinfo {year}
  {2024})}\BibitemShut {NoStop}%
\bibitem [{\citenamefont {Penrose}(1974)}]{penrose.74}%
  \BibitemOpen
  \bibfield  {author} {\bibinfo {author} {\bibfnamefont {R.}~\bibnamefont
  {Penrose}},\ }\bibfield  {title} {\bibinfo {title} {The role of aesthetics in
  pure and applied mathematical research},\ }\href
  {https://cir.nii.ac.jp/crid/1572543024090822912} {\bibfield  {journal}
  {\bibinfo  {journal} {Bull. Inst. Math. Appl.}\ }\textbf {\bibinfo {volume}
  {10}},\ \bibinfo {pages} {266} (\bibinfo {year} {1974})}\BibitemShut
  {NoStop}%
\bibitem [{\citenamefont {Hiller}(1985)}]{hiller.85}%
  \BibitemOpen
  \bibfield  {author} {\bibinfo {author} {\bibfnamefont {H.}~\bibnamefont
  {Hiller}},\ }\bibfield  {title} {\bibinfo {title} {{The crystallographic
  restriction in higher dimensions}},\ }\href
  {https://doi.org/10.1107/S0108767385001180} {\bibfield  {journal} {\bibinfo
  {journal} {Acta Crystallogr., Sect. A}\ }\textbf {\bibinfo {volume} {41}},\
  \bibinfo {pages} {541} (\bibinfo {year} {1985})}\BibitemShut {NoStop}%
\bibitem [{\citenamefont {Janot}(1997)}]{janot.97}%
  \BibitemOpen
  \bibfield  {author} {\bibinfo {author} {\bibfnamefont {C.}~\bibnamefont
  {Janot}},\ }\href {https://books.google.se/books?id=dpPePU-xs2oC} {\emph
  {\bibinfo {title} {{Quasicrystals: A Primer}}}},\ Monographs on the physics
  and chemistry of materials\ (\bibinfo  {publisher} {Clarendon Press},\
  \bibinfo {year} {1997})\BibitemShut {NoStop}%
\bibitem [{\citenamefont {Goldman}\ and\ \citenamefont
  {Kelton}(1993)}]{goldman.93}%
  \BibitemOpen
  \bibfield  {author} {\bibinfo {author} {\bibfnamefont {A.~I.}\ \bibnamefont
  {Goldman}}\ and\ \bibinfo {author} {\bibfnamefont {R.~F.}\ \bibnamefont
  {Kelton}},\ }\bibfield  {title} {\bibinfo {title} {Quasicrystals and
  crystalline approximants},\ }\href
  {https://doi.org/10.1103/RevModPhys.65.213} {\bibfield  {journal} {\bibinfo
  {journal} {Rev. Mod. Phys.}\ }\textbf {\bibinfo {volume} {65}},\ \bibinfo
  {pages} {213} (\bibinfo {year} {1993})}\BibitemShut {NoStop}%
\bibitem [{\citenamefont {Baake}\ and\ \citenamefont
  {Grimm}(2012)}]{baake.grimm.2012}%
  \BibitemOpen
  \bibfield  {author} {\bibinfo {author} {\bibfnamefont {M.}~\bibnamefont
  {Baake}}\ and\ \bibinfo {author} {\bibfnamefont {U.}~\bibnamefont {Grimm}},\
  }\bibfield  {title} {\bibinfo {title} {Mathematical diffraction of aperiodic
  structures},\ }\href {https://doi.org/10.1039/C2CS35120J} {\bibfield
  {journal} {\bibinfo  {journal} {Chem. Soc. Rev.}\ }\textbf {\bibinfo {volume}
  {41}},\ \bibinfo {pages} {6821} (\bibinfo {year} {2012})}\BibitemShut
  {NoStop}%
\bibitem [{\citenamefont {Kohmoto}\ \emph {et~al.}(1983)\citenamefont
  {Kohmoto}, \citenamefont {Kadanoff},\ and\ \citenamefont
  {Tang}}]{kohmoto.83}%
  \BibitemOpen
  \bibfield  {author} {\bibinfo {author} {\bibfnamefont {M.}~\bibnamefont
  {Kohmoto}}, \bibinfo {author} {\bibfnamefont {L.~P.}\ \bibnamefont
  {Kadanoff}},\ and\ \bibinfo {author} {\bibfnamefont {C.}~\bibnamefont
  {Tang}},\ }\bibfield  {title} {\bibinfo {title} {Localization problem in one
  dimension: Mapping and escape},\ }\href
  {https://doi.org/10.1103/PhysRevLett.50.1870} {\bibfield  {journal} {\bibinfo
   {journal} {Phys. Rev. Lett.}\ }\textbf {\bibinfo {volume} {50}},\ \bibinfo
  {pages} {1870} (\bibinfo {year} {1983})}\BibitemShut {NoStop}%
\bibitem [{\citenamefont {Ostlund}\ \emph {et~al.}(1983)\citenamefont
  {Ostlund}, \citenamefont {Pandit}, \citenamefont {Rand}, \citenamefont
  {Schellnhuber},\ and\ \citenamefont {Siggia}}]{ostlund.83}%
  \BibitemOpen
  \bibfield  {author} {\bibinfo {author} {\bibfnamefont {S.}~\bibnamefont
  {Ostlund}}, \bibinfo {author} {\bibfnamefont {R.}~\bibnamefont {Pandit}},
  \bibinfo {author} {\bibfnamefont {D.}~\bibnamefont {Rand}}, \bibinfo {author}
  {\bibfnamefont {H.~J.}\ \bibnamefont {Schellnhuber}},\ and\ \bibinfo {author}
  {\bibfnamefont {E.~D.}\ \bibnamefont {Siggia}},\ }\bibfield  {title}
  {\bibinfo {title} {{One-Dimensional Schr\"odinger Equation with an Almost
  Periodic Potential}},\ }\href {https://doi.org/10.1103/PhysRevLett.50.1873}
  {\bibfield  {journal} {\bibinfo  {journal} {Phys. Rev. Lett.}\ }\textbf
  {\bibinfo {volume} {50}},\ \bibinfo {pages} {1873} (\bibinfo {year}
  {1983})}\BibitemShut {NoStop}%
\bibitem [{\citenamefont {Ostlund}\ and\ \citenamefont
  {Pandit}(1984)}]{ostlund.pandit.84}%
  \BibitemOpen
  \bibfield  {author} {\bibinfo {author} {\bibfnamefont {S.}~\bibnamefont
  {Ostlund}}\ and\ \bibinfo {author} {\bibfnamefont {R.}~\bibnamefont
  {Pandit}},\ }\bibfield  {title} {\bibinfo {title} {{Renormalization-group
  analysis of the discrete quasiperiodic Schr\"odinger equation}},\ }\href
  {https://doi.org/10.1103/PhysRevB.29.1394} {\bibfield  {journal} {\bibinfo
  {journal} {Phys. Rev. B}\ }\textbf {\bibinfo {volume} {29}},\ \bibinfo
  {pages} {1394} (\bibinfo {year} {1984})}\BibitemShut {NoStop}%
\bibitem [{\citenamefont {Kohmoto}\ and\ \citenamefont
  {Oono}(1984)}]{kohmoto.oono.84}%
  \BibitemOpen
  \bibfield  {author} {\bibinfo {author} {\bibfnamefont {M.}~\bibnamefont
  {Kohmoto}}\ and\ \bibinfo {author} {\bibfnamefont {Y.}~\bibnamefont {Oono}},\
  }\bibfield  {title} {\bibinfo {title} {Cantor spectrum for an almost periodic
  schr{\"o}dinger equation and a dynamical map},\ }\href
  {https://doi.org/https://doi.org/10.1016/0375-9601(84)90928-9} {\bibfield
  {journal} {\bibinfo  {journal} {Physics Letters A}\ }\textbf {\bibinfo
  {volume} {102}},\ \bibinfo {pages} {145} (\bibinfo {year}
  {1984})}\BibitemShut {NoStop}%
\bibitem [{\citenamefont {Tang}\ and\ \citenamefont
  {Kohmoto}(1986)}]{tang.kohmoto.86}%
  \BibitemOpen
  \bibfield  {author} {\bibinfo {author} {\bibfnamefont {C.}~\bibnamefont
  {Tang}}\ and\ \bibinfo {author} {\bibfnamefont {M.}~\bibnamefont {Kohmoto}},\
  }\bibfield  {title} {\bibinfo {title} {Global scaling properties of the
  spectrum for a quasiperiodic schr\"odinger equation},\ }\href
  {https://doi.org/10.1103/PhysRevB.34.2041} {\bibfield  {journal} {\bibinfo
  {journal} {Phys. Rev. B}\ }\textbf {\bibinfo {volume} {34}},\ \bibinfo
  {pages} {2041} (\bibinfo {year} {1986})}\BibitemShut {NoStop}%
\bibitem [{\citenamefont {Kalugin}\ \emph {et~al.}(1986)\citenamefont
  {Kalugin}, \citenamefont {Yu.},\ and\ \citenamefont
  {Levitov}}]{kalugin.kitaev.levitov.86}%
  \BibitemOpen
  \bibfield  {author} {\bibinfo {author} {\bibfnamefont {P.~A.}\ \bibnamefont
  {Kalugin}}, \bibinfo {author} {\bibfnamefont {K.~A.}\ \bibnamefont {Yu.}},\
  and\ \bibinfo {author} {\bibfnamefont {L.~S.}\ \bibnamefont {Levitov}},\
  }\bibfield  {title} {\bibinfo {title} {Electron spectrum of a one-dimensional
  quasicrystal},\ }\href@noop {} {\bibfield  {journal} {\bibinfo  {journal}
  {JETP}\ }\textbf {\bibinfo {volume} {64}},\ \bibinfo {pages} {410} (\bibinfo
  {year} {1986})}\BibitemShut {NoStop}%
\bibitem [{\citenamefont {Evangelou}(1987)}]{evangelou.87}%
  \BibitemOpen
  \bibfield  {author} {\bibinfo {author} {\bibfnamefont {S.~N.}\ \bibnamefont
  {Evangelou}},\ }\bibfield  {title} {\bibinfo {title} {Multi-fractal spectra
  and wavefunctions of one-dimensional quasi-crystals},\ }\href
  {https://doi.org/10.1088/0022-3719/20/15/002} {\bibfield  {journal} {\bibinfo
   {journal} {J. Phys. C: Solid State Phys.}\ }\textbf {\bibinfo {volume}
  {20}},\ \bibinfo {pages} {L295} (\bibinfo {year} {1987})}\BibitemShut
  {NoStop}%
\bibitem [{\citenamefont {Kohmoto}\ \emph {et~al.}(1987)\citenamefont
  {Kohmoto}, \citenamefont {Sutherland},\ and\ \citenamefont
  {Tang}}]{kohmoto.sutherland.tang.87}%
  \BibitemOpen
  \bibfield  {author} {\bibinfo {author} {\bibfnamefont {M.}~\bibnamefont
  {Kohmoto}}, \bibinfo {author} {\bibfnamefont {B.}~\bibnamefont
  {Sutherland}},\ and\ \bibinfo {author} {\bibfnamefont {C.}~\bibnamefont
  {Tang}},\ }\bibfield  {title} {\bibinfo {title} {Critical wave functions and
  a cantor-set spectrum of a one-dimensional quasicrystal model},\ }\href
  {https://doi.org/10.1103/PhysRevB.35.1020} {\bibfield  {journal} {\bibinfo
  {journal} {Phys. Rev. B}\ }\textbf {\bibinfo {volume} {35}},\ \bibinfo
  {pages} {1020} (\bibinfo {year} {1987})}\BibitemShut {NoStop}%
\bibitem [{\citenamefont {Sutherland}\ and\ \citenamefont
  {Kohmoto}(1987)}]{sutherland.kohmoto.87}%
  \BibitemOpen
  \bibfield  {author} {\bibinfo {author} {\bibfnamefont {B.}~\bibnamefont
  {Sutherland}}\ and\ \bibinfo {author} {\bibfnamefont {M.}~\bibnamefont
  {Kohmoto}},\ }\bibfield  {title} {\bibinfo {title} {Resistance of a
  one-dimensional quasicrystal: Power-law growth},\ }\href
  {https://doi.org/10.1103/PhysRevB.36.5877} {\bibfield  {journal} {\bibinfo
  {journal} {Phys. Rev. B}\ }\textbf {\bibinfo {volume} {36}},\ \bibinfo
  {pages} {5877} (\bibinfo {year} {1987})}\BibitemShut {NoStop}%
\bibitem [{\citenamefont {Luck}(1989)}]{luck.89}%
  \BibitemOpen
  \bibfield  {author} {\bibinfo {author} {\bibfnamefont {J.~M.}\ \bibnamefont
  {Luck}},\ }\bibfield  {title} {\bibinfo {title} {Cantor spectra and scaling
  of gap widths in deterministic aperiodic systems},\ }\href
  {https://doi.org/10.1103/PhysRevB.39.5834} {\bibfield  {journal} {\bibinfo
  {journal} {Phys. Rev. B}\ }\textbf {\bibinfo {volume} {39}},\ \bibinfo
  {pages} {5834} (\bibinfo {year} {1989})}\BibitemShut {NoStop}%
\bibitem [{\citenamefont {Hiramoto}\ and\ \citenamefont
  {Kohmoto}(1992)}]{hiramoto.92}%
  \BibitemOpen
  \bibfield  {author} {\bibinfo {author} {\bibfnamefont {H.}~\bibnamefont
  {Hiramoto}}\ and\ \bibinfo {author} {\bibfnamefont {M.}~\bibnamefont
  {Kohmoto}},\ }\bibfield  {title} {\bibinfo {title} {Electronic spectral and
  wavefunction properties of one-dimensional quasiperiodic systems: A scaling
  approach},\ }\href {https://doi.org/10.1142/S0217979292000153} {\bibfield
  {journal} {\bibinfo  {journal} {Int. J. Mod. Phys. B}\ }\textbf {\bibinfo
  {volume} {06}},\ \bibinfo {pages} {281} (\bibinfo {year} {1992})}\BibitemShut
  {NoStop}%
\bibitem [{\citenamefont {Kalugin}\ and\ \citenamefont
  {Katz}(2014)}]{kalugin.katz.14}%
  \BibitemOpen
  \bibfield  {author} {\bibinfo {author} {\bibfnamefont {P.}~\bibnamefont
  {Kalugin}}\ and\ \bibinfo {author} {\bibfnamefont {A.}~\bibnamefont {Katz}},\
  }\bibfield  {title} {\bibinfo {title} {Electrons in deterministic
  quasicrystalline potentials and hidden conserved quantities},\ }\href
  {https://doi.org/10.1088/1751-8113/47/31/315206} {\bibfield  {journal}
  {\bibinfo  {journal} {J. Phys. A: Math. Theor.}\ }\textbf {\bibinfo {volume}
  {47}},\ \bibinfo {pages} {315206} (\bibinfo {year} {2014})}\BibitemShut
  {NoStop}%
\bibitem [{\citenamefont {Mac\'e}\ \emph {et~al.}(2016)\citenamefont {Mac\'e},
  \citenamefont {Jagannathan},\ and\ \citenamefont
  {Pi\'echon}}]{mace.jagannathan.piechon.16}%
  \BibitemOpen
  \bibfield  {author} {\bibinfo {author} {\bibfnamefont {N.}~\bibnamefont
  {Mac\'e}}, \bibinfo {author} {\bibfnamefont {A.}~\bibnamefont
  {Jagannathan}},\ and\ \bibinfo {author} {\bibfnamefont {F.}~\bibnamefont
  {Pi\'echon}},\ }\bibfield  {title} {\bibinfo {title} {Fractal dimensions of
  wave functions and local spectral measures on the fibonacci chain},\ }\href
  {https://doi.org/10.1103/PhysRevB.93.205153} {\bibfield  {journal} {\bibinfo
  {journal} {Phys. Rev. B}\ }\textbf {\bibinfo {volume} {93}},\ \bibinfo
  {pages} {205153} (\bibinfo {year} {2016})}\BibitemShut {NoStop}%
\bibitem [{\citenamefont {Mac\'e}\ \emph {et~al.}(2017)\citenamefont {Mac\'e},
  \citenamefont {Jagannathan}, \citenamefont {Kalugin}, \citenamefont
  {Mosseri},\ and\ \citenamefont {Pi\'echon}}]{mace.17}%
  \BibitemOpen
  \bibfield  {author} {\bibinfo {author} {\bibfnamefont {N.}~\bibnamefont
  {Mac\'e}}, \bibinfo {author} {\bibfnamefont {A.}~\bibnamefont {Jagannathan}},
  \bibinfo {author} {\bibfnamefont {P.}~\bibnamefont {Kalugin}}, \bibinfo
  {author} {\bibfnamefont {R.}~\bibnamefont {Mosseri}},\ and\ \bibinfo {author}
  {\bibfnamefont {F.}~\bibnamefont {Pi\'echon}},\ }\bibfield  {title} {\bibinfo
  {title} {Critical eigenstates and their properties in one- and
  two-dimensional quasicrystals},\ }\href
  {https://doi.org/10.1103/PhysRevB.96.045138} {\bibfield  {journal} {\bibinfo
  {journal} {Phys. Rev. B}\ }\textbf {\bibinfo {volume} {96}},\ \bibinfo
  {pages} {045138} (\bibinfo {year} {2017})}\BibitemShut {NoStop}%
\bibitem [{\citenamefont {Torquato}(2018)}]{torquato.18}%
  \BibitemOpen
  \bibfield  {author} {\bibinfo {author} {\bibfnamefont {S.}~\bibnamefont
  {Torquato}},\ }\bibfield  {title} {\bibinfo {title} {Hyperuniform states of
  matter},\ }\href
  {https://doi.org/https://doi.org/10.1016/j.physrep.2018.03.001} {\bibfield
  {journal} {\bibinfo  {journal} {Phys. Rep.}\ }\textbf {\bibinfo {volume}
  {745}},\ \bibinfo {pages} {1} (\bibinfo {year} {2018})}\BibitemShut {NoStop}%
\bibitem [{\citenamefont {Baake}\ and\ \citenamefont
  {Grimm}(2019)}]{baake.grimm.19}%
  \BibitemOpen
  \bibfield  {author} {\bibinfo {author} {\bibfnamefont {M.}~\bibnamefont
  {Baake}}\ and\ \bibinfo {author} {\bibfnamefont {U.}~\bibnamefont {Grimm}},\
  }\bibfield  {title} {\bibinfo {title} {Scaling of diffraction intensities
  near the origin: some rigorous results},\ }\href
  {https://doi.org/10.1088/1742-5468/ab02f2} {\bibfield  {journal} {\bibinfo
  {journal} {J. Stat. Mech.: Theory Exp.}\ }\textbf {\bibinfo {volume}
  {2019}}\bibinfo  {number} { (5)},\ \bibinfo {pages} {054003}}\BibitemShut
  {NoStop}%
\bibitem [{\citenamefont {Hafner}\ and\ \citenamefont
  {Kraj\ifmmode~\check{c}\else \v{c}\fi{}\'{\i}}(1992)}]{hafner.92}%
  \BibitemOpen
\bibfield  {number} {  }\bibfield  {author} {\bibinfo {author} {\bibfnamefont
  {J.}~\bibnamefont {Hafner}}\ and\ \bibinfo {author} {\bibfnamefont
  {M.}~\bibnamefont {Kraj\ifmmode~\check{c}\else \v{c}\fi{}\'{\i}}},\
  }\bibfield  {title} {\bibinfo {title} {Electronic structure and stability of
  quasicrystals: Quasiperiodic dispersion relations and pseudogaps},\ }\href
  {https://doi.org/10.1103/PhysRevLett.68.2321} {\bibfield  {journal} {\bibinfo
   {journal} {Phys. Rev. Lett.}\ }\textbf {\bibinfo {volume} {68}},\ \bibinfo
  {pages} {2321} (\bibinfo {year} {1992})}\BibitemShut {NoStop}%
\bibitem [{\citenamefont {Zijlstra}\ and\ \citenamefont
  {Janssen}(2000)}]{zijlstra.00}%
  \BibitemOpen
  \bibfield  {author} {\bibinfo {author} {\bibfnamefont {E.~S.}\ \bibnamefont
  {Zijlstra}}\ and\ \bibinfo {author} {\bibfnamefont {T.}~\bibnamefont
  {Janssen}},\ }\bibfield  {title} {\bibinfo {title} {Density of states and
  localization of electrons in a tight-binding model on the penrose tiling},\
  }\href {https://doi.org/10.1103/PhysRevB.61.3377} {\bibfield  {journal}
  {\bibinfo  {journal} {Phys. Rev. B}\ }\textbf {\bibinfo {volume} {61}},\
  \bibinfo {pages} {3377} (\bibinfo {year} {2000})}\BibitemShut {NoStop}%
\bibitem [{\citenamefont {Stadnik}\ \emph {et~al.}(2001)\citenamefont
  {Stadnik}, \citenamefont {Purdie}, \citenamefont {Baer},\ and\ \citenamefont
  {Lograsso}}]{stadnik.01}%
  \BibitemOpen
  \bibfield  {author} {\bibinfo {author} {\bibfnamefont {Z.~M.}\ \bibnamefont
  {Stadnik}}, \bibinfo {author} {\bibfnamefont {D.}~\bibnamefont {Purdie}},
  \bibinfo {author} {\bibfnamefont {Y.}~\bibnamefont {Baer}},\ and\ \bibinfo
  {author} {\bibfnamefont {T.~A.}\ \bibnamefont {Lograsso}},\ }\bibfield
  {title} {\bibinfo {title} {Absence of fine structure in the photoemission
  spectrum of the icosahedral al-pd-mn quasicrystal},\ }\href
  {https://doi.org/10.1103/PhysRevB.64.214202} {\bibfield  {journal} {\bibinfo
  {journal} {Phys. Rev. B}\ }\textbf {\bibinfo {volume} {64}},\ \bibinfo
  {pages} {214202} (\bibinfo {year} {2001})}\BibitemShut {NoStop}%
\bibitem [{\citenamefont {Rotenberg}\ \emph {et~al.}(2004)\citenamefont
  {Rotenberg}, \citenamefont {Theis},\ and\ \citenamefont
  {Horn}}]{rotenberg.04}%
  \BibitemOpen
  \bibfield  {author} {\bibinfo {author} {\bibfnamefont {E.}~\bibnamefont
  {Rotenberg}}, \bibinfo {author} {\bibfnamefont {W.}~\bibnamefont {Theis}},\
  and\ \bibinfo {author} {\bibfnamefont {K.}~\bibnamefont {Horn}},\ }\bibfield
  {title} {\bibinfo {title} {Electronic structure investigations of
  quasicrystals},\ }\href
  {https://doi.org/https://doi.org/10.1016/j.progsurf.2004.05.002} {\bibfield
  {journal} {\bibinfo  {journal} {Prog. Surf. Sci.}\ }\textbf {\bibinfo
  {volume} {75}},\ \bibinfo {pages} {237} (\bibinfo {year} {2004})},\ \bibinfo
  {note} {quasicrystals}\BibitemShut {NoStop}%
\bibitem [{\citenamefont {Bellissard}\ \emph {et~al.}(1989)\citenamefont
  {Bellissard}, \citenamefont {Iochum}, \citenamefont {Scoppola},\ and\
  \citenamefont {Testard}}]{bellissard.89}%
  \BibitemOpen
  \bibfield  {author} {\bibinfo {author} {\bibfnamefont {J.}~\bibnamefont
  {Bellissard}}, \bibinfo {author} {\bibfnamefont {B.}~\bibnamefont {Iochum}},
  \bibinfo {author} {\bibfnamefont {E.}~\bibnamefont {Scoppola}},\ and\
  \bibinfo {author} {\bibfnamefont {D.}~\bibnamefont {Testard}},\ }\bibfield
  {title} {\bibinfo {title} {Spectral properties of one dimensional
  quasi-crystals},\ }\href {https://doi.org/10.1007/BF01218415} {\bibfield
  {journal} {\bibinfo  {journal} {Commun. Math. Phys.}\ }\textbf {\bibinfo
  {volume} {125}},\ \bibinfo {pages} {527} (\bibinfo {year}
  {1989})}\BibitemShut {NoStop}%
\bibitem [{\citenamefont {Bellissard}\ \emph {et~al.}(1992)\citenamefont
  {Bellissard}, \citenamefont {Bovier},\ and\ \citenamefont
  {Ghez}}]{bellissard.92}%
  \BibitemOpen
  \bibfield  {author} {\bibinfo {author} {\bibfnamefont {J.}~\bibnamefont
  {Bellissard}}, \bibinfo {author} {\bibfnamefont {A.}~\bibnamefont {Bovier}},\
  and\ \bibinfo {author} {\bibfnamefont {J.-M.}\ \bibnamefont {Ghez}},\
  }\bibfield  {title} {\bibinfo {title} {{Gap labelling theorems for one
  dimensional discrete Schr{\"o}dinger operators}},\ }\href
  {https://doi.org/10.1142/S0129055X92000029} {\bibfield  {journal} {\bibinfo
  {journal} {Rev. Math. Phys.}\ }\textbf {\bibinfo {volume} {04}},\ \bibinfo
  {pages} {1} (\bibinfo {year} {1992})}\BibitemShut {NoStop}%
\bibitem [{\citenamefont {Mac{\'{e}}}\ \emph {et~al.}(2017)\citenamefont
  {Mac{\'{e}}}, \citenamefont {Jagannathan},\ and\ \citenamefont
  {Pi{\'{e}}chon}}]{mace.jagannathan.piechon.17}%
  \BibitemOpen
  \bibfield  {author} {\bibinfo {author} {\bibfnamefont {N.}~\bibnamefont
  {Mac{\'{e}}}}, \bibinfo {author} {\bibfnamefont {A.}~\bibnamefont
  {Jagannathan}},\ and\ \bibinfo {author} {\bibfnamefont {F.}~\bibnamefont
  {Pi{\'{e}}chon}},\ }\bibfield  {title} {\bibinfo {title} {{Gap structure of
  1D cut and project Hamiltonians}},\ }\href
  {https://doi.org/10.1088/1742-6596/809/1/012023} {\bibfield  {journal}
  {\bibinfo  {journal} {J. Phys. Conf. Ser}\ }\textbf {\bibinfo {volume}
  {809}},\ \bibinfo {pages} {012023} (\bibinfo {year} {2017})}\BibitemShut
  {NoStop}%
\bibitem [{\citenamefont {Yamamoto}\ and\ \citenamefont
  {Koshino}(2022)}]{yamamoto.22}%
  \BibitemOpen
  \bibfield  {author} {\bibinfo {author} {\bibfnamefont {K.}~\bibnamefont
  {Yamamoto}}\ and\ \bibinfo {author} {\bibfnamefont {M.}~\bibnamefont
  {Koshino}},\ }\bibfield  {title} {\bibinfo {title} {Topological gap labeling
  with third chern numbers in three-dimensional quasicrystals},\ }\href
  {https://doi.org/10.1103/PhysRevB.105.115410} {\bibfield  {journal} {\bibinfo
   {journal} {Phys. Rev. B}\ }\textbf {\bibinfo {volume} {105}},\ \bibinfo
  {pages} {115410} (\bibinfo {year} {2022})}\BibitemShut {NoStop}%
\bibitem [{\citenamefont {Kraus}\ \emph {et~al.}(2012)\citenamefont {Kraus},
  \citenamefont {Lahini}, \citenamefont {Ringel}, \citenamefont {Verbin},\ and\
  \citenamefont {Zilberberg}}]{kraus.12}%
  \BibitemOpen
  \bibfield  {author} {\bibinfo {author} {\bibfnamefont {Y.~E.}\ \bibnamefont
  {Kraus}}, \bibinfo {author} {\bibfnamefont {Y.}~\bibnamefont {Lahini}},
  \bibinfo {author} {\bibfnamefont {Z.}~\bibnamefont {Ringel}}, \bibinfo
  {author} {\bibfnamefont {M.}~\bibnamefont {Verbin}},\ and\ \bibinfo {author}
  {\bibfnamefont {O.}~\bibnamefont {Zilberberg}},\ }\bibfield  {title}
  {\bibinfo {title} {Topological states and adiabatic pumping in
  quasicrystals},\ }\href {https://doi.org/10.1103/PhysRevLett.109.106402}
  {\bibfield  {journal} {\bibinfo  {journal} {Phys. Rev. Lett.}\ }\textbf
  {\bibinfo {volume} {109}},\ \bibinfo {pages} {106402} (\bibinfo {year}
  {2012})}\BibitemShut {NoStop}%
\bibitem [{\citenamefont {Kraus}\ and\ \citenamefont
  {Zilberberg}(2012)}]{kraus.zilberberg.12}%
  \BibitemOpen
  \bibfield  {author} {\bibinfo {author} {\bibfnamefont {Y.~E.}\ \bibnamefont
  {Kraus}}\ and\ \bibinfo {author} {\bibfnamefont {O.}~\bibnamefont
  {Zilberberg}},\ }\bibfield  {title} {\bibinfo {title} {Topological
  equivalence between the fibonacci quasicrystal and the harper model},\ }\href
  {https://doi.org/10.1103/PhysRevLett.109.116404} {\bibfield  {journal}
  {\bibinfo  {journal} {Phys. Rev. Lett.}\ }\textbf {\bibinfo {volume} {109}},\
  \bibinfo {pages} {116404} (\bibinfo {year} {2012})}\BibitemShut {NoStop}%
\bibitem [{\citenamefont {Verbin}\ \emph {et~al.}(2013)\citenamefont {Verbin},
  \citenamefont {Zilberberg}, \citenamefont {Kraus}, \citenamefont {Lahini},\
  and\ \citenamefont {Silberberg}}]{verbin.13}%
  \BibitemOpen
  \bibfield  {author} {\bibinfo {author} {\bibfnamefont {M.}~\bibnamefont
  {Verbin}}, \bibinfo {author} {\bibfnamefont {O.}~\bibnamefont {Zilberberg}},
  \bibinfo {author} {\bibfnamefont {Y.~E.}\ \bibnamefont {Kraus}}, \bibinfo
  {author} {\bibfnamefont {Y.}~\bibnamefont {Lahini}},\ and\ \bibinfo {author}
  {\bibfnamefont {Y.}~\bibnamefont {Silberberg}},\ }\bibfield  {title}
  {\bibinfo {title} {Observation of topological phase transitions in photonic
  quasicrystals},\ }\href {https://doi.org/10.1103/PhysRevLett.110.076403}
  {\bibfield  {journal} {\bibinfo  {journal} {Phys. Rev. Lett.}\ }\textbf
  {\bibinfo {volume} {110}},\ \bibinfo {pages} {076403} (\bibinfo {year}
  {2013})}\BibitemShut {NoStop}%
\bibitem [{\citenamefont {Kraus}\ \emph {et~al.}(2013)\citenamefont {Kraus},
  \citenamefont {Ringel},\ and\ \citenamefont {Zilberberg}}]{kraus.13}%
  \BibitemOpen
  \bibfield  {author} {\bibinfo {author} {\bibfnamefont {Y.~E.}\ \bibnamefont
  {Kraus}}, \bibinfo {author} {\bibfnamefont {Z.}~\bibnamefont {Ringel}},\ and\
  \bibinfo {author} {\bibfnamefont {O.}~\bibnamefont {Zilberberg}},\ }\bibfield
   {title} {\bibinfo {title} {Four-dimensional quantum hall effect in a
  two-dimensional quasicrystal},\ }\href
  {https://doi.org/10.1103/PhysRevLett.111.226401} {\bibfield  {journal}
  {\bibinfo  {journal} {Phys. Rev. Lett.}\ }\textbf {\bibinfo {volume} {111}},\
  \bibinfo {pages} {226401} (\bibinfo {year} {2013})}\BibitemShut {NoStop}%
\bibitem [{\citenamefont {Verbin}\ \emph {et~al.}(2015)\citenamefont {Verbin},
  \citenamefont {Zilberberg}, \citenamefont {Lahini}, \citenamefont {Kraus},\
  and\ \citenamefont {Silberberg}}]{verbin.15}%
  \BibitemOpen
  \bibfield  {author} {\bibinfo {author} {\bibfnamefont {M.}~\bibnamefont
  {Verbin}}, \bibinfo {author} {\bibfnamefont {O.}~\bibnamefont {Zilberberg}},
  \bibinfo {author} {\bibfnamefont {Y.}~\bibnamefont {Lahini}}, \bibinfo
  {author} {\bibfnamefont {Y.~E.}\ \bibnamefont {Kraus}},\ and\ \bibinfo
  {author} {\bibfnamefont {Y.}~\bibnamefont {Silberberg}},\ }\bibfield  {title}
  {\bibinfo {title} {Topological pumping over a photonic fibonacci
  quasicrystal},\ }\href {https://doi.org/10.1103/PhysRevB.91.064201}
  {\bibfield  {journal} {\bibinfo  {journal} {Phys. Rev. B}\ }\textbf {\bibinfo
  {volume} {91}},\ \bibinfo {pages} {064201} (\bibinfo {year}
  {2015})}\BibitemShut {NoStop}%
\bibitem [{\citenamefont {Flicker}\ and\ \citenamefont {van
  Wezel}(2015)}]{flicker.wezel.15}%
  \BibitemOpen
  \bibfield  {author} {\bibinfo {author} {\bibfnamefont {F.}~\bibnamefont
  {Flicker}}\ and\ \bibinfo {author} {\bibfnamefont {J.}~\bibnamefont {van
  Wezel}},\ }\bibfield  {title} {\bibinfo {title} {Quasiperiodicity and 2d
  topology in 1d charge-ordered materials},\ }\href
  {https://doi.org/10.1209/0295-5075/111/37008} {\bibfield  {journal} {\bibinfo
   {journal} {Europhys. Lett.}\ }\textbf {\bibinfo {volume} {111}},\ \bibinfo
  {pages} {37008} (\bibinfo {year} {2015})}\BibitemShut {NoStop}%
\bibitem [{\citenamefont {Kraus}\ and\ \citenamefont
  {Zilberberg}(2016)}]{kraus.16}%
  \BibitemOpen
  \bibfield  {author} {\bibinfo {author} {\bibfnamefont {Y.~E.}\ \bibnamefont
  {Kraus}}\ and\ \bibinfo {author} {\bibfnamefont {O.}~\bibnamefont
  {Zilberberg}},\ }\bibfield  {title} {\bibinfo {title} {Quasiperiodicity and
  topology transcend dimensions},\ }\href {https://doi.org/10.1038/nphys3784}
  {\bibfield  {journal} {\bibinfo  {journal} {Nat. Phys.}\ }\textbf {\bibinfo
  {volume} {12}},\ \bibinfo {pages} {624} (\bibinfo {year} {2016})}\BibitemShut
  {NoStop}%
\bibitem [{\citenamefont {Dareau}\ \emph {et~al.}(2017)\citenamefont {Dareau},
  \citenamefont {Levy}, \citenamefont {Aguilera}, \citenamefont {Bouganne},
  \citenamefont {Akkermans}, \citenamefont {Gerbier},\ and\ \citenamefont
  {Beugnon}}]{dareau.17}%
  \BibitemOpen
  \bibfield  {author} {\bibinfo {author} {\bibfnamefont {A.}~\bibnamefont
  {Dareau}}, \bibinfo {author} {\bibfnamefont {E.}~\bibnamefont {Levy}},
  \bibinfo {author} {\bibfnamefont {M.~B.}\ \bibnamefont {Aguilera}}, \bibinfo
  {author} {\bibfnamefont {R.}~\bibnamefont {Bouganne}}, \bibinfo {author}
  {\bibfnamefont {E.}~\bibnamefont {Akkermans}}, \bibinfo {author}
  {\bibfnamefont {F.}~\bibnamefont {Gerbier}},\ and\ \bibinfo {author}
  {\bibfnamefont {J.}~\bibnamefont {Beugnon}},\ }\bibfield  {title} {\bibinfo
  {title} {Revealing the topology of quasicrystals with a diffraction
  experiment},\ }\href {https://doi.org/10.1103/PhysRevLett.119.215304}
  {\bibfield  {journal} {\bibinfo  {journal} {Phys. Rev. Lett.}\ }\textbf
  {\bibinfo {volume} {119}},\ \bibinfo {pages} {215304} (\bibinfo {year}
  {2017})}\BibitemShut {NoStop}%
\bibitem [{\citenamefont {Huang}\ and\ \citenamefont
  {Liu}(2018)}]{huang.liu.18}%
  \BibitemOpen
  \bibfield  {author} {\bibinfo {author} {\bibfnamefont {H.}~\bibnamefont
  {Huang}}\ and\ \bibinfo {author} {\bibfnamefont {F.}~\bibnamefont {Liu}},\
  }\bibfield  {title} {\bibinfo {title} {Quantum spin hall effect and spin bott
  index in a quasicrystal lattice},\ }\href
  {https://doi.org/10.1103/PhysRevLett.121.126401} {\bibfield  {journal}
  {\bibinfo  {journal} {Phys. Rev. Lett.}\ }\textbf {\bibinfo {volume} {121}},\
  \bibinfo {pages} {126401} (\bibinfo {year} {2018})}\BibitemShut {NoStop}%
\bibitem [{\citenamefont {Petrides}\ \emph {et~al.}(2018)\citenamefont
  {Petrides}, \citenamefont {Price},\ and\ \citenamefont
  {Zilberberg}}]{petrides.18}%
  \BibitemOpen
  \bibfield  {author} {\bibinfo {author} {\bibfnamefont {I.}~\bibnamefont
  {Petrides}}, \bibinfo {author} {\bibfnamefont {H.~M.}\ \bibnamefont
  {Price}},\ and\ \bibinfo {author} {\bibfnamefont {O.}~\bibnamefont
  {Zilberberg}},\ }\bibfield  {title} {\bibinfo {title} {Six-dimensional
  quantum hall effect and three-dimensional topological pumps},\ }\href
  {https://doi.org/10.1103/PhysRevB.98.125431} {\bibfield  {journal} {\bibinfo
  {journal} {Phys. Rev. B}\ }\textbf {\bibinfo {volume} {98}},\ \bibinfo
  {pages} {125431} (\bibinfo {year} {2018})}\BibitemShut {NoStop}%
\bibitem [{\citenamefont {Huang}\ and\ \citenamefont
  {Liu}(2019)}]{huang.liu.19}%
  \BibitemOpen
  \bibfield  {author} {\bibinfo {author} {\bibfnamefont {H.}~\bibnamefont
  {Huang}}\ and\ \bibinfo {author} {\bibfnamefont {F.}~\bibnamefont {Liu}},\
  }\bibfield  {title} {\bibinfo {title} {Comparison of quantum spin hall states
  in quasicrystals and crystals},\ }\href
  {https://doi.org/10.1103/PhysRevB.100.085119} {\bibfield  {journal} {\bibinfo
   {journal} {Phys. Rev. B}\ }\textbf {\bibinfo {volume} {100}},\ \bibinfo
  {pages} {085119} (\bibinfo {year} {2019})}\BibitemShut {NoStop}%
\bibitem [{\citenamefont {Chen}\ \emph {et~al.}(2020)\citenamefont {Chen},
  \citenamefont {Chen}, \citenamefont {Gao}, \citenamefont {Zhou},\ and\
  \citenamefont {Xu}}]{chen.20}%
  \BibitemOpen
  \bibfield  {author} {\bibinfo {author} {\bibfnamefont {R.}~\bibnamefont
  {Chen}}, \bibinfo {author} {\bibfnamefont {C.-Z.}\ \bibnamefont {Chen}},
  \bibinfo {author} {\bibfnamefont {J.-H.}\ \bibnamefont {Gao}}, \bibinfo
  {author} {\bibfnamefont {B.}~\bibnamefont {Zhou}},\ and\ \bibinfo {author}
  {\bibfnamefont {D.-H.}\ \bibnamefont {Xu}},\ }\bibfield  {title} {\bibinfo
  {title} {Higher-order topological insulators in quasicrystals},\ }\href
  {https://doi.org/10.1103/PhysRevLett.124.036803} {\bibfield  {journal}
  {\bibinfo  {journal} {Phys. Rev. Lett.}\ }\textbf {\bibinfo {volume} {124}},\
  \bibinfo {pages} {036803} (\bibinfo {year} {2020})}\BibitemShut {NoStop}%
\bibitem [{\citenamefont {Rai}\ \emph {et~al.}(2021)\citenamefont {Rai},
  \citenamefont {Schl\"omer}, \citenamefont {Matsumura}, \citenamefont {Haas},\
  and\ \citenamefont {Jagannathan}}]{rai.21}%
  \BibitemOpen
  \bibfield  {author} {\bibinfo {author} {\bibfnamefont {G.}~\bibnamefont
  {Rai}}, \bibinfo {author} {\bibfnamefont {H.}~\bibnamefont {Schl\"omer}},
  \bibinfo {author} {\bibfnamefont {C.}~\bibnamefont {Matsumura}}, \bibinfo
  {author} {\bibfnamefont {S.}~\bibnamefont {Haas}},\ and\ \bibinfo {author}
  {\bibfnamefont {A.}~\bibnamefont {Jagannathan}},\ }\bibfield  {title}
  {\bibinfo {title} {Bulk topological signatures of a quasicrystal},\ }\href
  {https://doi.org/10.1103/PhysRevB.104.184202} {\bibfield  {journal} {\bibinfo
   {journal} {Phys. Rev. B}\ }\textbf {\bibinfo {volume} {104}},\ \bibinfo
  {pages} {184202} (\bibinfo {year} {2021})}\BibitemShut {NoStop}%
\bibitem [{\citenamefont {Fan}\ and\ \citenamefont
  {Huang}(2021)}]{fan.huang.21}%
  \BibitemOpen
  \bibfield  {author} {\bibinfo {author} {\bibfnamefont {J.}~\bibnamefont
  {Fan}}\ and\ \bibinfo {author} {\bibfnamefont {H.}~\bibnamefont {Huang}},\
  }\bibfield  {title} {\bibinfo {title} {Topological states in quasicrystals},\
  }\href {https://doi.org/10.1007/s11467-021-1100-y} {\bibfield  {journal}
  {\bibinfo  {journal} {Front. Phys.}\ }\textbf {\bibinfo {volume} {17}},\
  \bibinfo {pages} {13203} (\bibinfo {year} {2021})}\BibitemShut {NoStop}%
\bibitem [{\citenamefont {Sykes}\ and\ \citenamefont
  {Barnett}(2022)}]{sykes.barnett.22}%
  \BibitemOpen
  \bibfield  {author} {\bibinfo {author} {\bibfnamefont {J.}~\bibnamefont
  {Sykes}}\ and\ \bibinfo {author} {\bibfnamefont {R.}~\bibnamefont
  {Barnett}},\ }\bibfield  {title} {\bibinfo {title} {{1D quasicrystals and
  topological markers}},\ }\href {https://doi.org/10.1088/2633-4356/ac75a6}
  {\bibfield  {journal} {\bibinfo  {journal} {Mater. Quantum. Technol.}\
  }\textbf {\bibinfo {volume} {2}},\ \bibinfo {pages} {025005} (\bibinfo {year}
  {2022})}\BibitemShut {NoStop}%
\bibitem [{\citenamefont {Tamura}\ \emph {et~al.}(2010)\citenamefont {Tamura},
  \citenamefont {Muro}, \citenamefont {Hiroto}, \citenamefont {Nishimoto},\
  and\ \citenamefont {Takabatake}}]{tamura.10}%
  \BibitemOpen
  \bibfield  {author} {\bibinfo {author} {\bibfnamefont {R.}~\bibnamefont
  {Tamura}}, \bibinfo {author} {\bibfnamefont {Y.}~\bibnamefont {Muro}},
  \bibinfo {author} {\bibfnamefont {T.}~\bibnamefont {Hiroto}}, \bibinfo
  {author} {\bibfnamefont {K.}~\bibnamefont {Nishimoto}},\ and\ \bibinfo
  {author} {\bibfnamefont {T.}~\bibnamefont {Takabatake}},\ }\bibfield  {title}
  {\bibinfo {title} {{Long-range magnetic order in the quasicrystalline
  approximant ${\text{Cd}}_{6}\text{Tb}$}},\ }\href
  {https://doi.org/10.1103/PhysRevB.82.220201} {\bibfield  {journal} {\bibinfo
  {journal} {Phys. Rev. B}\ }\textbf {\bibinfo {volume} {82}},\ \bibinfo
  {pages} {220201(R)} (\bibinfo {year} {2010})}\BibitemShut {NoStop}%
\bibitem [{\citenamefont {Khosravian}\ and\ \citenamefont
  {Lado}(2021)}]{khosravian.21}%
  \BibitemOpen
  \bibfield  {author} {\bibinfo {author} {\bibfnamefont {M.}~\bibnamefont
  {Khosravian}}\ and\ \bibinfo {author} {\bibfnamefont {J.~L.}\ \bibnamefont
  {Lado}},\ }\bibfield  {title} {\bibinfo {title} {{Quasiperiodic criticality
  and spin-triplet superconductivity in superconductor-antiferromagnet moir\'e
  patterns}},\ }\href {https://doi.org/10.1103/PhysRevResearch.3.013262}
  {\bibfield  {journal} {\bibinfo  {journal} {Phys. Rev. Res.}\ }\textbf
  {\bibinfo {volume} {3}},\ \bibinfo {pages} {013262} (\bibinfo {year}
  {2021})}\BibitemShut {NoStop}%
\bibitem [{\citenamefont {Ghanta}\ \emph {et~al.}(2023)\citenamefont {Ghanta},
  \citenamefont {H{\"a}ussermann},\ and\ \citenamefont {Rydh}}]{ghanta.23}%
  \BibitemOpen
  \bibfield  {author} {\bibinfo {author} {\bibfnamefont {S.}~\bibnamefont
  {Ghanta}}, \bibinfo {author} {\bibfnamefont {U.}~\bibnamefont
  {H{\"a}ussermann}},\ and\ \bibinfo {author} {\bibfnamefont {A.}~\bibnamefont
  {Rydh}},\ }\bibfield  {title} {\bibinfo {title} {{Synthesis, structure, and
  physical properties of a Y–Au-Ge 1/1 Tsai-type quasicrystal approximant and
  Y14(Au,Ge)51 with the Gd14Ag51 structure type}},\ }\href
  {https://doi.org/https://doi.org/10.1016/j.jssc.2023.124246} {\bibfield
  {journal} {\bibinfo  {journal} {Journal of Solid State Chemistry}\ }\textbf
  {\bibinfo {volume} {327}},\ \bibinfo {pages} {124246} (\bibinfo {year}
  {2023})}\BibitemShut {NoStop}%
\bibitem [{\citenamefont {Moustaj}\ \emph {et~al.}(2023)\citenamefont
  {Moustaj}, \citenamefont {R{\"o}ntgen}, \citenamefont {Morfonios},
  \citenamefont {Schmelcher},\ and\ \citenamefont {Smith}}]{moustaj.23}%
  \BibitemOpen
  \bibfield  {author} {\bibinfo {author} {\bibfnamefont {A.}~\bibnamefont
  {Moustaj}}, \bibinfo {author} {\bibfnamefont {M.}~\bibnamefont
  {R{\"o}ntgen}}, \bibinfo {author} {\bibfnamefont {C.~V.}\ \bibnamefont
  {Morfonios}}, \bibinfo {author} {\bibfnamefont {P.}~\bibnamefont
  {Schmelcher}},\ and\ \bibinfo {author} {\bibfnamefont {C.~M.}\ \bibnamefont
  {Smith}},\ }\bibfield  {title} {\bibinfo {title} {Spectral properties of two
  coupled fibonacci chains},\ }\href {https://doi.org/10.1088/1367-2630/acf0e0}
  {\bibfield  {journal} {\bibinfo  {journal} {New J. Phys.}\ }\textbf {\bibinfo
  {volume} {25}},\ \bibinfo {pages} {093019} (\bibinfo {year}
  {2023})}\BibitemShut {NoStop}%
\bibitem [{\citenamefont {Fukushima}\ \emph {et~al.}(2023)\citenamefont
  {Fukushima}, \citenamefont {Takemori}, \citenamefont {Sakai}, \citenamefont
  {Ichioka},\ and\ \citenamefont {Jagannathan}}]{fukushima.23}%
  \BibitemOpen
  \bibfield  {author} {\bibinfo {author} {\bibfnamefont {T.}~\bibnamefont
  {Fukushima}}, \bibinfo {author} {\bibfnamefont {N.}~\bibnamefont {Takemori}},
  \bibinfo {author} {\bibfnamefont {S.}~\bibnamefont {Sakai}}, \bibinfo
  {author} {\bibfnamefont {M.}~\bibnamefont {Ichioka}},\ and\ \bibinfo {author}
  {\bibfnamefont {A.}~\bibnamefont {Jagannathan}},\ }\bibfield  {title}
  {\bibinfo {title} {Supercurrent distribution in real-space and anomalous
  paramagnetic response in a superconducting quasicrystal},\ }\href
  {https://doi.org/10.1103/PhysRevResearch.5.043164} {\bibfield  {journal}
  {\bibinfo  {journal} {Phys. Rev. Res.}\ }\textbf {\bibinfo {volume} {5}},\
  \bibinfo {pages} {043164} (\bibinfo {year} {2023})}\BibitemShut {NoStop}%
\bibitem [{\citenamefont {Jiang}\ \emph {et~al.}(2023)\citenamefont {Jiang},
  \citenamefont {Zaccone}, \citenamefont {Setty},\ and\ \citenamefont
  {Baggioli}}]{jiang.23}%
  \BibitemOpen
  \bibfield  {author} {\bibinfo {author} {\bibfnamefont {C.}~\bibnamefont
  {Jiang}}, \bibinfo {author} {\bibfnamefont {A.}~\bibnamefont {Zaccone}},
  \bibinfo {author} {\bibfnamefont {C.}~\bibnamefont {Setty}},\ and\ \bibinfo
  {author} {\bibfnamefont {M.}~\bibnamefont {Baggioli}},\ }\bibfield  {title}
  {\bibinfo {title} {Glassy heat capacity from overdamped phasons and
  hypothetical phason-induced superconductivity in incommensurate structures},\
  }\href {https://doi.org/10.1103/PhysRevB.108.054203} {\bibfield  {journal}
  {\bibinfo  {journal} {Phys. Rev. B}\ }\textbf {\bibinfo {volume} {108}},\
  \bibinfo {pages} {054203} (\bibinfo {year} {2023})}\BibitemShut {NoStop}%
\bibitem [{\citenamefont {Fan}\ \emph {et~al.}(2021)\citenamefont {Fan},
  \citenamefont {Chern},\ and\ \citenamefont {Lin}}]{fan.21}%
  \BibitemOpen
  \bibfield  {author} {\bibinfo {author} {\bibfnamefont {Z.}~\bibnamefont
  {Fan}}, \bibinfo {author} {\bibfnamefont {G.-W.}\ \bibnamefont {Chern}},\
  and\ \bibinfo {author} {\bibfnamefont {S.-Z.}\ \bibnamefont {Lin}},\
  }\bibfield  {title} {\bibinfo {title} {Enhanced superconductivity in
  quasiperiodic crystals},\ }\href
  {https://doi.org/10.1103/PhysRevResearch.3.023195} {\bibfield  {journal}
  {\bibinfo  {journal} {Phys. Rev. Research}\ }\textbf {\bibinfo {volume}
  {3}},\ \bibinfo {pages} {023195} (\bibinfo {year} {2021})}\BibitemShut
  {NoStop}%
\bibitem [{\citenamefont {Zhang}\ and\ \citenamefont
  {Foster}(2022)}]{zhang.22}%
  \BibitemOpen
  \bibfield  {author} {\bibinfo {author} {\bibfnamefont {X.}~\bibnamefont
  {Zhang}}\ and\ \bibinfo {author} {\bibfnamefont {M.~S.}\ \bibnamefont
  {Foster}},\ }\bibfield  {title} {\bibinfo {title} {Enhanced amplitude for
  superconductivity due to spectrum-wide wave function criticality in
  quasiperiodic and power-law random hopping models},\ }\href
  {https://doi.org/10.1103/PhysRevB.106.L180503} {\bibfield  {journal}
  {\bibinfo  {journal} {Phys. Rev. B}\ }\textbf {\bibinfo {volume} {106}},\
  \bibinfo {pages} {L180503} (\bibinfo {year} {2022})}\BibitemShut {NoStop}%
\bibitem [{\citenamefont {Oliveira}\ \emph {et~al.}(2023)\citenamefont
  {Oliveira}, \citenamefont {Gonçalves}, \citenamefont {Ribeiro},
  \citenamefont {Castro},\ and\ \citenamefont {Amorim}}]{oliveira.23}%
  \BibitemOpen
  \bibfield  {author} {\bibinfo {author} {\bibfnamefont {R.}~\bibnamefont
  {Oliveira}}, \bibinfo {author} {\bibfnamefont {M.}~\bibnamefont
  {Gonçalves}}, \bibinfo {author} {\bibfnamefont {P.}~\bibnamefont {Ribeiro}},
  \bibinfo {author} {\bibfnamefont {E.~V.}\ \bibnamefont {Castro}},\ and\
  \bibinfo {author} {\bibfnamefont {B.}~\bibnamefont {Amorim}},\ }\href@noop {}
  {\bibinfo {title} {Incommensurability-induced enhancement of
  superconductivity in one dimensional critical systems}} (\bibinfo {year}
  {2023}),\ \Eprint {https://arxiv.org/abs/2303.17656} {arXiv:2303.17656
  [cond-mat.supr-con]} \BibitemShut {NoStop}%
\bibitem [{\citenamefont {Sun}\ \emph {et~al.}(2024)\citenamefont {Sun},
  \citenamefont {\ifmmode \check{C}\else \v{C}\fi{}ade\ifmmode~\check{z}\else
  \v{z}\fi{}}, \citenamefont {Yurkevich},\ and\ \citenamefont
  {Andreanov}}]{sun.24}%
  \BibitemOpen
  \bibfield  {author} {\bibinfo {author} {\bibfnamefont {M.}~\bibnamefont
  {Sun}}, \bibinfo {author} {\bibfnamefont {T.}~\bibnamefont {\ifmmode
  \check{C}\else \v{C}\fi{}ade\ifmmode~\check{z}\else \v{z}\fi{}}}, \bibinfo
  {author} {\bibfnamefont {I.}~\bibnamefont {Yurkevich}},\ and\ \bibinfo
  {author} {\bibfnamefont {A.}~\bibnamefont {Andreanov}},\ }\bibfield  {title}
  {\bibinfo {title} {Enhancement of superconductivity in the fibonacci chain},\
  }\href {https://doi.org/10.1103/PhysRevB.109.134504} {\bibfield  {journal}
  {\bibinfo  {journal} {Phys. Rev. B}\ }\textbf {\bibinfo {volume} {109}},\
  \bibinfo {pages} {134504} (\bibinfo {year} {2024})}\BibitemShut {NoStop}%
\bibitem [{\citenamefont {Wang}\ \emph {et~al.}(2024)\citenamefont {Wang},
  \citenamefont {Rai}, \citenamefont {Matsumura}, \citenamefont {Jagannathan},\
  and\ \citenamefont {Haas}}]{wang.24}%
  \BibitemOpen
  \bibfield  {author} {\bibinfo {author} {\bibfnamefont {Y.}~\bibnamefont
  {Wang}}, \bibinfo {author} {\bibfnamefont {G.}~\bibnamefont {Rai}}, \bibinfo
  {author} {\bibfnamefont {C.}~\bibnamefont {Matsumura}}, \bibinfo {author}
  {\bibfnamefont {A.}~\bibnamefont {Jagannathan}},\ and\ \bibinfo {author}
  {\bibfnamefont {S.}~\bibnamefont {Haas}},\ }\bibfield  {title} {\bibinfo
  {title} {{Superconductivity in the Fibonacci chain}},\ }\href
  {https://doi.org/10.1103/PhysRevB.109.214507} {\bibfield  {journal} {\bibinfo
   {journal} {Phys. Rev. B}\ }\textbf {\bibinfo {volume} {109}},\ \bibinfo
  {pages} {214507} (\bibinfo {year} {2024})}\BibitemShut {NoStop}%
\bibitem [{\citenamefont {Rai}\ \emph {et~al.}(2019)\citenamefont {Rai},
  \citenamefont {Haas},\ and\ \citenamefont {Jagannathan}}]{rai.19}%
  \BibitemOpen
  \bibfield  {author} {\bibinfo {author} {\bibfnamefont {G.}~\bibnamefont
  {Rai}}, \bibinfo {author} {\bibfnamefont {S.}~\bibnamefont {Haas}},\ and\
  \bibinfo {author} {\bibfnamefont {A.}~\bibnamefont {Jagannathan}},\
  }\bibfield  {title} {\bibinfo {title} {Proximity effect in a
  superconductor-quasicrystal hybrid ring},\ }\href
  {https://doi.org/10.1103/PhysRevB.100.165121} {\bibfield  {journal} {\bibinfo
   {journal} {Phys. Rev. B}\ }\textbf {\bibinfo {volume} {100}},\ \bibinfo
  {pages} {165121} (\bibinfo {year} {2019})}\BibitemShut {NoStop}%
\bibitem [{\citenamefont {Rai}\ \emph {et~al.}(2020)\citenamefont {Rai},
  \citenamefont {Haas},\ and\ \citenamefont {Jagannathan}}]{rai.20}%
  \BibitemOpen
  \bibfield  {author} {\bibinfo {author} {\bibfnamefont {G.}~\bibnamefont
  {Rai}}, \bibinfo {author} {\bibfnamefont {S.}~\bibnamefont {Haas}},\ and\
  \bibinfo {author} {\bibfnamefont {A.}~\bibnamefont {Jagannathan}},\
  }\bibfield  {title} {\bibinfo {title} {Superconducting proximity effect and
  order parameter fluctuations in disordered and quasiperiodic systems},\
  }\href {https://doi.org/10.1103/PhysRevB.102.134211} {\bibfield  {journal}
  {\bibinfo  {journal} {Phys. Rev. B}\ }\textbf {\bibinfo {volume} {102}},\
  \bibinfo {pages} {134211} (\bibinfo {year} {2020})}\BibitemShut {NoStop}%
\bibitem [{\citenamefont {Kobiałka}\ \emph {et~al.}(2024)\citenamefont
  {Kobiałka}, \citenamefont {Awoga}, \citenamefont {Leijnse}, \citenamefont
  {Domański}, \citenamefont {Holmvall},\ and\ \citenamefont
  {Black-Schaffer}}]{kobialka.24}%
  \BibitemOpen
  \bibfield  {author} {\bibinfo {author} {\bibfnamefont {A.}~\bibnamefont
  {Kobiałka}}, \bibinfo {author} {\bibfnamefont {O.~A.}\ \bibnamefont
  {Awoga}}, \bibinfo {author} {\bibfnamefont {M.}~\bibnamefont {Leijnse}},
  \bibinfo {author} {\bibfnamefont {T.}~\bibnamefont {Domański}}, \bibinfo
  {author} {\bibfnamefont {P.}~\bibnamefont {Holmvall}},\ and\ \bibinfo
  {author} {\bibfnamefont {A.~M.}\ \bibnamefont {Black-Schaffer}},\ }\href
  {https://arxiv.org/abs/2405.12178} {\bibinfo {title} {{Topological
  superconductivity in Fibonacci quasicrystals}}} (\bibinfo {year} {2024}),\
  \Eprint {https://arxiv.org/abs/2405.12178} {arXiv:2405.12178
  [cond-mat.mes-hall]} \BibitemShut {NoStop}%
\bibitem [{\citenamefont {Qiu}\ and\ \citenamefont {Hsueh}(2014)}]{qiu.14}%
  \BibitemOpen
  \bibfield  {author} {\bibinfo {author} {\bibfnamefont {R.}~\bibnamefont
  {Qiu}}\ and\ \bibinfo {author} {\bibfnamefont {W.}~\bibnamefont {Hsueh}},\
  }\bibfield  {title} {\bibinfo {title} {Giant persistent currents in
  quasiperiodic mesoscopic rings},\ }\href
  {https://doi.org/https://doi.org/10.1016/j.physleta.2014.01.032} {\bibfield
  {journal} {\bibinfo  {journal} {Phys. Lett. A}\ }\textbf {\bibinfo {volume}
  {378}},\ \bibinfo {pages} {851} (\bibinfo {year} {2014})}\BibitemShut
  {NoStop}%
\bibitem [{\citenamefont {Roy}\ \emph {et~al.}(2023)\citenamefont {Roy},
  \citenamefont {Ganguly},\ and\ \citenamefont {Maiti}}]{roy.23}%
  \BibitemOpen
  \bibfield  {author} {\bibinfo {author} {\bibfnamefont {S.}~\bibnamefont
  {Roy}}, \bibinfo {author} {\bibfnamefont {S.}~\bibnamefont {Ganguly}},\ and\
  \bibinfo {author} {\bibfnamefont {S.~K.}\ \bibnamefont {Maiti}},\ }\bibfield
  {title} {\bibinfo {title} {Interplay between hopping dimerization and
  quasi-periodicity on flux-driven circular current in an incommensurate
  su-schrieffer-heeger ring},\ }\href
  {https://doi.org/10.1038/s41598-023-31354-9} {\bibfield  {journal} {\bibinfo
  {journal} {Sci. Rep.}\ }\textbf {\bibinfo {volume} {13}},\ \bibinfo {pages}
  {4093} (\bibinfo {year} {2023})}\BibitemShut {NoStop}%
\bibitem [{\citenamefont {Mayou}\ \emph {et~al.}(1993)\citenamefont {Mayou},
  \citenamefont {Berger}, \citenamefont {Cyrot-Lackmann}, \citenamefont
  {Klein},\ and\ \citenamefont {Lanco}}]{mayou.93}%
  \BibitemOpen
  \bibfield  {author} {\bibinfo {author} {\bibfnamefont {D.}~\bibnamefont
  {Mayou}}, \bibinfo {author} {\bibfnamefont {C.}~\bibnamefont {Berger}},
  \bibinfo {author} {\bibfnamefont {F.}~\bibnamefont {Cyrot-Lackmann}},
  \bibinfo {author} {\bibfnamefont {T.}~\bibnamefont {Klein}},\ and\ \bibinfo
  {author} {\bibfnamefont {P.}~\bibnamefont {Lanco}},\ }\bibfield  {title}
  {\bibinfo {title} {Evidence for unconventional electronic transport in
  quasicrystals},\ }\href {https://doi.org/10.1103/PhysRevLett.70.3915}
  {\bibfield  {journal} {\bibinfo  {journal} {Phys. Rev. Lett.}\ }\textbf
  {\bibinfo {volume} {70}},\ \bibinfo {pages} {3915} (\bibinfo {year}
  {1993})}\BibitemShut {NoStop}%
\bibitem [{\citenamefont {Pierce}\ \emph {et~al.}(1994)\citenamefont {Pierce},
  \citenamefont {Guo},\ and\ \citenamefont {Poon}}]{pierce.guo.poon.94}%
  \BibitemOpen
  \bibfield  {author} {\bibinfo {author} {\bibfnamefont {F.~S.}\ \bibnamefont
  {Pierce}}, \bibinfo {author} {\bibfnamefont {Q.}~\bibnamefont {Guo}},\ and\
  \bibinfo {author} {\bibfnamefont {S.~J.}\ \bibnamefont {Poon}},\ }\bibfield
  {title} {\bibinfo {title} {{Enhanced Insulatorlike Electron Transport
  Behavior of Thermally Tuned Quasicrystalline States of Al-Pd-Re Alloys}},\
  }\href {https://doi.org/10.1103/PhysRevLett.73.2220} {\bibfield  {journal}
  {\bibinfo  {journal} {Phys. Rev. Lett.}\ }\textbf {\bibinfo {volume} {73}},\
  \bibinfo {pages} {2220} (\bibinfo {year} {1994})}\BibitemShut {NoStop}%
\bibitem [{\citenamefont {Roche}\ \emph {et~al.}(1997)\citenamefont {Roche},
  \citenamefont {Trambly~de Laissardière},\ and\ \citenamefont
  {Mayou}}]{roche.trambly.mayou.97}%
  \BibitemOpen
  \bibfield  {author} {\bibinfo {author} {\bibfnamefont {S.}~\bibnamefont
  {Roche}}, \bibinfo {author} {\bibfnamefont {G.}~\bibnamefont {Trambly~de
  Laissardière}},\ and\ \bibinfo {author} {\bibfnamefont {D.}~\bibnamefont
  {Mayou}},\ }\bibfield  {title} {\bibinfo {title} {{Electronic transport
  properties of quasicrystals}},\ }\href {https://doi.org/10.1063/1.531914}
  {\bibfield  {journal} {\bibinfo  {journal} {J. Math. Phys.}\ }\textbf
  {\bibinfo {volume} {38}},\ \bibinfo {pages} {1794} (\bibinfo {year}
  {1997})}\BibitemShut {NoStop}%
\bibitem [{\citenamefont {Jeon}\ and\ \citenamefont {Lee}(2021)}]{jeon.21}%
  \BibitemOpen
  \bibfield  {author} {\bibinfo {author} {\bibfnamefont {J.}~\bibnamefont
  {Jeon}}\ and\ \bibinfo {author} {\bibfnamefont {S.}~\bibnamefont {Lee}},\
  }\bibfield  {title} {\bibinfo {title} {Topological critical states and
  anomalous electronic transmittance in one-dimensional quasicrystals},\ }\href
  {https://doi.org/10.1103/PhysRevResearch.3.013168} {\bibfield  {journal}
  {\bibinfo  {journal} {Phys. Rev. Research}\ }\textbf {\bibinfo {volume}
  {3}},\ \bibinfo {pages} {013168} (\bibinfo {year} {2021})}\BibitemShut
  {NoStop}%
\bibitem [{\citenamefont {Jeon}\ \emph {et~al.}(2022)\citenamefont {Jeon},
  \citenamefont {Kim},\ and\ \citenamefont {Lee}}]{jeon.kim.lee.22}%
  \BibitemOpen
  \bibfield  {author} {\bibinfo {author} {\bibfnamefont {J.}~\bibnamefont
  {Jeon}}, \bibinfo {author} {\bibfnamefont {S.~K.}\ \bibnamefont {Kim}},\ and\
  \bibinfo {author} {\bibfnamefont {S.}~\bibnamefont {Lee}},\ }\bibfield
  {title} {\bibinfo {title} {Fractalized magnon transport on a quasicrystal
  with enhanced stability},\ }\href
  {https://doi.org/10.1103/PhysRevB.106.134431} {\bibfield  {journal} {\bibinfo
   {journal} {Phys. Rev. B}\ }\textbf {\bibinfo {volume} {106}},\ \bibinfo
  {pages} {134431} (\bibinfo {year} {2022})}\BibitemShut {NoStop}%
\bibitem [{\citenamefont {Wang}\ and\ \citenamefont
  {Zhao}(2023)}]{wang.zhao.23}%
  \BibitemOpen
  \bibfield  {author} {\bibinfo {author} {\bibfnamefont {B.~X.}\ \bibnamefont
  {Wang}}\ and\ \bibinfo {author} {\bibfnamefont {C.~Y.}\ \bibnamefont
  {Zhao}},\ }\bibfield  {title} {\bibinfo {title} {Topological phonon polariton
  enhanced radiative heat transfer in bichromatic nanoparticle arrays mimicking
  aubry-andr\'e-harper model},\ }\href
  {https://doi.org/10.1103/PhysRevB.107.125409} {\bibfield  {journal} {\bibinfo
   {journal} {Phys. Rev. B}\ }\textbf {\bibinfo {volume} {107}},\ \bibinfo
  {pages} {125409} (\bibinfo {year} {2023})}\BibitemShut {NoStop}%
\bibitem [{\citenamefont {Arrigoni}\ \emph {et~al.}(2004)\citenamefont
  {Arrigoni}, \citenamefont {Fradkin},\ and\ \citenamefont
  {Kivelson}}]{arrigoni.fradkin.kivelson.04}%
  \BibitemOpen
  \bibfield  {author} {\bibinfo {author} {\bibfnamefont {E.}~\bibnamefont
  {Arrigoni}}, \bibinfo {author} {\bibfnamefont {E.}~\bibnamefont {Fradkin}},\
  and\ \bibinfo {author} {\bibfnamefont {S.~A.}\ \bibnamefont {Kivelson}},\
  }\bibfield  {title} {\bibinfo {title} {Mechanism of high-temperature
  superconductivity in a striped hubbard model},\ }\href
  {https://doi.org/10.1103/PhysRevB.69.214519} {\bibfield  {journal} {\bibinfo
  {journal} {Phys. Rev. B}\ }\textbf {\bibinfo {volume} {69}},\ \bibinfo
  {pages} {214519} (\bibinfo {year} {2004})}\BibitemShut {NoStop}%
\bibitem [{\citenamefont {Martin}\ \emph {et~al.}(2005)\citenamefont {Martin},
  \citenamefont {Podolsky},\ and\ \citenamefont
  {Kivelson}}]{martin.podolsky.kivelson.05}%
  \BibitemOpen
  \bibfield  {author} {\bibinfo {author} {\bibfnamefont {I.}~\bibnamefont
  {Martin}}, \bibinfo {author} {\bibfnamefont {D.}~\bibnamefont {Podolsky}},\
  and\ \bibinfo {author} {\bibfnamefont {S.~A.}\ \bibnamefont {Kivelson}},\
  }\bibfield  {title} {\bibinfo {title} {Enhancement of superconductivity by
  local inhomogeneities},\ }\href {https://doi.org/10.1103/PhysRevB.72.060502}
  {\bibfield  {journal} {\bibinfo  {journal} {Phys. Rev. B}\ }\textbf {\bibinfo
  {volume} {72}},\ \bibinfo {pages} {060502(R)} (\bibinfo {year}
  {2005})}\BibitemShut {NoStop}%
\bibitem [{\citenamefont {Feigel'man}\ \emph {et~al.}(2007)\citenamefont
  {Feigel'man}, \citenamefont {Ioffe}, \citenamefont {Kravtsov},\ and\
  \citenamefont {Yuzbashyan}}]{feigelman.07}%
  \BibitemOpen
  \bibfield  {author} {\bibinfo {author} {\bibfnamefont {M.~V.}\ \bibnamefont
  {Feigel'man}}, \bibinfo {author} {\bibfnamefont {L.~B.}\ \bibnamefont
  {Ioffe}}, \bibinfo {author} {\bibfnamefont {V.~E.}\ \bibnamefont
  {Kravtsov}},\ and\ \bibinfo {author} {\bibfnamefont {E.~A.}\ \bibnamefont
  {Yuzbashyan}},\ }\bibfield  {title} {\bibinfo {title} {Eigenfunction
  fractality and pseudogap state near the superconductor-insulator
  transition},\ }\href {https://doi.org/10.1103/PhysRevLett.98.027001}
  {\bibfield  {journal} {\bibinfo  {journal} {Phys. Rev. Lett.}\ }\textbf
  {\bibinfo {volume} {98}},\ \bibinfo {pages} {027001} (\bibinfo {year}
  {2007})}\BibitemShut {NoStop}%
\bibitem [{\citenamefont {Feigel'man}\ \emph {et~al.}(2010)\citenamefont
  {Feigel'man}, \citenamefont {Ioffe}, \citenamefont {Kravtsov},\ and\
  \citenamefont {Cuevas}}]{feigelman.10}%
  \BibitemOpen
  \bibfield  {author} {\bibinfo {author} {\bibfnamefont {M.}~\bibnamefont
  {Feigel'man}}, \bibinfo {author} {\bibfnamefont {L.}~\bibnamefont {Ioffe}},
  \bibinfo {author} {\bibfnamefont {V.}~\bibnamefont {Kravtsov}},\ and\
  \bibinfo {author} {\bibfnamefont {E.}~\bibnamefont {Cuevas}},\ }\bibfield
  {title} {\bibinfo {title} {Fractal superconductivity near localization
  threshold},\ }\href
  {https://doi.org/https://doi.org/10.1016/j.aop.2010.04.001} {\bibfield
  {journal} {\bibinfo  {journal} {Ann. Phys. (NY)}\ }\textbf {\bibinfo {volume}
  {325}},\ \bibinfo {pages} {1390} (\bibinfo {year} {2010})},\ \bibinfo {note}
  {july 2010 Special Issue}\BibitemShut {NoStop}%
\bibitem [{\citenamefont {Burmistrov}\ \emph {et~al.}(2012)\citenamefont
  {Burmistrov}, \citenamefont {Gornyi},\ and\ \citenamefont
  {Mirlin}}]{burmistrov.12}%
  \BibitemOpen
  \bibfield  {author} {\bibinfo {author} {\bibfnamefont {I.~S.}\ \bibnamefont
  {Burmistrov}}, \bibinfo {author} {\bibfnamefont {I.~V.}\ \bibnamefont
  {Gornyi}},\ and\ \bibinfo {author} {\bibfnamefont {A.~D.}\ \bibnamefont
  {Mirlin}},\ }\bibfield  {title} {\bibinfo {title} {Enhancement of the
  critical temperature of superconductors by anderson localization},\ }\href
  {https://doi.org/10.1103/PhysRevLett.108.017002} {\bibfield  {journal}
  {\bibinfo  {journal} {Phys. Rev. Lett.}\ }\textbf {\bibinfo {volume} {108}},\
  \bibinfo {pages} {017002} (\bibinfo {year} {2012})}\BibitemShut {NoStop}%
\bibitem [{\citenamefont {Mayoh}\ and\ \citenamefont
  {Garc\'{\i}a-Garc\'{\i}a}(2015)}]{mayoh.15}%
  \BibitemOpen
  \bibfield  {author} {\bibinfo {author} {\bibfnamefont {J.}~\bibnamefont
  {Mayoh}}\ and\ \bibinfo {author} {\bibfnamefont {A.~M.}\ \bibnamefont
  {Garc\'{\i}a-Garc\'{\i}a}},\ }\bibfield  {title} {\bibinfo {title} {Global
  critical temperature in disordered superconductors with weak
  multifractality},\ }\href {https://doi.org/10.1103/PhysRevB.92.174526}
  {\bibfield  {journal} {\bibinfo  {journal} {Phys. Rev. B}\ }\textbf {\bibinfo
  {volume} {92}},\ \bibinfo {pages} {174526} (\bibinfo {year}
  {2015})}\BibitemShut {NoStop}%
\bibitem [{\citenamefont {Zhao}\ \emph {et~al.}(2019)\citenamefont {Zhao},
  \citenamefont {Lin}, \citenamefont {Xiao}, \citenamefont {Huang},
  \citenamefont {Yao}, \citenamefont {Yan}, \citenamefont {Xing}, \citenamefont
  {Zhang}, \citenamefont {Li}, \citenamefont {Hoshino}, \citenamefont {Wang},
  \citenamefont {Zhou}, \citenamefont {Gu}, \citenamefont {Bahramy},
  \citenamefont {Yao}, \citenamefont {Nagaosa}, \citenamefont {Xue},
  \citenamefont {Law}, \citenamefont {Chen},\ and\ \citenamefont
  {Ji}}]{zhao.19}%
  \BibitemOpen
  \bibfield  {author} {\bibinfo {author} {\bibfnamefont {K.}~\bibnamefont
  {Zhao}}, \bibinfo {author} {\bibfnamefont {H.}~\bibnamefont {Lin}}, \bibinfo
  {author} {\bibfnamefont {X.}~\bibnamefont {Xiao}}, \bibinfo {author}
  {\bibfnamefont {W.}~\bibnamefont {Huang}}, \bibinfo {author} {\bibfnamefont
  {W.}~\bibnamefont {Yao}}, \bibinfo {author} {\bibfnamefont {M.}~\bibnamefont
  {Yan}}, \bibinfo {author} {\bibfnamefont {Y.}~\bibnamefont {Xing}}, \bibinfo
  {author} {\bibfnamefont {Q.}~\bibnamefont {Zhang}}, \bibinfo {author}
  {\bibfnamefont {Z.-X.}\ \bibnamefont {Li}}, \bibinfo {author} {\bibfnamefont
  {S.}~\bibnamefont {Hoshino}}, \bibinfo {author} {\bibfnamefont
  {J.}~\bibnamefont {Wang}}, \bibinfo {author} {\bibfnamefont {S.}~\bibnamefont
  {Zhou}}, \bibinfo {author} {\bibfnamefont {L.}~\bibnamefont {Gu}}, \bibinfo
  {author} {\bibfnamefont {M.~S.}\ \bibnamefont {Bahramy}}, \bibinfo {author}
  {\bibfnamefont {H.}~\bibnamefont {Yao}}, \bibinfo {author} {\bibfnamefont
  {N.}~\bibnamefont {Nagaosa}}, \bibinfo {author} {\bibfnamefont {Q.-K.}\
  \bibnamefont {Xue}}, \bibinfo {author} {\bibfnamefont {K.~T.}\ \bibnamefont
  {Law}}, \bibinfo {author} {\bibfnamefont {X.}~\bibnamefont {Chen}},\ and\
  \bibinfo {author} {\bibfnamefont {S.-H.}\ \bibnamefont {Ji}},\ }\bibfield
  {title} {\bibinfo {title} {Disorder-induced multifractal superconductivity in
  monolayer niobium dichalcogenides},\ }\href
  {https://doi.org/10.1038/s41567-019-0570-0} {\bibfield  {journal} {\bibinfo
  {journal} {Nat. Phys.}\ }\textbf {\bibinfo {volume} {15}},\ \bibinfo {pages}
  {904} (\bibinfo {year} {2019})}\BibitemShut {NoStop}%
\bibitem [{\citenamefont {Burmistrov}\ \emph {et~al.}(2021)\citenamefont
  {Burmistrov}, \citenamefont {Gornyi},\ and\ \citenamefont
  {Mirlin}}]{burmistrov.21}%
  \BibitemOpen
  \bibfield  {author} {\bibinfo {author} {\bibfnamefont {I.}~\bibnamefont
  {Burmistrov}}, \bibinfo {author} {\bibfnamefont {I.}~\bibnamefont {Gornyi}},\
  and\ \bibinfo {author} {\bibfnamefont {A.}~\bibnamefont {Mirlin}},\
  }\bibfield  {title} {\bibinfo {title} {Multifractally-enhanced
  superconductivity in thin films},\ }\href
  {https://doi.org/https://doi.org/10.1016/j.aop.2021.168499} {\bibfield
  {journal} {\bibinfo  {journal} {Ann. Phys. (NY)}\ }\textbf {\bibinfo {volume}
  {435}},\ \bibinfo {pages} {168499} (\bibinfo {year} {2021})},\ \bibinfo
  {note} {special Issue on Localisation 2020}\BibitemShut {NoStop}%
\bibitem [{\citenamefont {Schreiber}\ and\ \citenamefont
  {Grussbach}(1991)}]{schreiber.91}%
  \BibitemOpen
  \bibfield  {author} {\bibinfo {author} {\bibfnamefont {M.}~\bibnamefont
  {Schreiber}}\ and\ \bibinfo {author} {\bibfnamefont {H.}~\bibnamefont
  {Grussbach}},\ }\bibfield  {title} {\bibinfo {title} {Multifractal wave
  functions at the anderson transition},\ }\href
  {https://doi.org/10.1103/PhysRevLett.67.607} {\bibfield  {journal} {\bibinfo
  {journal} {Phys. Rev. Lett.}\ }\textbf {\bibinfo {volume} {67}},\ \bibinfo
  {pages} {607} (\bibinfo {year} {1991})}\BibitemShut {NoStop}%
\bibitem [{\citenamefont {Evers}\ and\ \citenamefont
  {Mirlin}(2008)}]{evers.mirlin.08}%
  \BibitemOpen
  \bibfield  {author} {\bibinfo {author} {\bibfnamefont {F.}~\bibnamefont
  {Evers}}\ and\ \bibinfo {author} {\bibfnamefont {A.~D.}\ \bibnamefont
  {Mirlin}},\ }\bibfield  {title} {\bibinfo {title} {Anderson transitions},\
  }\href {https://doi.org/10.1103/RevModPhys.80.1355} {\bibfield  {journal}
  {\bibinfo  {journal} {Rev. Mod. Phys.}\ }\textbf {\bibinfo {volume} {80}},\
  \bibinfo {pages} {1355} (\bibinfo {year} {2008})}\BibitemShut {NoStop}%
\bibitem [{\citenamefont {Jagannathan}\ and\ \citenamefont
  {Tarzia}(2020)}]{jagannathan.tarzia.20}%
  \BibitemOpen
  \bibfield  {author} {\bibinfo {author} {\bibfnamefont {A.}~\bibnamefont
  {Jagannathan}}\ and\ \bibinfo {author} {\bibfnamefont {M.}~\bibnamefont
  {Tarzia}},\ }\bibfield  {title} {\bibinfo {title} {Re-entrance and
  localization phenomena in disordered fibonacci chains},\ }\href
  {https://doi.org/10.1140/epjb/e2020-100504-7} {\bibfield  {journal} {\bibinfo
   {journal} {Eur. Phys. J. B}\ }\textbf {\bibinfo {volume} {93}},\ \bibinfo
  {pages} {46} (\bibinfo {year} {2020})}\BibitemShut {NoStop}%
\bibitem [{\citenamefont {Nagao}\ \emph {et~al.}(2015)\citenamefont {Nagao},
  \citenamefont {Inuzuka}, \citenamefont {Nishimoto},\ and\ \citenamefont
  {Edagawa}}]{nagao.15}%
  \BibitemOpen
  \bibfield  {author} {\bibinfo {author} {\bibfnamefont {K.}~\bibnamefont
  {Nagao}}, \bibinfo {author} {\bibfnamefont {T.}~\bibnamefont {Inuzuka}},
  \bibinfo {author} {\bibfnamefont {K.}~\bibnamefont {Nishimoto}},\ and\
  \bibinfo {author} {\bibfnamefont {K.}~\bibnamefont {Edagawa}},\ }\bibfield
  {title} {\bibinfo {title} {Experimental observation of quasicrystal growth},\
  }\href {https://doi.org/10.1103/PhysRevLett.115.075501} {\bibfield  {journal}
  {\bibinfo  {journal} {Phys. Rev. Lett.}\ }\textbf {\bibinfo {volume} {115}},\
  \bibinfo {pages} {075501} (\bibinfo {year} {2015})}\BibitemShut {NoStop}%
\bibitem [{\citenamefont {Wolf}\ \emph {et~al.}(2021)\citenamefont {Wolf},
  \citenamefont {Bolfarini}, \citenamefont {Kiminami},\ and\ \citenamefont
  {Botta}}]{wolf.21}%
  \BibitemOpen
  \bibfield  {author} {\bibinfo {author} {\bibfnamefont {W.}~\bibnamefont
  {Wolf}}, \bibinfo {author} {\bibfnamefont {C.}~\bibnamefont {Bolfarini}},
  \bibinfo {author} {\bibfnamefont {C.~S.}\ \bibnamefont {Kiminami}},\ and\
  \bibinfo {author} {\bibfnamefont {W.~J.}\ \bibnamefont {Botta}},\ }\bibfield
  {title} {\bibinfo {title} {{Recent developments on fabrication of Al‐matrix
  composites reinforced with quasicrystals: From metastable to conventional
  processing}},\ }\href {https://doi.org/10.1557/s43578-020-00083-4} {\bibfield
   {journal} {\bibinfo  {journal} {J. Mater. Res.}\ }\textbf {\bibinfo {volume}
  {36}},\ \bibinfo {pages} {281} (\bibinfo {year} {2021})}\BibitemShut
  {NoStop}%
\bibitem [{\citenamefont {Eigler}\ and\ \citenamefont
  {Schweizer}(1990)}]{eigler.90}%
  \BibitemOpen
  \bibfield  {author} {\bibinfo {author} {\bibfnamefont {D.~M.}\ \bibnamefont
  {Eigler}}\ and\ \bibinfo {author} {\bibfnamefont {E.~K.}\ \bibnamefont
  {Schweizer}},\ }\bibfield  {title} {\bibinfo {title} {Positioning single
  atoms with a scanning tunnelling microscope},\ }\href
  {https://doi.org/10.1038/344524a0} {\bibfield  {journal} {\bibinfo  {journal}
  {Nature}\ }\textbf {\bibinfo {volume} {344}},\ \bibinfo {pages} {524}
  (\bibinfo {year} {1990})}\BibitemShut {NoStop}%
\bibitem [{\citenamefont {Stroscio}\ and\ \citenamefont
  {Eigler}(1991)}]{stroscio.91}%
  \BibitemOpen
  \bibfield  {author} {\bibinfo {author} {\bibfnamefont {J.~A.}\ \bibnamefont
  {Stroscio}}\ and\ \bibinfo {author} {\bibfnamefont {D.~M.}\ \bibnamefont
  {Eigler}},\ }\bibfield  {title} {\bibinfo {title} {Atomic and molecular
  manipulation with the scanning tunneling microscope},\ }\href
  {https://doi.org/10.1126/science.254.5036.1319} {\bibfield  {journal}
  {\bibinfo  {journal} {Science}\ }\textbf {\bibinfo {volume} {254}},\ \bibinfo
  {pages} {1319} (\bibinfo {year} {1991})}\BibitemShut {NoStop}%
\bibitem [{\citenamefont {Gomes}\ \emph {et~al.}(2012)\citenamefont {Gomes},
  \citenamefont {Mar}, \citenamefont {Ko}, \citenamefont {Guinea},\ and\
  \citenamefont {Manoharan}}]{gomes.12}%
  \BibitemOpen
  \bibfield  {author} {\bibinfo {author} {\bibfnamefont {K.~K.}\ \bibnamefont
  {Gomes}}, \bibinfo {author} {\bibfnamefont {W.}~\bibnamefont {Mar}}, \bibinfo
  {author} {\bibfnamefont {W.}~\bibnamefont {Ko}}, \bibinfo {author}
  {\bibfnamefont {F.}~\bibnamefont {Guinea}},\ and\ \bibinfo {author}
  {\bibfnamefont {H.~C.}\ \bibnamefont {Manoharan}},\ }\bibfield  {title}
  {\bibinfo {title} {Designer dirac fermions and topological phases in
  molecular graphene},\ }\href {https://doi.org/10.1038/nature10941} {\bibfield
   {journal} {\bibinfo  {journal} {Nature}\ }\textbf {\bibinfo {volume}
  {483}},\ \bibinfo {pages} {306} (\bibinfo {year} {2012})}\BibitemShut
  {NoStop}%
\bibitem [{\citenamefont {Polini}\ \emph {et~al.}(2013)\citenamefont {Polini},
  \citenamefont {Guinea}, \citenamefont {Lewenstein}, \citenamefont
  {Manoharan},\ and\ \citenamefont {Pellegrini}}]{polini.13}%
  \BibitemOpen
  \bibfield  {author} {\bibinfo {author} {\bibfnamefont {M.}~\bibnamefont
  {Polini}}, \bibinfo {author} {\bibfnamefont {F.}~\bibnamefont {Guinea}},
  \bibinfo {author} {\bibfnamefont {M.}~\bibnamefont {Lewenstein}}, \bibinfo
  {author} {\bibfnamefont {H.~C.}\ \bibnamefont {Manoharan}},\ and\ \bibinfo
  {author} {\bibfnamefont {V.}~\bibnamefont {Pellegrini}},\ }\bibfield  {title}
  {\bibinfo {title} {Artificial honeycomb lattices for electrons, atoms and
  photons},\ }\href {https://doi.org/10.1038/nnano.2013.161} {\bibfield
  {journal} {\bibinfo  {journal} {Nat. Nanotech.}\ }\textbf {\bibinfo {volume}
  {8}},\ \bibinfo {pages} {625} (\bibinfo {year} {2013})}\BibitemShut {NoStop}%
\bibitem [{\citenamefont {Drost}\ \emph {et~al.}(2017)\citenamefont {Drost},
  \citenamefont {Ojanen}, \citenamefont {Harju},\ and\ \citenamefont
  {Liljeroth}}]{drost.17}%
  \BibitemOpen
  \bibfield  {author} {\bibinfo {author} {\bibfnamefont {R.}~\bibnamefont
  {Drost}}, \bibinfo {author} {\bibfnamefont {T.}~\bibnamefont {Ojanen}},
  \bibinfo {author} {\bibfnamefont {A.}~\bibnamefont {Harju}},\ and\ \bibinfo
  {author} {\bibfnamefont {P.}~\bibnamefont {Liljeroth}},\ }\bibfield  {title}
  {\bibinfo {title} {Topological states in engineered atomic lattices},\ }\href
  {https://doi.org/10.1038/nphys4080} {\bibfield  {journal} {\bibinfo
  {journal} {Nat. Phys.}\ }\textbf {\bibinfo {volume} {13}},\ \bibinfo {pages}
  {668} (\bibinfo {year} {2017})}\BibitemShut {NoStop}%
\bibitem [{\citenamefont {Khajetoorians}\ \emph {et~al.}(2019)\citenamefont
  {Khajetoorians}, \citenamefont {Wegner}, \citenamefont {Otte},\ and\
  \citenamefont {Swart}}]{khajetoorians.19}%
  \BibitemOpen
  \bibfield  {author} {\bibinfo {author} {\bibfnamefont {A.~A.}\ \bibnamefont
  {Khajetoorians}}, \bibinfo {author} {\bibfnamefont {D.}~\bibnamefont
  {Wegner}}, \bibinfo {author} {\bibfnamefont {A.~F.}\ \bibnamefont {Otte}},\
  and\ \bibinfo {author} {\bibfnamefont {I.}~\bibnamefont {Swart}},\ }\bibfield
   {title} {\bibinfo {title} {Creating designer quantum states of matter
  atom-by-atom},\ }\href {https://doi.org/10.1038/s42254-019-0108-5} {\bibfield
   {journal} {\bibinfo  {journal} {Nat. Rev. Phys.}\ }\textbf {\bibinfo
  {volume} {1}},\ \bibinfo {pages} {703} (\bibinfo {year} {2019})}\BibitemShut
  {NoStop}%
\bibitem [{\citenamefont {Schneider}\ \emph {et~al.}(2020)\citenamefont
  {Schneider}, \citenamefont {Brinker}, \citenamefont {Steinbrecher},
  \citenamefont {Hermenau}, \citenamefont {Posske}, \citenamefont {dos
  Santos~Dias}, \citenamefont {Lounis}, \citenamefont {Wiesendanger},\ and\
  \citenamefont {Wiebe}}]{schneider.20}%
  \BibitemOpen
  \bibfield  {author} {\bibinfo {author} {\bibfnamefont {L.}~\bibnamefont
  {Schneider}}, \bibinfo {author} {\bibfnamefont {S.}~\bibnamefont {Brinker}},
  \bibinfo {author} {\bibfnamefont {M.}~\bibnamefont {Steinbrecher}}, \bibinfo
  {author} {\bibfnamefont {J.}~\bibnamefont {Hermenau}}, \bibinfo {author}
  {\bibfnamefont {T.}~\bibnamefont {Posske}}, \bibinfo {author} {\bibfnamefont
  {M.}~\bibnamefont {dos Santos~Dias}}, \bibinfo {author} {\bibfnamefont
  {S.}~\bibnamefont {Lounis}}, \bibinfo {author} {\bibfnamefont
  {R.}~\bibnamefont {Wiesendanger}},\ and\ \bibinfo {author} {\bibfnamefont
  {J.}~\bibnamefont {Wiebe}},\ }\bibfield  {title} {\bibinfo {title}
  {Controlling in-gap end states by linking nonmagnetic atoms and
  artificially-constructed spin chains on superconductors},\ }\href
  {https://doi.org/10.1038/s41467-020-18540-3} {\bibfield  {journal} {\bibinfo
  {journal} {Nat. Commun.}\ }\textbf {\bibinfo {volume} {11}},\ \bibinfo
  {pages} {4707} (\bibinfo {year} {2020})}\BibitemShut {NoStop}%
\bibitem [{\citenamefont {Huda}\ \emph {et~al.}(2020)\citenamefont {Huda},
  \citenamefont {Kezilebieke}, \citenamefont {Ojanen}, \citenamefont {Drost},\
  and\ \citenamefont {Liljeroth}}]{huda.20}%
  \BibitemOpen
  \bibfield  {author} {\bibinfo {author} {\bibfnamefont {M.~N.}\ \bibnamefont
  {Huda}}, \bibinfo {author} {\bibfnamefont {S.}~\bibnamefont {Kezilebieke}},
  \bibinfo {author} {\bibfnamefont {T.}~\bibnamefont {Ojanen}}, \bibinfo
  {author} {\bibfnamefont {R.}~\bibnamefont {Drost}},\ and\ \bibinfo {author}
  {\bibfnamefont {P.}~\bibnamefont {Liljeroth}},\ }\bibfield  {title} {\bibinfo
  {title} {Tuneable topological domain wall states in engineered atomic
  chains},\ }\href {https://doi.org/10.1038/s41535-020-0219-3} {\bibfield
  {journal} {\bibinfo  {journal} {npj Quantum Mater.}\ }\textbf {\bibinfo
  {volume} {5}},\ \bibinfo {pages} {17} (\bibinfo {year} {2020})}\BibitemShut
  {NoStop}%
\bibitem [{\citenamefont {{K{\"u}ster}}\ \emph {et~al.}(2022)\citenamefont
  {{K{\"u}ster}}, \citenamefont {{Brinker}}, \citenamefont {{Hess}},
  \citenamefont {{Loss}}, \citenamefont {{Parkin}}, \citenamefont
  {{Klinovaja}}, \citenamefont {{Lounis}},\ and\ \citenamefont
  {{Sessi}}}]{kuster.22}%
  \BibitemOpen
  \bibfield  {author} {\bibinfo {author} {\bibfnamefont {F.}~\bibnamefont
  {{K{\"u}ster}}}, \bibinfo {author} {\bibfnamefont {S.}~\bibnamefont
  {{Brinker}}}, \bibinfo {author} {\bibfnamefont {R.}~\bibnamefont {{Hess}}},
  \bibinfo {author} {\bibfnamefont {D.}~\bibnamefont {{Loss}}}, \bibinfo
  {author} {\bibfnamefont {S.~S.~P.}\ \bibnamefont {{Parkin}}}, \bibinfo
  {author} {\bibfnamefont {J.}~\bibnamefont {{Klinovaja}}}, \bibinfo {author}
  {\bibfnamefont {S.}~\bibnamefont {{Lounis}}},\ and\ \bibinfo {author}
  {\bibfnamefont {P.}~\bibnamefont {{Sessi}}},\ }\bibfield  {title} {\bibinfo
  {title} {{Non-Majorana modes in diluted spin chains proximitized to a
  superconductor}},\ }\href {https://doi.org/10.1073/pnas.2210589119}
  {\bibfield  {journal} {\bibinfo  {journal} {Proc. Natl. Acad. Sci. U.S.A.}\
  }\textbf {\bibinfo {volume} {119}},\ \bibinfo {eid} {e2210589119} (\bibinfo
  {year} {2022})}\BibitemShut {NoStop}%
\bibitem [{\citenamefont {Freeney}\ \emph {et~al.}(2022)\citenamefont
  {Freeney}, \citenamefont {Slot}, \citenamefont {Gardenier}, \citenamefont
  {Swart},\ and\ \citenamefont {Vanmaekelbergh}}]{freeney.22}%
  \BibitemOpen
  \bibfield  {author} {\bibinfo {author} {\bibfnamefont {S.~E.}\ \bibnamefont
  {Freeney}}, \bibinfo {author} {\bibfnamefont {M.~R.}\ \bibnamefont {Slot}},
  \bibinfo {author} {\bibfnamefont {T.~S.}\ \bibnamefont {Gardenier}}, \bibinfo
  {author} {\bibfnamefont {I.}~\bibnamefont {Swart}},\ and\ \bibinfo {author}
  {\bibfnamefont {D.}~\bibnamefont {Vanmaekelbergh}},\ }\bibfield  {title}
  {\bibinfo {title} {Electronic quantum materials simulated with artificial
  model lattices},\ }\href {https://doi.org/10.1021/acsnanoscienceau.1c00054}
  {\bibfield  {journal} {\bibinfo  {journal} {ACS Nanosci. Au}\ }\textbf
  {\bibinfo {volume} {2}},\ \bibinfo {pages} {198} (\bibinfo {year}
  {2022})}\BibitemShut {NoStop}%
\bibitem [{\citenamefont {Guidoni}\ \emph {et~al.}(1997)\citenamefont
  {Guidoni}, \citenamefont {Trich\'e}, \citenamefont {Verkerk},\ and\
  \citenamefont {Grynberg}}]{guidoni.97}%
  \BibitemOpen
  \bibfield  {author} {\bibinfo {author} {\bibfnamefont {L.}~\bibnamefont
  {Guidoni}}, \bibinfo {author} {\bibfnamefont {C.}~\bibnamefont {Trich\'e}},
  \bibinfo {author} {\bibfnamefont {P.}~\bibnamefont {Verkerk}},\ and\ \bibinfo
  {author} {\bibfnamefont {G.}~\bibnamefont {Grynberg}},\ }\bibfield  {title}
  {\bibinfo {title} {Quasiperiodic optical lattices},\ }\href
  {https://doi.org/10.1103/PhysRevLett.79.3363} {\bibfield  {journal} {\bibinfo
   {journal} {Phys. Rev. Lett.}\ }\textbf {\bibinfo {volume} {79}},\ \bibinfo
  {pages} {3363} (\bibinfo {year} {1997})}\BibitemShut {NoStop}%
\bibitem [{\citenamefont {Ledieu}\ \emph {et~al.}(2004)\citenamefont {Ledieu},
  \citenamefont {Hoeft}, \citenamefont {Reid}, \citenamefont {Smerdon},
  \citenamefont {Diehl}, \citenamefont {Lograsso}, \citenamefont {Ross},\ and\
  \citenamefont {McGrath}}]{ledieu.04}%
  \BibitemOpen
  \bibfield  {author} {\bibinfo {author} {\bibfnamefont {J.}~\bibnamefont
  {Ledieu}}, \bibinfo {author} {\bibfnamefont {J.~T.}\ \bibnamefont {Hoeft}},
  \bibinfo {author} {\bibfnamefont {D.~E.}\ \bibnamefont {Reid}}, \bibinfo
  {author} {\bibfnamefont {J.~A.}\ \bibnamefont {Smerdon}}, \bibinfo {author}
  {\bibfnamefont {R.~D.}\ \bibnamefont {Diehl}}, \bibinfo {author}
  {\bibfnamefont {T.~A.}\ \bibnamefont {Lograsso}}, \bibinfo {author}
  {\bibfnamefont {A.~R.}\ \bibnamefont {Ross}},\ and\ \bibinfo {author}
  {\bibfnamefont {R.}~\bibnamefont {McGrath}},\ }\bibfield  {title} {\bibinfo
  {title} {Pseudomorphic growth of a single element quasiperiodic ultrathin
  film on a quasicrystal substrate},\ }\href
  {https://doi.org/10.1103/PhysRevLett.92.135507} {\bibfield  {journal}
  {\bibinfo  {journal} {Phys. Rev. Lett.}\ }\textbf {\bibinfo {volume} {92}},\
  \bibinfo {pages} {135507} (\bibinfo {year} {2004})}\BibitemShut {NoStop}%
\bibitem [{\citenamefont {Sharma}\ \emph {et~al.}(2005)\citenamefont {Sharma},
  \citenamefont {Shimoda}, \citenamefont {Ross}, \citenamefont {Lograsso},\
  and\ \citenamefont {Tsai}}]{sharma.05}%
  \BibitemOpen
  \bibfield  {author} {\bibinfo {author} {\bibfnamefont {H.~R.}\ \bibnamefont
  {Sharma}}, \bibinfo {author} {\bibfnamefont {M.}~\bibnamefont {Shimoda}},
  \bibinfo {author} {\bibfnamefont {A.~R.}\ \bibnamefont {Ross}}, \bibinfo
  {author} {\bibfnamefont {T.~A.}\ \bibnamefont {Lograsso}},\ and\ \bibinfo
  {author} {\bibfnamefont {A.~P.}\ \bibnamefont {Tsai}},\ }\bibfield  {title}
  {\bibinfo {title} {Real-space observation of quasicrystalline sn monolayer
  formed on the fivefold surface of icosahedral
  $\mathrm{Al}\mathrm{Cu}\mathrm{Fe}$ quasicrystal},\ }\href
  {https://doi.org/10.1103/PhysRevB.72.045428} {\bibfield  {journal} {\bibinfo
  {journal} {Phys. Rev. B}\ }\textbf {\bibinfo {volume} {72}},\ \bibinfo
  {pages} {045428} (\bibinfo {year} {2005})}\BibitemShut {NoStop}%
\bibitem [{\citenamefont {Ledieu}\ \emph {et~al.}(2008)\citenamefont {Ledieu},
  \citenamefont {Leung}, \citenamefont {Wearing}, \citenamefont {McGrath},
  \citenamefont {Lograsso}, \citenamefont {Wu},\ and\ \citenamefont
  {Fourn\'ee}}]{ledieu.08}%
  \BibitemOpen
  \bibfield  {author} {\bibinfo {author} {\bibfnamefont {J.}~\bibnamefont
  {Ledieu}}, \bibinfo {author} {\bibfnamefont {L.}~\bibnamefont {Leung}},
  \bibinfo {author} {\bibfnamefont {L.~H.}\ \bibnamefont {Wearing}}, \bibinfo
  {author} {\bibfnamefont {R.}~\bibnamefont {McGrath}}, \bibinfo {author}
  {\bibfnamefont {T.~A.}\ \bibnamefont {Lograsso}}, \bibinfo {author}
  {\bibfnamefont {D.}~\bibnamefont {Wu}},\ and\ \bibinfo {author}
  {\bibfnamefont {V.}~\bibnamefont {Fourn\'ee}},\ }\bibfield  {title} {\bibinfo
  {title} {Self-assembly, structure, and electronic properties of a
  quasiperiodic lead monolayer},\ }\href
  {https://doi.org/10.1103/PhysRevB.77.073409} {\bibfield  {journal} {\bibinfo
  {journal} {Phys. Rev. B}\ }\textbf {\bibinfo {volume} {77}},\ \bibinfo
  {pages} {073409} (\bibinfo {year} {2008})}\BibitemShut {NoStop}%
\bibitem [{\citenamefont {Smerdon}\ \emph {et~al.}(2008)\citenamefont
  {Smerdon}, \citenamefont {Parle}, \citenamefont {Wearing}, \citenamefont
  {Lograsso}, \citenamefont {Ross},\ and\ \citenamefont
  {McGrath}}]{smerdon.08}%
  \BibitemOpen
  \bibfield  {author} {\bibinfo {author} {\bibfnamefont {J.~A.}\ \bibnamefont
  {Smerdon}}, \bibinfo {author} {\bibfnamefont {J.~K.}\ \bibnamefont {Parle}},
  \bibinfo {author} {\bibfnamefont {L.~H.}\ \bibnamefont {Wearing}}, \bibinfo
  {author} {\bibfnamefont {T.~A.}\ \bibnamefont {Lograsso}}, \bibinfo {author}
  {\bibfnamefont {A.~R.}\ \bibnamefont {Ross}},\ and\ \bibinfo {author}
  {\bibfnamefont {R.}~\bibnamefont {McGrath}},\ }\bibfield  {title} {\bibinfo
  {title} {{Nucleation and growth of a quasicrystalline monolayer: Bi
  adsorption on the fivefold surface of
  $i{\text{-Al}}_{70}{\text{Pd}}_{21}{\text{Mn}}_{9}$}},\ }\href
  {https://doi.org/10.1103/PhysRevB.78.075407} {\bibfield  {journal} {\bibinfo
  {journal} {Phys. Rev. B}\ }\textbf {\bibinfo {volume} {78}},\ \bibinfo
  {pages} {075407} (\bibinfo {year} {2008})}\BibitemShut {NoStop}%
\bibitem [{\citenamefont {Talapin}\ \emph {et~al.}(2009)\citenamefont
  {Talapin}, \citenamefont {Shevchenko}, \citenamefont {Bodnarchuk},
  \citenamefont {Ye}, \citenamefont {Chen},\ and\ \citenamefont
  {Murray}}]{talapin.09}%
  \BibitemOpen
  \bibfield  {author} {\bibinfo {author} {\bibfnamefont {D.~V.}\ \bibnamefont
  {Talapin}}, \bibinfo {author} {\bibfnamefont {E.~V.}\ \bibnamefont
  {Shevchenko}}, \bibinfo {author} {\bibfnamefont {M.~I.}\ \bibnamefont
  {Bodnarchuk}}, \bibinfo {author} {\bibfnamefont {X.}~\bibnamefont {Ye}},
  \bibinfo {author} {\bibfnamefont {J.}~\bibnamefont {Chen}},\ and\ \bibinfo
  {author} {\bibfnamefont {C.~B.}\ \bibnamefont {Murray}},\ }\bibfield  {title}
  {\bibinfo {title} {Quasicrystalline order in self-assembled binary
  nanoparticle superlattices},\ }\href {https://doi.org/10.1038/nature08439}
  {\bibfield  {journal} {\bibinfo  {journal} {Nature}\ }\textbf {\bibinfo
  {volume} {461}},\ \bibinfo {pages} {964} (\bibinfo {year}
  {2009})}\BibitemShut {NoStop}%
\bibitem [{\citenamefont {F{\"o}rster}\ \emph {et~al.}(2013)\citenamefont
  {F{\"o}rster}, \citenamefont {Meinel}, \citenamefont {Hammer}, \citenamefont
  {Trautmann},\ and\ \citenamefont {Widdra}}]{forster.13}%
  \BibitemOpen
  \bibfield  {author} {\bibinfo {author} {\bibfnamefont {S.}~\bibnamefont
  {F{\"o}rster}}, \bibinfo {author} {\bibfnamefont {K.}~\bibnamefont {Meinel}},
  \bibinfo {author} {\bibfnamefont {R.}~\bibnamefont {Hammer}}, \bibinfo
  {author} {\bibfnamefont {M.}~\bibnamefont {Trautmann}},\ and\ \bibinfo
  {author} {\bibfnamefont {W.}~\bibnamefont {Widdra}},\ }\bibfield  {title}
  {\bibinfo {title} {Quasicrystalline structure formation in a classical
  crystalline thin-film system},\ }\href {https://doi.org/10.1038/nature12514}
  {\bibfield  {journal} {\bibinfo  {journal} {Nature}\ }\textbf {\bibinfo
  {volume} {502}},\ \bibinfo {pages} {215} (\bibinfo {year}
  {2013})}\BibitemShut {NoStop}%
\bibitem [{\citenamefont {Jia}\ \emph {et~al.}(2016)\citenamefont {Jia},
  \citenamefont {Yang}, \citenamefont {Yang},\ and\ \citenamefont
  {Ebendorff-Heidepriem}}]{jia.16}%
  \BibitemOpen
  \bibfield  {author} {\bibinfo {author} {\bibfnamefont {P.}~\bibnamefont
  {Jia}}, \bibinfo {author} {\bibfnamefont {Z.}~\bibnamefont {Yang}}, \bibinfo
  {author} {\bibfnamefont {J.}~\bibnamefont {Yang}},\ and\ \bibinfo {author}
  {\bibfnamefont {H.}~\bibnamefont {Ebendorff-Heidepriem}},\ }\bibfield
  {title} {\bibinfo {title} {Quasiperiodic nanohole arrays on optical fibers as
  plasmonic sensors: Fabrication and sensitivity determination},\ }\href
  {https://doi.org/10.1021/acssensors.6b00436} {\bibfield  {journal} {\bibinfo
  {journal} {ACS Sens.}\ }\textbf {\bibinfo {volume} {1}},\ \bibinfo {pages}
  {1078} (\bibinfo {year} {2016})}\BibitemShut {NoStop}%
\bibitem [{\citenamefont {Bandres}\ \emph {et~al.}(2016)\citenamefont
  {Bandres}, \citenamefont {Rechtsman},\ and\ \citenamefont
  {Segev}}]{bandres.16}%
  \BibitemOpen
  \bibfield  {author} {\bibinfo {author} {\bibfnamefont {M.~A.}\ \bibnamefont
  {Bandres}}, \bibinfo {author} {\bibfnamefont {M.~C.}\ \bibnamefont
  {Rechtsman}},\ and\ \bibinfo {author} {\bibfnamefont {M.}~\bibnamefont
  {Segev}},\ }\bibfield  {title} {\bibinfo {title} {Topological photonic
  quasicrystals: Fractal topological spectrum and protected transport},\ }\href
  {https://doi.org/10.1103/PhysRevX.6.011016} {\bibfield  {journal} {\bibinfo
  {journal} {Phys. Rev. X}\ }\textbf {\bibinfo {volume} {6}},\ \bibinfo {pages}
  {011016} (\bibinfo {year} {2016})}\BibitemShut {NoStop}%
\bibitem [{\citenamefont {Collins}\ \emph {et~al.}(2017)\citenamefont
  {Collins}, \citenamefont {Witte}, \citenamefont {Silverman}, \citenamefont
  {Green},\ and\ \citenamefont {Gomes}}]{collins.17}%
  \BibitemOpen
  \bibfield  {author} {\bibinfo {author} {\bibfnamefont {L.~C.}\ \bibnamefont
  {Collins}}, \bibinfo {author} {\bibfnamefont {T.~G.}\ \bibnamefont {Witte}},
  \bibinfo {author} {\bibfnamefont {R.}~\bibnamefont {Silverman}}, \bibinfo
  {author} {\bibfnamefont {D.~B.}\ \bibnamefont {Green}},\ and\ \bibinfo
  {author} {\bibfnamefont {K.~K.}\ \bibnamefont {Gomes}},\ }\bibfield  {title}
  {\bibinfo {title} {Imaging quasiperiodic electronic states in a synthetic
  penrose tiling},\ }\href {https://doi.org/10.1038/ncomms15961} {\bibfield
  {journal} {\bibinfo  {journal} {Nat. Commun.}\ }\textbf {\bibinfo {volume}
  {8}},\ \bibinfo {pages} {15961} (\bibinfo {year} {2017})}\BibitemShut
  {NoStop}%
\bibitem [{\citenamefont {Yan}\ and\ \citenamefont {Liljeroth}(2019)}]{yan.19}%
  \BibitemOpen
  \bibfield  {author} {\bibinfo {author} {\bibfnamefont {L.}~\bibnamefont
  {Yan}}\ and\ \bibinfo {author} {\bibfnamefont {P.}~\bibnamefont
  {Liljeroth}},\ }\bibfield  {title} {\bibinfo {title} {Engineered electronic
  states in atomically precise artificial lattices and graphene nanoribbons},\
  }\href {https://doi.org/10.1080/23746149.2019.1651672} {\bibfield  {journal}
  {\bibinfo  {journal} {Adv. Phys.: X}\ }\textbf {\bibinfo {volume} {4}},\
  \bibinfo {pages} {1651672} (\bibinfo {year} {2019})}\BibitemShut {NoStop}%
\bibitem [{\citenamefont {Xin}\ \emph {et~al.}(2022)\citenamefont {Xin},
  \citenamefont {Zhang}, \citenamefont {Wan}, \citenamefont {Liu},
  \citenamefont {Ding}, \citenamefont {Li}, \citenamefont {Wang},\ and\
  \citenamefont {Dong}}]{xin.22}%
  \BibitemOpen
  \bibfield  {author} {\bibinfo {author} {\bibfnamefont {X.}~\bibnamefont
  {Xin}}, \bibinfo {author} {\bibfnamefont {S.}~\bibnamefont {Zhang}}, \bibinfo
  {author} {\bibfnamefont {P.}~\bibnamefont {Wan}}, \bibinfo {author}
  {\bibfnamefont {L.}~\bibnamefont {Liu}}, \bibinfo {author} {\bibfnamefont
  {W.}~\bibnamefont {Ding}}, \bibinfo {author} {\bibfnamefont {J.}~\bibnamefont
  {Li}}, \bibinfo {author} {\bibfnamefont {Q.}~\bibnamefont {Wang}},\ and\
  \bibinfo {author} {\bibfnamefont {C.}~\bibnamefont {Dong}},\ }\bibfield
  {title} {\bibinfo {title} {{Single-phase quasicrystalline Al–Cu–Fe thin
  film prepared by direct current magnetron sputtering on stainless steel}},\
  }\href {https://doi.org/https://doi.org/10.1016/j.tsf.2022.139272} {\bibfield
   {journal} {\bibinfo  {journal} {Thin Solid Films}\ }\textbf {\bibinfo
  {volume} {753}},\ \bibinfo {pages} {139272} (\bibinfo {year}
  {2022})}\BibitemShut {NoStop}%
\bibitem [{\citenamefont {Dang}\ \emph {et~al.}(2023)\citenamefont {Dang},
  \citenamefont {Qi}, \citenamefont {Zhao}, \citenamefont {Fan},\ and\
  \citenamefont {Lu}}]{dang.23}%
  \BibitemOpen
  \bibfield  {author} {\bibinfo {author} {\bibfnamefont {H.}~\bibnamefont
  {Dang}}, \bibinfo {author} {\bibfnamefont {D.}~\bibnamefont {Qi}}, \bibinfo
  {author} {\bibfnamefont {M.}~\bibnamefont {Zhao}}, \bibinfo {author}
  {\bibfnamefont {C.}~\bibnamefont {Fan}},\ and\ \bibinfo {author}
  {\bibfnamefont {C.~S.}\ \bibnamefont {Lu}},\ }\bibfield  {title} {\bibinfo
  {title} {Thermal-induced interfacial behavior of a thin one-dimensional
  hexagonal quasicrystal film},\ }\href
  {https://doi.org/10.1007/s10483-023-2989-7} {\bibfield  {journal} {\bibinfo
  {journal} {Appl. Math. Mech. -Engl. Ed.}\ }\textbf {\bibinfo {volume} {44}},\
  \bibinfo {pages} {841} (\bibinfo {year} {2023})}\BibitemShut {NoStop}%
\bibitem [{\citenamefont {Lubin}\ \emph {et~al.}(2012)\citenamefont {Lubin},
  \citenamefont {Zhou}, \citenamefont {Hryn}, \citenamefont {Huntington},\ and\
  \citenamefont {Odom}}]{lubin.12}%
  \BibitemOpen
  \bibfield  {author} {\bibinfo {author} {\bibfnamefont {S.~M.}\ \bibnamefont
  {Lubin}}, \bibinfo {author} {\bibfnamefont {W.}~\bibnamefont {Zhou}},
  \bibinfo {author} {\bibfnamefont {A.~J.}\ \bibnamefont {Hryn}}, \bibinfo
  {author} {\bibfnamefont {M.~D.}\ \bibnamefont {Huntington}},\ and\ \bibinfo
  {author} {\bibfnamefont {T.~W.}\ \bibnamefont {Odom}},\ }\bibfield  {title}
  {\bibinfo {title} {High-rotational symmetry lattices fabricated by moiré
  nanolithography},\ }\href {https://doi.org/10.1021/nl302535p} {\bibfield
  {journal} {\bibinfo  {journal} {Nano Lett.}\ }\textbf {\bibinfo {volume}
  {12}},\ \bibinfo {pages} {4948} (\bibinfo {year} {2012})}\BibitemShut
  {NoStop}%
\bibitem [{\citenamefont {Ahn}\ \emph {et~al.}(2018)\citenamefont {Ahn},
  \citenamefont {Moon}, \citenamefont {Kim}, \citenamefont {Kim}, \citenamefont
  {Shin}, \citenamefont {Kim}, \citenamefont {Cha}, \citenamefont {Kahng},
  \citenamefont {Kim}, \citenamefont {Koshino}, \citenamefont {Son},
  \citenamefont {Yang},\ and\ \citenamefont {Ahn}}]{joon-ahn.18}%
  \BibitemOpen
  \bibfield  {author} {\bibinfo {author} {\bibfnamefont {S.~J.}\ \bibnamefont
  {Ahn}}, \bibinfo {author} {\bibfnamefont {P.}~\bibnamefont {Moon}}, \bibinfo
  {author} {\bibfnamefont {T.-H.}\ \bibnamefont {Kim}}, \bibinfo {author}
  {\bibfnamefont {H.-W.}\ \bibnamefont {Kim}}, \bibinfo {author} {\bibfnamefont
  {H.-C.}\ \bibnamefont {Shin}}, \bibinfo {author} {\bibfnamefont {E.~H.}\
  \bibnamefont {Kim}}, \bibinfo {author} {\bibfnamefont {H.~W.}\ \bibnamefont
  {Cha}}, \bibinfo {author} {\bibfnamefont {S.-J.}\ \bibnamefont {Kahng}},
  \bibinfo {author} {\bibfnamefont {P.}~\bibnamefont {Kim}}, \bibinfo {author}
  {\bibfnamefont {M.}~\bibnamefont {Koshino}}, \bibinfo {author} {\bibfnamefont
  {Y.-W.}\ \bibnamefont {Son}}, \bibinfo {author} {\bibfnamefont {C.-W.}\
  \bibnamefont {Yang}},\ and\ \bibinfo {author} {\bibfnamefont {J.~R.}\
  \bibnamefont {Ahn}},\ }\bibfield  {title} {\bibinfo {title} {Dirac electrons
  in a dodecagonal graphene quasicrystal},\ }\href
  {https://doi.org/10.1126/science.aar8412} {\bibfield  {journal} {\bibinfo
  {journal} {Science}\ }\textbf {\bibinfo {volume} {361}},\ \bibinfo {pages}
  {782} (\bibinfo {year} {2018})}\BibitemShut {NoStop}%
\bibitem [{\citenamefont {Mahmood}\ \emph {et~al.}(2021)\citenamefont
  {Mahmood}, \citenamefont {Ramirez},\ and\ \citenamefont
  {Hillier}}]{mahmood.21}%
  \BibitemOpen
  \bibfield  {author} {\bibinfo {author} {\bibfnamefont {R.}~\bibnamefont
  {Mahmood}}, \bibinfo {author} {\bibfnamefont {A.~V.}\ \bibnamefont
  {Ramirez}},\ and\ \bibinfo {author} {\bibfnamefont {A.~C.}\ \bibnamefont
  {Hillier}},\ }\bibfield  {title} {\bibinfo {title} {Creating two-dimensional
  quasicrystal, supercell, and moiré lattices with laser interference
  lithography: Implications for photonic bandgap materials},\ }\href
  {https://doi.org/10.1021/acsanm.1c00210} {\bibfield  {journal} {\bibinfo
  {journal} {ACS Appl. Nano Mater.}\ }\textbf {\bibinfo {volume} {4}},\
  \bibinfo {pages} {8851} (\bibinfo {year} {2021})}\BibitemShut {NoStop}%
\bibitem [{\citenamefont {Uri}\ \emph {et~al.}(2023)\citenamefont {Uri},
  \citenamefont {de~la Barrera}, \citenamefont {Randeria}, \citenamefont
  {Rodan-Legrain}, \citenamefont {Devakul}, \citenamefont {Crowley},
  \citenamefont {Paul}, \citenamefont {Watanabe}, \citenamefont {Taniguchi},
  \citenamefont {Lifshitz}, \citenamefont {Fu}, \citenamefont {Ashoori},\ and\
  \citenamefont {Jarillo-Herrero}}]{uri.23}%
  \BibitemOpen
  \bibfield  {author} {\bibinfo {author} {\bibfnamefont {A.}~\bibnamefont
  {Uri}}, \bibinfo {author} {\bibfnamefont {S.~C.}\ \bibnamefont {de~la
  Barrera}}, \bibinfo {author} {\bibfnamefont {M.~T.}\ \bibnamefont
  {Randeria}}, \bibinfo {author} {\bibfnamefont {D.}~\bibnamefont
  {Rodan-Legrain}}, \bibinfo {author} {\bibfnamefont {T.}~\bibnamefont
  {Devakul}}, \bibinfo {author} {\bibfnamefont {P.~J.~D.}\ \bibnamefont
  {Crowley}}, \bibinfo {author} {\bibfnamefont {N.}~\bibnamefont {Paul}},
  \bibinfo {author} {\bibfnamefont {K.}~\bibnamefont {Watanabe}}, \bibinfo
  {author} {\bibfnamefont {T.}~\bibnamefont {Taniguchi}}, \bibinfo {author}
  {\bibfnamefont {R.}~\bibnamefont {Lifshitz}}, \bibinfo {author}
  {\bibfnamefont {L.}~\bibnamefont {Fu}}, \bibinfo {author} {\bibfnamefont
  {R.~C.}\ \bibnamefont {Ashoori}},\ and\ \bibinfo {author} {\bibfnamefont
  {P.}~\bibnamefont {Jarillo-Herrero}},\ }\bibfield  {title} {\bibinfo {title}
  {Superconductivity and strong interactions in a tunable moiré
  quasicrystal},\ }\href {https://doi.org/10.1038/s41586-023-06294-z}
  {\bibfield  {journal} {\bibinfo  {journal} {Nature}\ }\textbf {\bibinfo
  {volume} {620}},\ \bibinfo {pages} {762} (\bibinfo {year}
  {2023})}\BibitemShut {NoStop}%
\bibitem [{\citenamefont {Ghadimi}\ and\ \citenamefont
  {Yang}(2024)}]{ghadimi.24}%
  \BibitemOpen
  \bibfield  {author} {\bibinfo {author} {\bibfnamefont {R.}~\bibnamefont
  {Ghadimi}}\ and\ \bibinfo {author} {\bibfnamefont {B.-J.}\ \bibnamefont
  {Yang}},\ }\href {https://arxiv.org/abs/2408.05714} {\bibinfo {title}
  {Quasiperiodic pairing in graphene quasicrystals}} (\bibinfo {year} {2024}),\
  \Eprint {https://arxiv.org/abs/2408.05714} {arXiv:2408.05714
  [cond-mat.supr-con]} \BibitemShut {NoStop}%
\bibitem [{\citenamefont {Merlin}\ \emph {et~al.}(1985)\citenamefont {Merlin},
  \citenamefont {Bajema}, \citenamefont {Clarke}, \citenamefont {Juang},\ and\
  \citenamefont {Bhattacharya}}]{merlin.85}%
  \BibitemOpen
  \bibfield  {author} {\bibinfo {author} {\bibfnamefont {R.}~\bibnamefont
  {Merlin}}, \bibinfo {author} {\bibfnamefont {K.}~\bibnamefont {Bajema}},
  \bibinfo {author} {\bibfnamefont {R.}~\bibnamefont {Clarke}}, \bibinfo
  {author} {\bibfnamefont {F.~Y.}\ \bibnamefont {Juang}},\ and\ \bibinfo
  {author} {\bibfnamefont {P.~K.}\ \bibnamefont {Bhattacharya}},\ }\bibfield
  {title} {\bibinfo {title} {Quasiperiodic gaas-alas heterostructures},\ }\href
  {https://doi.org/10.1103/PhysRevLett.55.1768} {\bibfield  {journal} {\bibinfo
   {journal} {Phys. Rev. Lett.}\ }\textbf {\bibinfo {volume} {55}},\ \bibinfo
  {pages} {1768} (\bibinfo {year} {1985})}\BibitemShut {NoStop}%
\bibitem [{\citenamefont {Todd}\ \emph {et~al.}(1986)\citenamefont {Todd},
  \citenamefont {Merlin}, \citenamefont {Clarke}, \citenamefont {Mohanty},\
  and\ \citenamefont {Axe}}]{todd.86}%
  \BibitemOpen
  \bibfield  {author} {\bibinfo {author} {\bibfnamefont {J.}~\bibnamefont
  {Todd}}, \bibinfo {author} {\bibfnamefont {R.}~\bibnamefont {Merlin}},
  \bibinfo {author} {\bibfnamefont {R.}~\bibnamefont {Clarke}}, \bibinfo
  {author} {\bibfnamefont {K.~M.}\ \bibnamefont {Mohanty}},\ and\ \bibinfo
  {author} {\bibfnamefont {J.~D.}\ \bibnamefont {Axe}},\ }\bibfield  {title}
  {\bibinfo {title} {Synchrotron x-ray study of a fibonacci superlattice},\
  }\href {https://doi.org/10.1103/PhysRevLett.57.1157} {\bibfield  {journal}
  {\bibinfo  {journal} {Phys. Rev. Lett.}\ }\textbf {\bibinfo {volume} {57}},\
  \bibinfo {pages} {1157} (\bibinfo {year} {1986})}\BibitemShut {NoStop}%
\bibitem [{\citenamefont {Bajema}\ and\ \citenamefont
  {Merlin}(1987)}]{bajema.merlin.87}%
  \BibitemOpen
  \bibfield  {author} {\bibinfo {author} {\bibfnamefont {K.}~\bibnamefont
  {Bajema}}\ and\ \bibinfo {author} {\bibfnamefont {R.}~\bibnamefont
  {Merlin}},\ }\bibfield  {title} {\bibinfo {title} {Raman scattering by
  acoustic phonons in fibonacci gaas-aias superlattices},\ }\href
  {https://doi.org/10.1103/PhysRevB.36.4555} {\bibfield  {journal} {\bibinfo
  {journal} {Phys. Rev. B}\ }\textbf {\bibinfo {volume} {36}},\ \bibinfo
  {pages} {4555} (\bibinfo {year} {1987})}\BibitemShut {NoStop}%
\bibitem [{\citenamefont {Cohn}\ \emph {et~al.}(1988)\citenamefont {Cohn},
  \citenamefont {Lin}, \citenamefont {Lamelas}, \citenamefont {He},
  \citenamefont {Clarke},\ and\ \citenamefont {Uher}}]{cohn.88}%
  \BibitemOpen
  \bibfield  {author} {\bibinfo {author} {\bibfnamefont {J.~L.}\ \bibnamefont
  {Cohn}}, \bibinfo {author} {\bibfnamefont {J.~J.}\ \bibnamefont {Lin}},
  \bibinfo {author} {\bibfnamefont {F.~J.}\ \bibnamefont {Lamelas}}, \bibinfo
  {author} {\bibfnamefont {H.}~\bibnamefont {He}}, \bibinfo {author}
  {\bibfnamefont {R.}~\bibnamefont {Clarke}},\ and\ \bibinfo {author}
  {\bibfnamefont {C.}~\bibnamefont {Uher}},\ }\bibfield  {title} {\bibinfo
  {title} {Upper critical fields of periodic and quasiperiodic nb-ta
  superlattices},\ }\href {https://doi.org/10.1103/PhysRevB.38.2326} {\bibfield
   {journal} {\bibinfo  {journal} {Phys. Rev. B}\ }\textbf {\bibinfo {volume}
  {38}},\ \bibinfo {pages} {2326} (\bibinfo {year} {1988})}\BibitemShut
  {NoStop}%
\bibitem [{\citenamefont {Zhu}\ \emph {et~al.}(1997)\citenamefont {Zhu},
  \citenamefont {Zhu}, \citenamefont {Qin}, \citenamefont {Wang}, \citenamefont
  {Ge},\ and\ \citenamefont {Ming}}]{zhu.97}%
  \BibitemOpen
  \bibfield  {author} {\bibinfo {author} {\bibfnamefont {S.-n.}\ \bibnamefont
  {Zhu}}, \bibinfo {author} {\bibfnamefont {Y.-y.}\ \bibnamefont {Zhu}},
  \bibinfo {author} {\bibfnamefont {Y.-q.}\ \bibnamefont {Qin}}, \bibinfo
  {author} {\bibfnamefont {H.-f.}\ \bibnamefont {Wang}}, \bibinfo {author}
  {\bibfnamefont {C.-z.}\ \bibnamefont {Ge}},\ and\ \bibinfo {author}
  {\bibfnamefont {N.-b.}\ \bibnamefont {Ming}},\ }\bibfield  {title} {\bibinfo
  {title} {{Experimental Realization of Second Harmonic Generation in a
  Fibonacci Optical Superlattice of LiTa${\mathrm{O}}_{3}$}},\ }\href
  {https://doi.org/10.1103/PhysRevLett.78.2752} {\bibfield  {journal} {\bibinfo
   {journal} {Phys. Rev. Lett.}\ }\textbf {\bibinfo {volume} {78}},\ \bibinfo
  {pages} {2752} (\bibinfo {year} {1997})}\BibitemShut {NoStop}%
\bibitem [{\citenamefont {Dal~Negro}\ \emph {et~al.}(2003)\citenamefont
  {Dal~Negro}, \citenamefont {Oton}, \citenamefont {Gaburro}, \citenamefont
  {Pavesi}, \citenamefont {Johnson}, \citenamefont {Lagendijk}, \citenamefont
  {Righini}, \citenamefont {Colocci},\ and\ \citenamefont
  {Wiersma}}]{negro.03}%
  \BibitemOpen
  \bibfield  {author} {\bibinfo {author} {\bibfnamefont {L.}~\bibnamefont
  {Dal~Negro}}, \bibinfo {author} {\bibfnamefont {C.~J.}\ \bibnamefont {Oton}},
  \bibinfo {author} {\bibfnamefont {Z.}~\bibnamefont {Gaburro}}, \bibinfo
  {author} {\bibfnamefont {L.}~\bibnamefont {Pavesi}}, \bibinfo {author}
  {\bibfnamefont {P.}~\bibnamefont {Johnson}}, \bibinfo {author} {\bibfnamefont
  {A.}~\bibnamefont {Lagendijk}}, \bibinfo {author} {\bibfnamefont
  {R.}~\bibnamefont {Righini}}, \bibinfo {author} {\bibfnamefont
  {M.}~\bibnamefont {Colocci}},\ and\ \bibinfo {author} {\bibfnamefont {D.~S.}\
  \bibnamefont {Wiersma}},\ }\bibfield  {title} {\bibinfo {title} {Light
  transport through the band-edge states of fibonacci quasicrystals},\ }\href
  {https://doi.org/10.1103/PhysRevLett.90.055501} {\bibfield  {journal}
  {\bibinfo  {journal} {Phys. Rev. Lett.}\ }\textbf {\bibinfo {volume} {90}},\
  \bibinfo {pages} {055501} (\bibinfo {year} {2003})}\BibitemShut {NoStop}%
\bibitem [{\citenamefont {Yuan}\ \emph {et~al.}(2018)\citenamefont {Yuan},
  \citenamefont {Lin}, \citenamefont {Xiao},\ and\ \citenamefont
  {Fan}}]{yuan.18}%
  \BibitemOpen
  \bibfield  {author} {\bibinfo {author} {\bibfnamefont {L.}~\bibnamefont
  {Yuan}}, \bibinfo {author} {\bibfnamefont {Q.}~\bibnamefont {Lin}}, \bibinfo
  {author} {\bibfnamefont {M.}~\bibnamefont {Xiao}},\ and\ \bibinfo {author}
  {\bibfnamefont {S.}~\bibnamefont {Fan}},\ }\bibfield  {title} {\bibinfo
  {title} {Synthetic dimension in photonics},\ }\href
  {https://doi.org/10.1364/OPTICA.5.001396} {\bibfield  {journal} {\bibinfo
  {journal} {Optica}\ }\textbf {\bibinfo {volume} {5}},\ \bibinfo {pages}
  {1396} (\bibinfo {year} {2018})}\BibitemShut {NoStop}%
\bibitem [{\citenamefont {Steurer}\ and\ \citenamefont
  {Sutter-Widmer}(2007)}]{steurer.sutter.07}%
  \BibitemOpen
  \bibfield  {author} {\bibinfo {author} {\bibfnamefont {W.}~\bibnamefont
  {Steurer}}\ and\ \bibinfo {author} {\bibfnamefont {D.}~\bibnamefont
  {Sutter-Widmer}},\ }\bibfield  {title} {\bibinfo {title} {Photonic and
  phononic quasicrystals},\ }\href
  {https://doi.org/10.1088/0022-3727/40/13/r01} {\bibfield  {journal} {\bibinfo
   {journal} {J. Phys. D: Appl. Phys.}\ }\textbf {\bibinfo {volume} {40}},\
  \bibinfo {pages} {R229} (\bibinfo {year} {2007})}\BibitemShut {NoStop}%
\bibitem [{\citenamefont {Tanese}\ \emph {et~al.}(2014)\citenamefont {Tanese},
  \citenamefont {Gurevich}, \citenamefont {Baboux}, \citenamefont {Jacqmin},
  \citenamefont {Lema\^{\i}tre}, \citenamefont {Galopin}, \citenamefont
  {Sagnes}, \citenamefont {Amo}, \citenamefont {Bloch},\ and\ \citenamefont
  {Akkermans}}]{tanese.14}%
  \BibitemOpen
  \bibfield  {author} {\bibinfo {author} {\bibfnamefont {D.}~\bibnamefont
  {Tanese}}, \bibinfo {author} {\bibfnamefont {E.}~\bibnamefont {Gurevich}},
  \bibinfo {author} {\bibfnamefont {F.}~\bibnamefont {Baboux}}, \bibinfo
  {author} {\bibfnamefont {T.}~\bibnamefont {Jacqmin}}, \bibinfo {author}
  {\bibfnamefont {A.}~\bibnamefont {Lema\^{\i}tre}}, \bibinfo {author}
  {\bibfnamefont {E.}~\bibnamefont {Galopin}}, \bibinfo {author} {\bibfnamefont
  {I.}~\bibnamefont {Sagnes}}, \bibinfo {author} {\bibfnamefont
  {A.}~\bibnamefont {Amo}}, \bibinfo {author} {\bibfnamefont {J.}~\bibnamefont
  {Bloch}},\ and\ \bibinfo {author} {\bibfnamefont {E.}~\bibnamefont
  {Akkermans}},\ }\bibfield  {title} {\bibinfo {title} {Fractal energy spectrum
  of a polariton gas in a fibonacci quasiperiodic potential},\ }\href
  {https://doi.org/10.1103/PhysRevLett.112.146404} {\bibfield  {journal}
  {\bibinfo  {journal} {Phys. Rev. Lett.}\ }\textbf {\bibinfo {volume} {112}},\
  \bibinfo {pages} {146404} (\bibinfo {year} {2014})}\BibitemShut {NoStop}%
\bibitem [{\citenamefont {Baboux}\ \emph {et~al.}(2017)\citenamefont {Baboux},
  \citenamefont {Levy}, \citenamefont {Lema\^{\i}tre}, \citenamefont {G\'omez},
  \citenamefont {Galopin}, \citenamefont {Le~Gratiet}, \citenamefont {Sagnes},
  \citenamefont {Amo}, \citenamefont {Bloch},\ and\ \citenamefont
  {Akkermans}}]{baboux.17}%
  \BibitemOpen
  \bibfield  {author} {\bibinfo {author} {\bibfnamefont {F.}~\bibnamefont
  {Baboux}}, \bibinfo {author} {\bibfnamefont {E.}~\bibnamefont {Levy}},
  \bibinfo {author} {\bibfnamefont {A.}~\bibnamefont {Lema\^{\i}tre}}, \bibinfo
  {author} {\bibfnamefont {C.}~\bibnamefont {G\'omez}}, \bibinfo {author}
  {\bibfnamefont {E.}~\bibnamefont {Galopin}}, \bibinfo {author} {\bibfnamefont
  {L.}~\bibnamefont {Le~Gratiet}}, \bibinfo {author} {\bibfnamefont
  {I.}~\bibnamefont {Sagnes}}, \bibinfo {author} {\bibfnamefont
  {A.}~\bibnamefont {Amo}}, \bibinfo {author} {\bibfnamefont {J.}~\bibnamefont
  {Bloch}},\ and\ \bibinfo {author} {\bibfnamefont {E.}~\bibnamefont
  {Akkermans}},\ }\bibfield  {title} {\bibinfo {title} {Measuring topological
  invariants from generalized edge states in polaritonic quasicrystals},\
  }\href {https://doi.org/10.1103/PhysRevB.95.161114} {\bibfield  {journal}
  {\bibinfo  {journal} {Phys. Rev. B}\ }\textbf {\bibinfo {volume} {95}},\
  \bibinfo {pages} {161114(R)} (\bibinfo {year} {2017})}\BibitemShut {NoStop}%
\bibitem [{\citenamefont {Goblot}\ \emph {et~al.}(2020)\citenamefont {Goblot},
  \citenamefont {Štrkalj}, \citenamefont {Pernet}, \citenamefont {Lado},
  \citenamefont {Dorow}, \citenamefont {Lemaître}, \citenamefont {Le~Gratiet},
  \citenamefont {Harouri}, \citenamefont {Sagnes}, \citenamefont {Ravets},
  \citenamefont {Amo}, \citenamefont {Bloch},\ and\ \citenamefont
  {Zilberberg}}]{goblot.20}%
  \BibitemOpen
  \bibfield  {author} {\bibinfo {author} {\bibfnamefont {V.}~\bibnamefont
  {Goblot}}, \bibinfo {author} {\bibfnamefont {A.}~\bibnamefont {Štrkalj}},
  \bibinfo {author} {\bibfnamefont {N.}~\bibnamefont {Pernet}}, \bibinfo
  {author} {\bibfnamefont {J.~L.}\ \bibnamefont {Lado}}, \bibinfo {author}
  {\bibfnamefont {C.}~\bibnamefont {Dorow}}, \bibinfo {author} {\bibfnamefont
  {A.}~\bibnamefont {Lemaître}}, \bibinfo {author} {\bibfnamefont
  {L.}~\bibnamefont {Le~Gratiet}}, \bibinfo {author} {\bibfnamefont
  {A.}~\bibnamefont {Harouri}}, \bibinfo {author} {\bibfnamefont
  {I.}~\bibnamefont {Sagnes}}, \bibinfo {author} {\bibfnamefont
  {S.}~\bibnamefont {Ravets}}, \bibinfo {author} {\bibfnamefont
  {A.}~\bibnamefont {Amo}}, \bibinfo {author} {\bibfnamefont {J.}~\bibnamefont
  {Bloch}},\ and\ \bibinfo {author} {\bibfnamefont {O.}~\bibnamefont
  {Zilberberg}},\ }\bibfield  {title} {\bibinfo {title} {Emergence of
  criticality through a cascade of delocalization transitions in quasiperiodic
  chains},\ }\href {https://doi.org/10.1038/s41567-020-0908-7} {\bibfield
  {journal} {\bibinfo  {journal} {Nat. Phys.}\ }\textbf {\bibinfo {volume}
  {16}},\ \bibinfo {pages} {832} (\bibinfo {year} {2020})}\BibitemShut
  {NoStop}%
\bibitem [{\citenamefont {Singh}\ \emph {et~al.}(2015)\citenamefont {Singh},
  \citenamefont {Saha}, \citenamefont {Parameswaran},\ and\ \citenamefont
  {Weld}}]{singh.15}%
  \BibitemOpen
  \bibfield  {author} {\bibinfo {author} {\bibfnamefont {K.}~\bibnamefont
  {Singh}}, \bibinfo {author} {\bibfnamefont {K.}~\bibnamefont {Saha}},
  \bibinfo {author} {\bibfnamefont {S.~A.}\ \bibnamefont {Parameswaran}},\ and\
  \bibinfo {author} {\bibfnamefont {D.~M.}\ \bibnamefont {Weld}},\ }\bibfield
  {title} {\bibinfo {title} {Fibonacci optical lattices for tunable quantum
  quasicrystals},\ }\href {https://doi.org/10.1103/PhysRevA.92.063426}
  {\bibfield  {journal} {\bibinfo  {journal} {Phys. Rev. A}\ }\textbf {\bibinfo
  {volume} {92}},\ \bibinfo {pages} {063426} (\bibinfo {year}
  {2015})}\BibitemShut {NoStop}%
\bibitem [{\citenamefont {Reisner}\ \emph {et~al.}(2023)\citenamefont
  {Reisner}, \citenamefont {Tahmi}, \citenamefont {Pi\'echon}, \citenamefont
  {Kuhl},\ and\ \citenamefont {Mortessagne}}]{reisner.23}%
  \BibitemOpen
  \bibfield  {author} {\bibinfo {author} {\bibfnamefont {M.}~\bibnamefont
  {Reisner}}, \bibinfo {author} {\bibfnamefont {Y.}~\bibnamefont {Tahmi}},
  \bibinfo {author} {\bibfnamefont {F.}~\bibnamefont {Pi\'echon}}, \bibinfo
  {author} {\bibfnamefont {U.}~\bibnamefont {Kuhl}},\ and\ \bibinfo {author}
  {\bibfnamefont {F.}~\bibnamefont {Mortessagne}},\ }\bibfield  {title}
  {\bibinfo {title} {Experimental observation of multifractality in fibonacci
  chains},\ }\href {https://doi.org/10.1103/PhysRevB.108.064210} {\bibfield
  {journal} {\bibinfo  {journal} {Phys. Rev. B}\ }\textbf {\bibinfo {volume}
  {108}},\ \bibinfo {pages} {064210} (\bibinfo {year} {2023})}\BibitemShut
  {NoStop}%
\bibitem [{\citenamefont {Franca}\ \emph {et~al.}(2024)\citenamefont {Franca},
  \citenamefont {Seidemann}, \citenamefont {Hassler}, \citenamefont {van~den
  Brink},\ and\ \citenamefont {Fulga}}]{franca.24}%
  \BibitemOpen
  \bibfield  {author} {\bibinfo {author} {\bibfnamefont {S.}~\bibnamefont
  {Franca}}, \bibinfo {author} {\bibfnamefont {T.}~\bibnamefont {Seidemann}},
  \bibinfo {author} {\bibfnamefont {F.}~\bibnamefont {Hassler}}, \bibinfo
  {author} {\bibfnamefont {J.}~\bibnamefont {van~den Brink}},\ and\ \bibinfo
  {author} {\bibfnamefont {I.~C.}\ \bibnamefont {Fulga}},\ }\bibfield  {title}
  {\bibinfo {title} {Impedance spectroscopy of chiral symmetric topoelectrical
  circuits},\ }\href {https://doi.org/10.1103/PhysRevB.109.L241103} {\bibfield
  {journal} {\bibinfo  {journal} {Phys. Rev. B}\ }\textbf {\bibinfo {volume}
  {109}},\ \bibinfo {pages} {L241103} (\bibinfo {year} {2024})}\BibitemShut
  {NoStop}%
\bibitem [{\citenamefont {Lisiecki}\ \emph {et~al.}(2019)\citenamefont
  {Lisiecki}, \citenamefont {Rych\l{}y}, \citenamefont
  {Ku\ifmmode~\acute{s}\else \'{s}\fi{}wik}, \citenamefont
  {G\l{}owi\ifmmode~\acute{n}\else \'{n}\fi{}ski}, \citenamefont {K\l{}os},
  \citenamefont {Gro\ss{}}, \citenamefont {Tr\"ager}, \citenamefont {Bykova},
  \citenamefont {Weigand}, \citenamefont {Zelent}, \citenamefont {Goering},
  \citenamefont {Sch\"utz}, \citenamefont {Krawczyk}, \citenamefont
  {Stobiecki}, \citenamefont {Dubowik},\ and\ \citenamefont
  {Gr\"afe}}]{lisiecki.19}%
  \BibitemOpen
  \bibfield  {author} {\bibinfo {author} {\bibfnamefont {F.}~\bibnamefont
  {Lisiecki}}, \bibinfo {author} {\bibfnamefont {J.}~\bibnamefont {Rych\l{}y}},
  \bibinfo {author} {\bibfnamefont {P.}~\bibnamefont {Ku\ifmmode~\acute{s}\else
  \'{s}\fi{}wik}}, \bibinfo {author} {\bibfnamefont {H.}~\bibnamefont
  {G\l{}owi\ifmmode~\acute{n}\else \'{n}\fi{}ski}}, \bibinfo {author}
  {\bibfnamefont {J.~W.}\ \bibnamefont {K\l{}os}}, \bibinfo {author}
  {\bibfnamefont {F.}~\bibnamefont {Gro\ss{}}}, \bibinfo {author}
  {\bibfnamefont {N.}~\bibnamefont {Tr\"ager}}, \bibinfo {author}
  {\bibfnamefont {I.}~\bibnamefont {Bykova}}, \bibinfo {author} {\bibfnamefont
  {M.}~\bibnamefont {Weigand}}, \bibinfo {author} {\bibfnamefont
  {M.}~\bibnamefont {Zelent}}, \bibinfo {author} {\bibfnamefont {E.~J.}\
  \bibnamefont {Goering}}, \bibinfo {author} {\bibfnamefont {G.}~\bibnamefont
  {Sch\"utz}}, \bibinfo {author} {\bibfnamefont {M.}~\bibnamefont {Krawczyk}},
  \bibinfo {author} {\bibfnamefont {F.}~\bibnamefont {Stobiecki}}, \bibinfo
  {author} {\bibfnamefont {J.}~\bibnamefont {Dubowik}},\ and\ \bibinfo {author}
  {\bibfnamefont {J.}~\bibnamefont {Gr\"afe}},\ }\bibfield  {title} {\bibinfo
  {title} {Magnons in a quasicrystal: Propagation, extinction, and localization
  of spin waves in fibonacci structures},\ }\href
  {https://doi.org/10.1103/PhysRevApplied.11.054061} {\bibfield  {journal}
  {\bibinfo  {journal} {Phys. Rev. Appl.}\ }\textbf {\bibinfo {volume} {11}},\
  \bibinfo {pages} {054061} (\bibinfo {year} {2019})}\BibitemShut {NoStop}%
\bibitem [{\citenamefont {Josephson}(1962)}]{josephson.62}%
  \BibitemOpen
  \bibfield  {author} {\bibinfo {author} {\bibfnamefont {B.}~\bibnamefont
  {Josephson}},\ }\bibfield  {title} {\bibinfo {title} {Possible new effects in
  superconductive tunnelling},\ }\href
  {https://doi.org/https://doi.org/10.1016/0031-9163(62)91369-0} {\bibfield
  {journal} {\bibinfo  {journal} {Phys. Lett.}\ }\textbf {\bibinfo {volume}
  {1}},\ \bibinfo {pages} {251} (\bibinfo {year} {1962})}\BibitemShut {NoStop}%
\bibitem [{\citenamefont {Josephson}(1974)}]{josephson.74}%
  \BibitemOpen
  \bibfield  {author} {\bibinfo {author} {\bibfnamefont {B.~D.}\ \bibnamefont
  {Josephson}},\ }\bibfield  {title} {\bibinfo {title} {The discovery of
  tunnelling supercurrents},\ }\href
  {https://doi.org/10.1103/RevModPhys.46.251} {\bibfield  {journal} {\bibinfo
  {journal} {Rev. Mod. Phys.}\ }\textbf {\bibinfo {volume} {46}},\ \bibinfo
  {pages} {251} (\bibinfo {year} {1974})}\BibitemShut {NoStop}%
\bibitem [{\citenamefont {Likharev}(1979)}]{likharev.79}%
  \BibitemOpen
  \bibfield  {author} {\bibinfo {author} {\bibfnamefont {K.~K.}\ \bibnamefont
  {Likharev}},\ }\bibfield  {title} {\bibinfo {title} {Superconducting weak
  links},\ }\href {https://doi.org/10.1103/RevModPhys.51.101} {\bibfield
  {journal} {\bibinfo  {journal} {Rev. Mod. Phys.}\ }\textbf {\bibinfo {volume}
  {51}},\ \bibinfo {pages} {101} (\bibinfo {year} {1979})}\BibitemShut
  {NoStop}%
\bibitem [{\citenamefont {Golubov}\ \emph {et~al.}(2004)\citenamefont
  {Golubov}, \citenamefont {Kupriyanov},\ and\ \citenamefont
  {Il'ichev}}]{golubov.04}%
  \BibitemOpen
  \bibfield  {author} {\bibinfo {author} {\bibfnamefont {A.~A.}\ \bibnamefont
  {Golubov}}, \bibinfo {author} {\bibfnamefont {M.~Y.}\ \bibnamefont
  {Kupriyanov}},\ and\ \bibinfo {author} {\bibfnamefont {E.}~\bibnamefont
  {Il'ichev}},\ }\bibfield  {title} {\bibinfo {title} {The current-phase
  relation in josephson junctions},\ }\href
  {https://doi.org/10.1103/RevModPhys.76.411} {\bibfield  {journal} {\bibinfo
  {journal} {Rev. Mod. Phys.}\ }\textbf {\bibinfo {volume} {76}},\ \bibinfo
  {pages} {411} (\bibinfo {year} {2004})}\BibitemShut {NoStop}%
\bibitem [{\citenamefont {Ishii}(1970)}]{ishii.70}%
  \BibitemOpen
  \bibfield  {author} {\bibinfo {author} {\bibfnamefont {C.}~\bibnamefont
  {Ishii}},\ }\bibfield  {title} {\bibinfo {title} {{Josephson Currents through
  Junctions with Normal Metal Barriers}},\ }\href
  {https://doi.org/10.1143/PTP.44.1525} {\bibfield  {journal} {\bibinfo
  {journal} {Prog. Theor. Phys.}\ }\textbf {\bibinfo {volume} {44}},\ \bibinfo
  {pages} {1525} (\bibinfo {year} {1970})}\BibitemShut {NoStop}%
\bibitem [{\citenamefont {Affleck}\ \emph {et~al.}(2000)\citenamefont
  {Affleck}, \citenamefont {Caux},\ and\ \citenamefont
  {Zagoskin}}]{affleck.00}%
  \BibitemOpen
  \bibfield  {author} {\bibinfo {author} {\bibfnamefont {I.}~\bibnamefont
  {Affleck}}, \bibinfo {author} {\bibfnamefont {J.-S.}\ \bibnamefont {Caux}},\
  and\ \bibinfo {author} {\bibfnamefont {A.~M.}\ \bibnamefont {Zagoskin}},\
  }\bibfield  {title} {\bibinfo {title} {Andreev scattering and josephson
  current in a one-dimensional electron liquid},\ }\href
  {https://doi.org/10.1103/PhysRevB.62.1433} {\bibfield  {journal} {\bibinfo
  {journal} {Phys. Rev. B}\ }\textbf {\bibinfo {volume} {62}},\ \bibinfo
  {pages} {1433} (\bibinfo {year} {2000})}\BibitemShut {NoStop}%
\bibitem [{\citenamefont {Beenakker}(1992)}]{beenakker.92}%
  \BibitemOpen
  \bibfield  {author} {\bibinfo {author} {\bibfnamefont {C.~W.~J.}\
  \bibnamefont {Beenakker}},\ }\bibfield  {title} {\bibinfo {title} {Three
  ``universal'' mesoscopic josephson effects},\ }in\ \href@noop {} {\emph
  {\bibinfo {booktitle} {Transport Phenomena in Mesoscopic Systems}}},\
  \bibinfo {editor} {edited by\ \bibinfo {editor} {\bibfnamefont
  {H.}~\bibnamefont {Fukuyama}}\ and\ \bibinfo {editor} {\bibfnamefont
  {T.}~\bibnamefont {Ando}}}\ (\bibinfo  {publisher} {Springer Berlin
  Heidelberg},\ \bibinfo {address} {Berlin, Heidelberg},\ \bibinfo {year}
  {1992})\ pp.\ \bibinfo {pages} {235--253}\BibitemShut {NoStop}%
\bibitem [{\citenamefont {Nikoli\ifmmode~\acute{c}\else \'{c}\fi{}}\ \emph
  {et~al.}(2001)\citenamefont {Nikoli\ifmmode~\acute{c}\else \'{c}\fi{}},
  \citenamefont {Freericks},\ and\ \citenamefont {Miller}}]{nikolic.01}%
  \BibitemOpen
  \bibfield  {author} {\bibinfo {author} {\bibfnamefont {B.~K.}\ \bibnamefont
  {Nikoli\ifmmode~\acute{c}\else \'{c}\fi{}}}, \bibinfo {author} {\bibfnamefont
  {J.~K.}\ \bibnamefont {Freericks}},\ and\ \bibinfo {author} {\bibfnamefont
  {P.}~\bibnamefont {Miller}},\ }\bibfield  {title} {\bibinfo {title}
  {Intrinsic reduction of josephson critical current in short ballistic sns
  weak links},\ }\href {https://doi.org/10.1103/PhysRevB.64.212507} {\bibfield
  {journal} {\bibinfo  {journal} {Phys. Rev. B}\ }\textbf {\bibinfo {volume}
  {64}},\ \bibinfo {pages} {212507} (\bibinfo {year} {2001})}\BibitemShut
  {NoStop}%
\bibitem [{\citenamefont {Sonin}(2024)}]{sonin.24}%
  \BibitemOpen
  \bibfield  {author} {\bibinfo {author} {\bibfnamefont {E.}~\bibnamefont
  {Sonin}},\ }\bibfield  {title} {\bibinfo {title} {Andreev reflection, andreev
  states, and long ballistic sns junction},\ }\bibfield  {journal} {\bibinfo
  {journal} {J. Low Temp. Phys.}\ }\href
  {https://doi.org/10.1007/s10909-024-03137-7} {10.1007/s10909-024-03137-7}
  (\bibinfo {year} {2024})\BibitemShut {NoStop}%
\bibitem [{\citenamefont {Sandberg}(2022)}]{sandberg.22}%
  \BibitemOpen
  \bibfield  {author} {\bibinfo {author} {\bibfnamefont {A.}~\bibnamefont
  {Sandberg}},\ }\emph {\bibinfo {title} {Quasicrystal nanowires in
  SNS-junctions}},\ \href@noop {} {Master's thesis},\ \bibinfo  {school}
  {Uppsala University} (\bibinfo {year} {2022})\BibitemShut {NoStop}%
\bibitem [{\citenamefont {Su}\ \emph {et~al.}(1979)\citenamefont {Su},
  \citenamefont {Schrieffer},\ and\ \citenamefont {Heeger}}]{su.shrieffer.79}%
  \BibitemOpen
  \bibfield  {author} {\bibinfo {author} {\bibfnamefont {W.~P.}\ \bibnamefont
  {Su}}, \bibinfo {author} {\bibfnamefont {J.~R.}\ \bibnamefont {Schrieffer}},\
  and\ \bibinfo {author} {\bibfnamefont {A.~J.}\ \bibnamefont {Heeger}},\
  }\bibfield  {title} {\bibinfo {title} {Solitons in polyacetylene},\ }\href
  {https://doi.org/10.1103/PhysRevLett.42.1698} {\bibfield  {journal} {\bibinfo
   {journal} {Phys. Rev. Lett.}\ }\textbf {\bibinfo {volume} {42}},\ \bibinfo
  {pages} {1698} (\bibinfo {year} {1979})}\BibitemShut {NoStop}%
\bibitem [{\citenamefont {Harper}(1955)}]{harper.55}%
  \BibitemOpen
  \bibfield  {author} {\bibinfo {author} {\bibfnamefont {P.~G.}\ \bibnamefont
  {Harper}},\ }\bibfield  {title} {\bibinfo {title} {Single band motion of
  conduction electrons in a uniform magnetic field},\ }\href
  {https://doi.org/10.1088/0370-1298/68/10/304} {\bibfield  {journal} {\bibinfo
   {journal} {Proc. Phys. Soc. London A}\ }\textbf {\bibinfo {volume} {68}},\
  \bibinfo {pages} {874} (\bibinfo {year} {1955})}\BibitemShut {NoStop}%
\bibitem [{\citenamefont {Aubry}\ and\ \citenamefont
  {Andr{\'e}}(1980)}]{aubry.andre.80}%
  \BibitemOpen
  \bibfield  {author} {\bibinfo {author} {\bibfnamefont {S.}~\bibnamefont
  {Aubry}}\ and\ \bibinfo {author} {\bibfnamefont {G.}~\bibnamefont
  {Andr{\'e}}},\ }\bibfield  {title} {\bibinfo {title} {Analyticity breaking
  and anderson localization in incommensurate lattices},\ }\href@noop {}
  {\bibfield  {journal} {\bibinfo  {journal} {Ann. Israel Phys. Soc}\ }\textbf
  {\bibinfo {volume} {3}},\ \bibinfo {pages} {18} (\bibinfo {year}
  {1980})}\BibitemShut {NoStop}%
\bibitem [{\citenamefont {Gordon}\ \emph {et~al.}(1997)\citenamefont {Gordon},
  \citenamefont {Jitomirskaya}, \citenamefont {Last},\ and\ \citenamefont
  {Simon}}]{gordon.97}%
  \BibitemOpen
  \bibfield  {author} {\bibinfo {author} {\bibfnamefont {A.~Y.}\ \bibnamefont
  {Gordon}}, \bibinfo {author} {\bibfnamefont {S.}~\bibnamefont
  {Jitomirskaya}}, \bibinfo {author} {\bibfnamefont {Y.}~\bibnamefont {Last}},\
  and\ \bibinfo {author} {\bibfnamefont {B.}~\bibnamefont {Simon}},\ }\bibfield
   {title} {\bibinfo {title} {{Duality and singular continuous spectrum in the
  almost Mathieu equation}},\ }\href {https://doi.org/10.1007/BF02392693}
  {\bibfield  {journal} {\bibinfo  {journal} {Acta Math.}\ }\textbf {\bibinfo
  {volume} {178}},\ \bibinfo {pages} {169 } (\bibinfo {year}
  {1997})}\BibitemShut {NoStop}%
\bibitem [{Note1()}]{Note1}%
  \BibitemOpen
  \bibinfo {note} {{{We diagonalize the Hamiltonian both with and without
  self-consistency using a homogeneous bulk magnitude $|\Delta _j|$. Our
  criterion for self-consistency is that from self-consistency iteration number
  $m$ to ${m+i}$, the global relative error $\epsilon _{\protect \mathrm {G}} =
  \left \| \Delta _{m+1}-\Delta _{m} \right \|_2/\left \|\Delta _{m} \right
  \|_2 < 10^{-7}$, where we use a Polyak convergence accelerator~\cite
  {SuperConga:2023}. We generally find negligible influence of self-consistency
  apart from a small quantitative difference, which can be reproduced without
  self-consistency by simply adjusting $\Delta _0$.}}}\BibitemShut {Stop}%
\bibitem [{\citenamefont {Zhu}(2016)}]{zhu.16}%
  \BibitemOpen
  \bibfield  {author} {\bibinfo {author} {\bibfnamefont {J.-X.}\ \bibnamefont
  {Zhu}},\ }\href {https://doi.org/10.1007/978-3-319-31314-6} {\emph {\bibinfo
  {title} {Bogoliubov-de Gennes method and its applications}}},\ \bibinfo
  {series} {Lecture Notes in Physics}, Vol.\ \bibinfo {volume} {924}\ (\bibinfo
   {publisher} {Springer International Publishing},\ \bibinfo {address}
  {Switzerland},\ \bibinfo {year} {2016})\BibitemShut {NoStop}%
\bibitem [{\citenamefont {Shumeiko}\ \emph {et~al.}(1997)\citenamefont
  {Shumeiko}, \citenamefont {Bratus},\ and\ \citenamefont
  {Wendin}}]{shumeiko.97}%
  \BibitemOpen
  \bibfield  {author} {\bibinfo {author} {\bibfnamefont {V.~S.}\ \bibnamefont
  {Shumeiko}}, \bibinfo {author} {\bibfnamefont {E.~N.}\ \bibnamefont
  {Bratus}},\ and\ \bibinfo {author} {\bibfnamefont {G.}~\bibnamefont
  {Wendin}},\ }\bibfield  {title} {\bibinfo {title} {{Scattering theory of
  superconductive tunneling in quantum junctions}},\ }\href
  {https://doi.org/10.1063/1.593475} {\bibfield  {journal} {\bibinfo  {journal}
  {Low Temp. Phys.}\ }\textbf {\bibinfo {volume} {23}},\ \bibinfo {pages} {181}
  (\bibinfo {year} {1997})}\BibitemShut {NoStop}%
\bibitem [{\citenamefont {Bena}(2012)}]{bena.12}%
  \BibitemOpen
  \bibfield  {author} {\bibinfo {author} {\bibfnamefont {C.}~\bibnamefont
  {Bena}},\ }\bibfield  {title} {\bibinfo {title} {Metamorphosis and taxonomy
  of andreev bound states},\ }\href
  {https://doi.org/10.1140/epjb/e2012-30133-0} {\bibfield  {journal} {\bibinfo
  {journal} {Eur. Phys. J. B}\ }\textbf {\bibinfo {volume} {85}},\ \bibinfo
  {pages} {196} (\bibinfo {year} {2012})}\BibitemShut {NoStop}%
\bibitem [{\citenamefont {Jackiw}\ and\ \citenamefont
  {Rebbi}(1976)}]{jackiw.rebbi.76}%
  \BibitemOpen
  \bibfield  {author} {\bibinfo {author} {\bibfnamefont {R.}~\bibnamefont
  {Jackiw}}\ and\ \bibinfo {author} {\bibfnamefont {C.}~\bibnamefont {Rebbi}},\
  }\bibfield  {title} {\bibinfo {title} {Solitons with fermion number
  \textonehalf{}},\ }\href {https://doi.org/10.1103/PhysRevD.13.3398}
  {\bibfield  {journal} {\bibinfo  {journal} {Phys. Rev. D}\ }\textbf {\bibinfo
  {volume} {13}},\ \bibinfo {pages} {3398} (\bibinfo {year}
  {1976})}\BibitemShut {NoStop}%
\bibitem [{\citenamefont {Furusaki}\ \emph {et~al.}(1992)\citenamefont
  {Furusaki}, \citenamefont {Takayanagi},\ and\ \citenamefont
  {Tsukada}}]{furusaki.92}%
  \BibitemOpen
  \bibfield  {author} {\bibinfo {author} {\bibfnamefont {A.}~\bibnamefont
  {Furusaki}}, \bibinfo {author} {\bibfnamefont {H.}~\bibnamefont
  {Takayanagi}},\ and\ \bibinfo {author} {\bibfnamefont {M.}~\bibnamefont
  {Tsukada}},\ }\bibfield  {title} {\bibinfo {title} {Josephson effect of the
  superconducting quantum point contact},\ }\href
  {https://doi.org/10.1103/PhysRevB.45.10563} {\bibfield  {journal} {\bibinfo
  {journal} {Phys. Rev. B}\ }\textbf {\bibinfo {volume} {45}},\ \bibinfo
  {pages} {10563} (\bibinfo {year} {1992})}\BibitemShut {NoStop}%
\bibitem [{\citenamefont {Kouwenhoven}\ \emph {et~al.}(1990)\citenamefont
  {Kouwenhoven}, \citenamefont {Hekking}, \citenamefont {van Wees},
  \citenamefont {Harmans}, \citenamefont {Timmering},\ and\ \citenamefont
  {Foxon}}]{kouwenhoven.90}%
  \BibitemOpen
  \bibfield  {author} {\bibinfo {author} {\bibfnamefont {L.~P.}\ \bibnamefont
  {Kouwenhoven}}, \bibinfo {author} {\bibfnamefont {F.~W.~J.}\ \bibnamefont
  {Hekking}}, \bibinfo {author} {\bibfnamefont {B.~J.}\ \bibnamefont {van
  Wees}}, \bibinfo {author} {\bibfnamefont {C.~J. P.~M.}\ \bibnamefont
  {Harmans}}, \bibinfo {author} {\bibfnamefont {C.~E.}\ \bibnamefont
  {Timmering}},\ and\ \bibinfo {author} {\bibfnamefont {C.~T.}\ \bibnamefont
  {Foxon}},\ }\bibfield  {title} {\bibinfo {title} {Transport through a finite
  one-dimensional crystal},\ }\href
  {https://doi.org/10.1103/PhysRevLett.65.361} {\bibfield  {journal} {\bibinfo
  {journal} {Phys. Rev. Lett.}\ }\textbf {\bibinfo {volume} {65}},\ \bibinfo
  {pages} {361} (\bibinfo {year} {1990})}\BibitemShut {NoStop}%
\bibitem [{\citenamefont {Chatterjee}\ \emph {et~al.}(2022)\citenamefont
  {Chatterjee}, \citenamefont {Ansaloni}, \citenamefont {Rasmussen},
  \citenamefont {Brovang}, \citenamefont {Fedele}, \citenamefont
  {Bohuslavskyi}, \citenamefont {Krause},\ and\ \citenamefont
  {Kuemmeth}}]{chatterjee.22}%
  \BibitemOpen
  \bibfield  {author} {\bibinfo {author} {\bibfnamefont {A.}~\bibnamefont
  {Chatterjee}}, \bibinfo {author} {\bibfnamefont {F.}~\bibnamefont
  {Ansaloni}}, \bibinfo {author} {\bibfnamefont {T.}~\bibnamefont {Rasmussen}},
  \bibinfo {author} {\bibfnamefont {B.}~\bibnamefont {Brovang}}, \bibinfo
  {author} {\bibfnamefont {F.}~\bibnamefont {Fedele}}, \bibinfo {author}
  {\bibfnamefont {H.}~\bibnamefont {Bohuslavskyi}}, \bibinfo {author}
  {\bibfnamefont {O.}~\bibnamefont {Krause}},\ and\ \bibinfo {author}
  {\bibfnamefont {F.}~\bibnamefont {Kuemmeth}},\ }\bibfield  {title} {\bibinfo
  {title} {Autonomous estimation of high-dimensional coulomb diamonds from
  sparse measurements},\ }\href
  {https://doi.org/10.1103/PhysRevApplied.18.064040} {\bibfield  {journal}
  {\bibinfo  {journal} {Phys. Rev. Appl.}\ }\textbf {\bibinfo {volume} {18}},\
  \bibinfo {pages} {064040} (\bibinfo {year} {2022})}\BibitemShut {NoStop}%
\bibitem [{\citenamefont {Kuzmanovski}\ \emph {et~al.}(2023)\citenamefont
  {Kuzmanovski}, \citenamefont {Souto}, \citenamefont {Wong},\ and\
  \citenamefont {Balatsky}}]{kuzmanovski.23}%
  \BibitemOpen
  \bibfield  {author} {\bibinfo {author} {\bibfnamefont {D.}~\bibnamefont
  {Kuzmanovski}}, \bibinfo {author} {\bibfnamefont {R.~S.}\ \bibnamefont
  {Souto}}, \bibinfo {author} {\bibfnamefont {P.~J.}\ \bibnamefont {Wong}},\
  and\ \bibinfo {author} {\bibfnamefont {A.~V.}\ \bibnamefont {Balatsky}},\
  }\href@noop {} {\bibinfo {title} {Mobile topological su-schrieffer-heeger
  soliton in a josephson metamaterial}} (\bibinfo {year} {2023}),\ \Eprint
  {https://arxiv.org/abs/2312.03456} {arXiv:2312.03456 [cond-mat.supr-con]}
  \BibitemShut {NoStop}%
\bibitem [{\citenamefont {Splitthoff}\ \emph {et~al.}(2024)\citenamefont
  {Splitthoff}, \citenamefont {Belo}, \citenamefont {Jin}, \citenamefont {Li},
  \citenamefont {Greplova},\ and\ \citenamefont {Andersen}}]{splitthoff.24}%
  \BibitemOpen
  \bibfield  {author} {\bibinfo {author} {\bibfnamefont {L.~J.}\ \bibnamefont
  {Splitthoff}}, \bibinfo {author} {\bibfnamefont {M.~C.}\ \bibnamefont
  {Belo}}, \bibinfo {author} {\bibfnamefont {G.}~\bibnamefont {Jin}}, \bibinfo
  {author} {\bibfnamefont {Y.}~\bibnamefont {Li}}, \bibinfo {author}
  {\bibfnamefont {E.}~\bibnamefont {Greplova}},\ and\ \bibinfo {author}
  {\bibfnamefont {C.~K.}\ \bibnamefont {Andersen}},\ }\href
  {https://arxiv.org/abs/2404.07371} {\bibinfo {title} {{Gate-tunable phase
  transition in a bosonic Su-Schrieffer-Heeger chain}}} (\bibinfo {year}
  {2024}),\ \Eprint {https://arxiv.org/abs/2404.07371} {arXiv:2404.07371
  [quant-ph]} \BibitemShut {NoStop}%
\bibitem [{\citenamefont {Zubchenko}\ \emph {et~al.}(2024)\citenamefont
  {Zubchenko}, \citenamefont {Middlebrooks}, \citenamefont {Rasmussen},
  \citenamefont {Lausen}, \citenamefont {Kuemmeth}, \citenamefont
  {Chatterjee},\ and\ \citenamefont {Zwolak}}]{zubchenko.24}%
  \BibitemOpen
  \bibfield  {author} {\bibinfo {author} {\bibfnamefont {A.}~\bibnamefont
  {Zubchenko}}, \bibinfo {author} {\bibfnamefont {D.}~\bibnamefont
  {Middlebrooks}}, \bibinfo {author} {\bibfnamefont {T.}~\bibnamefont
  {Rasmussen}}, \bibinfo {author} {\bibfnamefont {L.}~\bibnamefont {Lausen}},
  \bibinfo {author} {\bibfnamefont {F.}~\bibnamefont {Kuemmeth}}, \bibinfo
  {author} {\bibfnamefont {A.}~\bibnamefont {Chatterjee}},\ and\ \bibinfo
  {author} {\bibfnamefont {J.~P.}\ \bibnamefont {Zwolak}},\ }\href
  {https://arxiv.org/abs/2407.20061} {\bibinfo {title} {Autonomous
  bootstrapping of quantum dot devices}} (\bibinfo {year} {2024}),\ \Eprint
  {https://arxiv.org/abs/2407.20061} {arXiv:2407.20061 [cond-mat.mes-hall]}
  \BibitemShut {NoStop}%
\bibitem [{\citenamefont {Amundsen}\ \emph {et~al.}(2024)\citenamefont
  {Amundsen}, \citenamefont {{Juri\v{c}i\'{c}}},\ and\ \citenamefont
  {Ouassou}}]{amundsen.24}%
  \BibitemOpen
  \bibfield  {author} {\bibinfo {author} {\bibfnamefont {M.}~\bibnamefont
  {Amundsen}}, \bibinfo {author} {\bibfnamefont {V.}~\bibnamefont
  {{Juri\v{c}i\'{c}}}},\ and\ \bibinfo {author} {\bibfnamefont {J.~A.}\
  \bibnamefont {Ouassou}},\ }\href@noop {} {\bibinfo {title} {Josephson effect
  in a fractal geometry}} (\bibinfo {year} {2024}),\ \Eprint
  {https://arxiv.org/abs/2404.01373} {arXiv:2404.01373 [cond-mat.supr-con]}
  \BibitemShut {NoStop}%
\bibitem [{\citenamefont {Holmvall}\ \emph {et~al.}(2023)\citenamefont
  {Holmvall}, \citenamefont {Wall~Wennerdal}, \citenamefont {Håkansson},
  \citenamefont {Stadler}, \citenamefont {Shevtsov}, \citenamefont
  {Löfwander},\ and\ \citenamefont {Fogelström}}]{SuperConga:2023}%
  \BibitemOpen
  \bibfield  {author} {\bibinfo {author} {\bibfnamefont {P.}~\bibnamefont
  {Holmvall}}, \bibinfo {author} {\bibfnamefont {N.}~\bibnamefont
  {Wall~Wennerdal}}, \bibinfo {author} {\bibfnamefont {M.}~\bibnamefont
  {Håkansson}}, \bibinfo {author} {\bibfnamefont {P.}~\bibnamefont {Stadler}},
  \bibinfo {author} {\bibfnamefont {O.}~\bibnamefont {Shevtsov}}, \bibinfo
  {author} {\bibfnamefont {T.}~\bibnamefont {Löfwander}},\ and\ \bibinfo
  {author} {\bibfnamefont {M.}~\bibnamefont {Fogelström}},\ }\bibfield
  {title} {\bibinfo {title} {{SuperConga: An open-source framework for
  mesoscopic superconductivity}},\ }\href {https://doi.org/10.1063/5.0100324}
  {\bibfield  {journal} {\bibinfo  {journal} {Appl. Phys. Rev.}\ }\textbf
  {\bibinfo {volume} {10}},\ \bibinfo {pages} {011317} (\bibinfo {year}
  {2023})}\BibitemShut {NoStop}%
\end{thebibliography}%

\end{document}